\documentclass[twocolumn]{aastex63}

\submitjournal{ApJ}
\usepackage{amsmath}
\usepackage{booktabs}
\usepackage{float}
\usepackage{longtable}

\begin{document}
\title{Searching for Anomalies in the ZTF Catalog of Periodic Variable Stars}

\author[0000-0003-2776-082X]{Ho-Sang Chan}
\affiliation{Department of Physics and Institute of Theoretical Physics, The Chinese University of Hong Kong, Shatin, N.T., Hong Kong}
\affiliation{Center for Computational Astrophysics, Flatiron Institute, New York, NY 10010, USA}
\author[0000-0002-5814-4061]{V. Ashley Villar}
\affiliation{Department of Astronomy \& Astrophysics, The Pennsylvania State University, University Park, PA 16802, USA}
\affiliation{Institute for Computational \& Data Sciences, The Pennsylvania State University, University Park, PA, USA}
\affiliation{Institute for Gravitation and the Cosmos, The Pennsylvania State University, University Park, PA 16802, USA
}
\author[0000-0002-0814-3378]{Siu-Hei Cheung}
\affiliation{Department of Physics and Institute of Theoretical Physics, The Chinese University of Hong Kong, Shatin, N.T., Hong Kong}
\affiliation{Center for Computational Astrophysics, Flatiron Institute, New York, NY 10010, USA}
\author{Shirley Ho}
\affiliation{Center for Computational Astrophysics, Flatiron Institute, New York, NY 10010, USA; New York University, New York, NY 10011; Princeton University, Princeton, NJ 08540; Carnegie Mellon University, Pittsburgh, PA 15213}
\author[0000-0002-7296-6547]{Anna J. G. O’Grady}
\affiliation{David A. Dunlap Department of Astronomy and Astrophysics, University of Toronto, 50 St. George Street, Toronto, Ontario, M5S 3H4 Canada}
\affiliation{Dunlap Institute for Astronomy and Astrophysics, University of Toronto, 50 St. George Street, Toronto, Ontario, M5S 3H4, Canada}
\author[0000-0001-7081-0082]{Maria R. Drout}
\affiliation{David A. Dunlap Department of Astronomy and Astrophysics, University of Toronto, 50 St. George Street, Toronto, Ontario, M5S 3H4 Canada}
\affiliation{The Observatories of the Carnegie Institution for Science, 813 Santa Barbara St., Pasadena, CA 91101, USA}
\author[0000-0002-6718-9472]{Mathieu Renzo}
\affiliation{Columbia Astrophysics Laboratory, Columbia University, New York, New York 10027, USA}
\affiliation{Center for Computational Astrophysics, Flatiron Institute, New York, NY 10010, USA}

\begin{abstract}
Periodic variables illuminate the physical processes of stars throughout their lifetime. Wide-field surveys continue to increase our discovery rates of periodic variable stars. Automated approaches are essential to identify interesting periodic variable stars for multi-wavelength and spectroscopic follow-up. Here, we present a novel unsupervised machine learning approach to hunt for anomalous periodic variables using phase-folded light curves presented in the Zwicky Transient Facility Catalogue of Periodic Variable Stars by \citet{Chen_2020}. We use a convolutional variational autoencoder to learn a low dimensional latent representation, and we search for anomalies within this latent dimension via an isolation forest. We identify anomalies with irregular variability. Most of the top anomalies are likely highly variable Red Giants or Asymptotic Giant Branch stars concentrated in the Milky Way galactic disk; a fraction of the identified anomalies are more consistent with Young Stellar Objects. Detailed spectroscopic follow-up observations are encouraged to reveal the nature of these anomalies.
\end{abstract}


\section{Introduction} \label{sec:intro}
Astronomical variables consist of extragalactic and galactic sources which vary on timescales of seconds to years. A division between intrinsic and extrinsic variables is typically made according to the origin of their variability. The former is due to the physical changes of the object itself, while the latter involves geometrical effects such as eclipsing \citep{Eyer_2008}. Some periodic variable stars, such as Cepheids, exhibit a strong correlation between their pulsation period and luminosity, which make them excellent candidates for cosmic distance measurements \citep{Zgirski_2017} and calibrations \citep{10.1093/mnras/stx1600}. Cepheid variables are also used to measure the Hubble constant \citep{Pierce1994}, and to test theories of stellar pulsation \citep{Pietrzynski2010}. In addition, variable star populations are used to constrain the metallicity and distance of globular clusters \citep{2012A&A...548A..92K}, and tidal effects of cataclysmic variable star binaries have been used to constrain modified gravity theories \citep{2021ApJ...910...23B}. Throughout the past few decades, optical surveys, such as the Optical Gravitational Lensing Experiment (OGLE; \citealt{2008AcA....58...69U, 2015AcA....65....1U}), the Palomar Transient Factory (PTF; \citealt{Law_2009}), Gaia \citep{gaiacollaboration}, and the Zwicky Transient Facility (ZTF; \citealt{Chen_2020}), have made great progress in expanding our understanding of periodic variable stars. There are now nearly a hundred classification catalogs distinguishing different types of periodic variable stars \citep{Samus2017}, and millions of known variable stars (see, for example, \citealt{2019MNRAS.486.1907J}). \\

Wide-field, untargeted surveys continue to exponentially increase our discovery rates of galactic and extra-galactic transients, including periodic variable stars \citep{Henrion2013}. ZTF, a wide-field survey that saw first light in 2017, is conducted with the Palomar $48$ inch Schmidt telescope which has a $47$ degree field of view \citep{2019PASP..131a8002B}. ZTF has produced more than a hundred million light curves across the $gri$ filters. The second data release of the ZTF spans a total observation period of March 2018 - June 2019. The public Northern-equatorial sky survey covers $\approx 20,000$ deg$^{2}$ at a typical cadence of three days. ZTF is, in many ways, a path finder for the upcoming Legacy Survey of Space and Time (LSST) conducted by the Vera Rubin Observatory is expected to commence in early 2024 \citep{2019eeu..confE..23G}. With LSSt, billions of variable stars will be observed \citep{Jacklin_2017, Ivezi__2019} at a cadence of several days across $ugriz$ filters. \\

In this work, we focus on the periodic variables discovered with these wide-field surveys such as ZTF. In particular, \citet{Chen_2020} utilize the ZTF Data Release 2 archive to search for and to classify new periodic variables down to a $r$-band magnitude of $\sim 20.6$. They find a total of $781,602$ periodic variables, of which $621,702$ are newly discovered. These variables and their properties are published as the ZTF Catalog of Periodic Variable Stars (ZTF CPVS). They measure several observable features, including the variable star period, the phase difference between $g$- and $r$-bands, amplitude, absolute Wesenheit magnitude, and adjusted $R^{2}$ (which represents how well data are fitted by the Fourier reconstruction). They then classify the periodic variable stars with linear cuts in this measured feature space. They report a resulting misclassification rate of $2\%$, and a period accuracy of $99\%$, when compared against the ATLAS \citep{2018AJ....156..241H}, WISE \citep{2018ApJS..237...28C}, ASAS-AN \citep{2018MNRAS.477.3145J}, and the CATALINA \citep{Drake_2014, 2017MNRAS.469.3688D} catalog. Here, we will use this catalog to search for the most unusual periodic variables discovered with ZTF.

\subsection{Anomaly Detection through Machine Learning in Astronomy} \label{subsec:machine}
Given the large number of events being detected from wide-field surveys, it is reasonable to expect \textit{anomalous} periodic variables which defy expectations. These may potentially contain information about new physics. Indeed, the discoveries of anomalous periodic variables have been challenging our understanding to the chemical composition of stars \citep{10.1093/mnras/stab2065, Niemczura2017}, the stellar evolution models \citep{2017A&A...604A..29G}, Galactic metallicity \citep{2018SerAJ.197...13J}, the physics of accretion and mass transfer \citep{10.1093/mnras/stx811, 2021arXiv210615756K}, the physics of Active Galactic Nuclei \citep{2021arXiv210607660S}, and even the formation of low mass white dwarf \citep{Masuda_2019}. \\

Anomaly detection has long played a key role in scientific discovery within astrophysics. Examples include the discovery of dark matter \citep{1970ApJ...159..379R}, Type Iax supernova \citep{Li_2001}, anomalies in the CMB temperature anisotropies \citep{PhysRevD.77.023534}, galaxies lacking dark matter \citep{vanDokkum2018} and quasi-stellar objects \citep{Massey_2019}. In recent years, machine learning has arisen as an automatic and promising method to look for \textit{out-of-distribution} \citep{10.1145/1541880.1541882} anomalous astronomical objects. \citet{FUSTES20131530} used a Self-Organizing Map plus post-processed spectral information to better understand outliers found in the Gaia astronomical surveys. By using a Wasserstein generative adversarial network and t-Distributed Stochastic Neighbour Embedding as a dimensionality reduction algorithm, \citet{10.1093/mnras/staa1647} identified anomalous galaxies. \citet{2021arXiv210502434S} used also a Wasserstein generative adversarial network, but with a convolutional autoencoder (C-VAE), as a dimensionality reduction algorithm to search for abnormal galaxy images in the Hyper Suprime-Cam survey. \\

In addition, the physical process of astronomical objects would often be reflected in their optical variation. Therefore, searching anomalous astronomical objects through their light curves could be an alternative and robust means of anomaly detection. By training a Random Forest on the raw light curves extracted from the SDSS surveys, \citet{10.1093/mnras/stw3021} found several rare events that are of great interest to the astronomical community, including H$\delta$-strong galaxies, outflows and shocks, supernovae, galaxy-lenses, and double-peaked emission lines. \citet{Nun_2016} performed dimensionality reduction on the raw light curves provided by the MACHO catalog, and they search for anomalies through a weighted combination of the five following outliers detection algorithms: the k-Nearest Neighbors, Random Forest, Joint Probability approaches, Local Correlation Integral, and Learned Probability Distribution. They identify anomalies that belong to rare classes: novae, He-burning, and Red Giant stars, while other outlier light curves identified have no available information associated with them. \citet{Zhang_2018} proposed to look for anomalous transient via a Long Short-term Memory neural network. \citet{10.1093/mnras/stz1528} suggested searching for anomalous X-ray transients using a variational autoencoder. \citet{2021MNRAS.502.5147M} extracted features from light curves given by the ZTF Data Release 3 and searched for anomalies using the isolation forest, Local Outlier Factor, Gaussian Mixture Model, and One-class Support Vector Machines. From this search, they identified 23 non-cataloged astrophysical events of interest. \citet{10.1093/mnras/stz2362} pre-processed raw light curves from the Open Supernova Catalog \citep{2017ApJ...835...64G}, and they searched for anomalies directly using an isolation forest. In addition, they performed dimensionality reduction using the t-Distributed Stochastic Neighbour Embedding. They found 27 peculiar objects, including super-luminous supernovae, abnormal Type Ia supernovae, unusual Type II supernovae, Active Galactic Nucleus, and binary microlensing events. Finally, \citet{2021arXiv210312102V} searched for anomalous extragalactic transients in the latent space created by a variational recurrent autoencoder neural network, finding that such a method can detect rare superluminous and pair-instability supernovae. \\

Previous studies on periodic variables that utilize machine learning focus primarily on classifications \citep{Jamal_2020, 10.1093/mnras/stab1248, 2018NatAs...2..151N}, and deep generative modeling \citep{2020arXiv200507773M}. Although \citet{2021MNRAS.502.5147M} searched for anomalous transient detected with ZTF, they did not specifically focus on periodic variables. Here, we aim to provide an anomaly detection algorithm to effectively search for anomalous periodic variables detected with ZTF. Our methods presented here are general enough to be applied to LSST-like datastreams in the future. The article is organized as follows: Section \ref{sec:method} introduces the procedure of our data pre-processing, machine learning pipelines, and the method of anomaly searches. We present and discuss our results in Section \ref{sec:results}, and in Section \ref{sec:conclusion} we conclude our study.

\section{Methodology} \label{sec:method}
\begin{figure*}[ht!]
\gridline{\fig{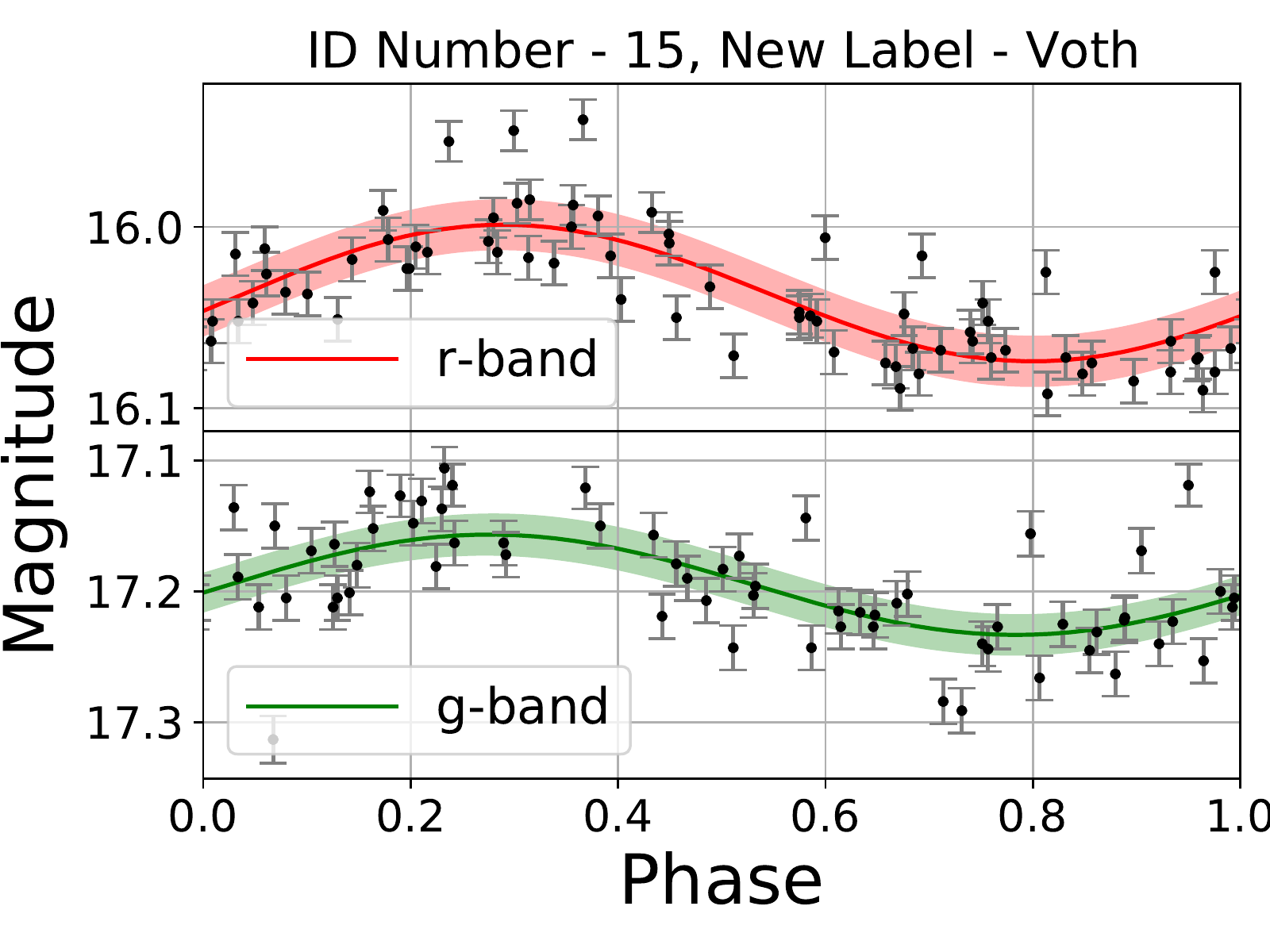}{0.3\textwidth}{(a) BY Draconis (BYDra)}
          \fig{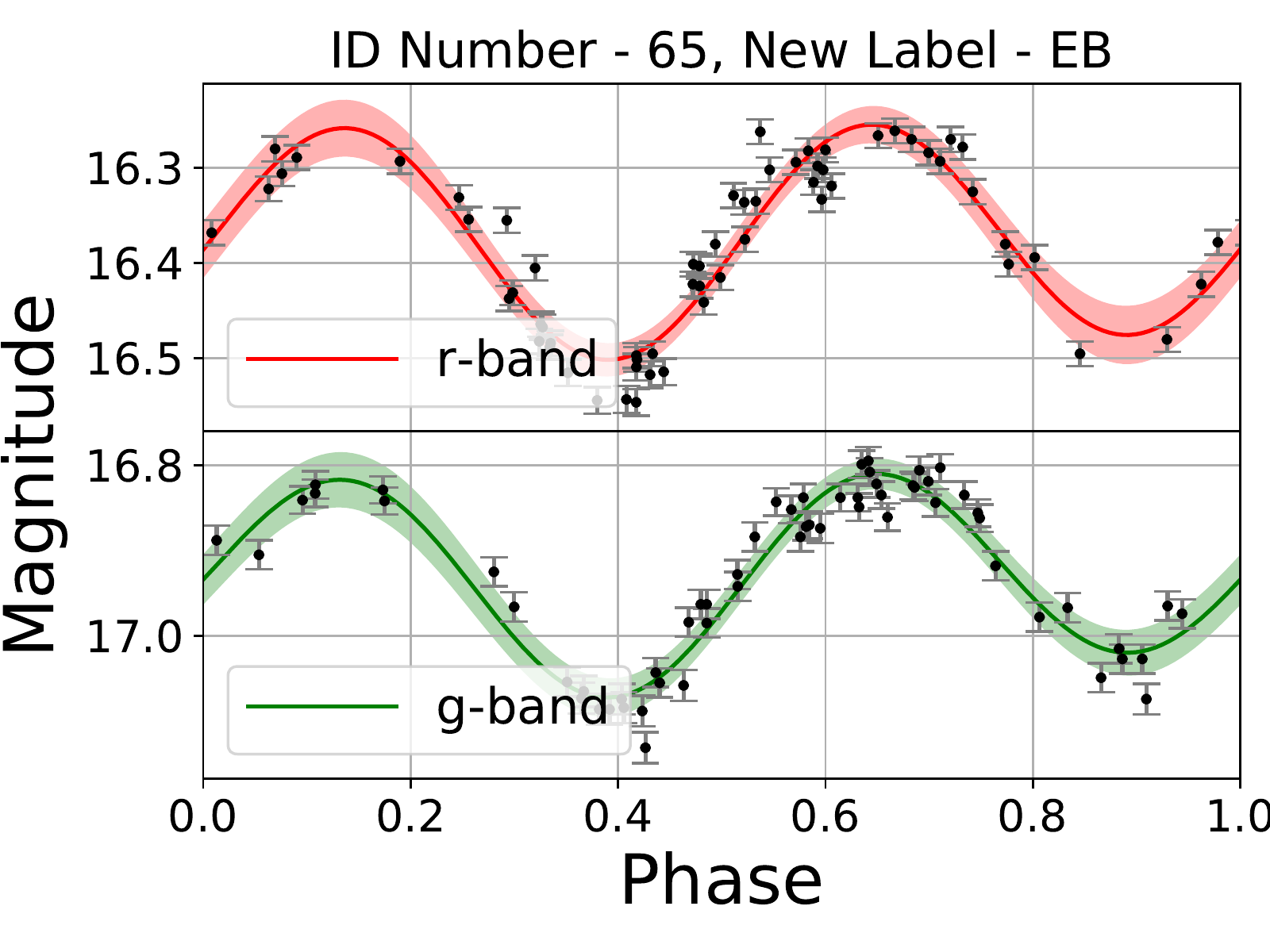}{0.3\textwidth}{(b) Eclipsing W UMa (EW)}
          \fig{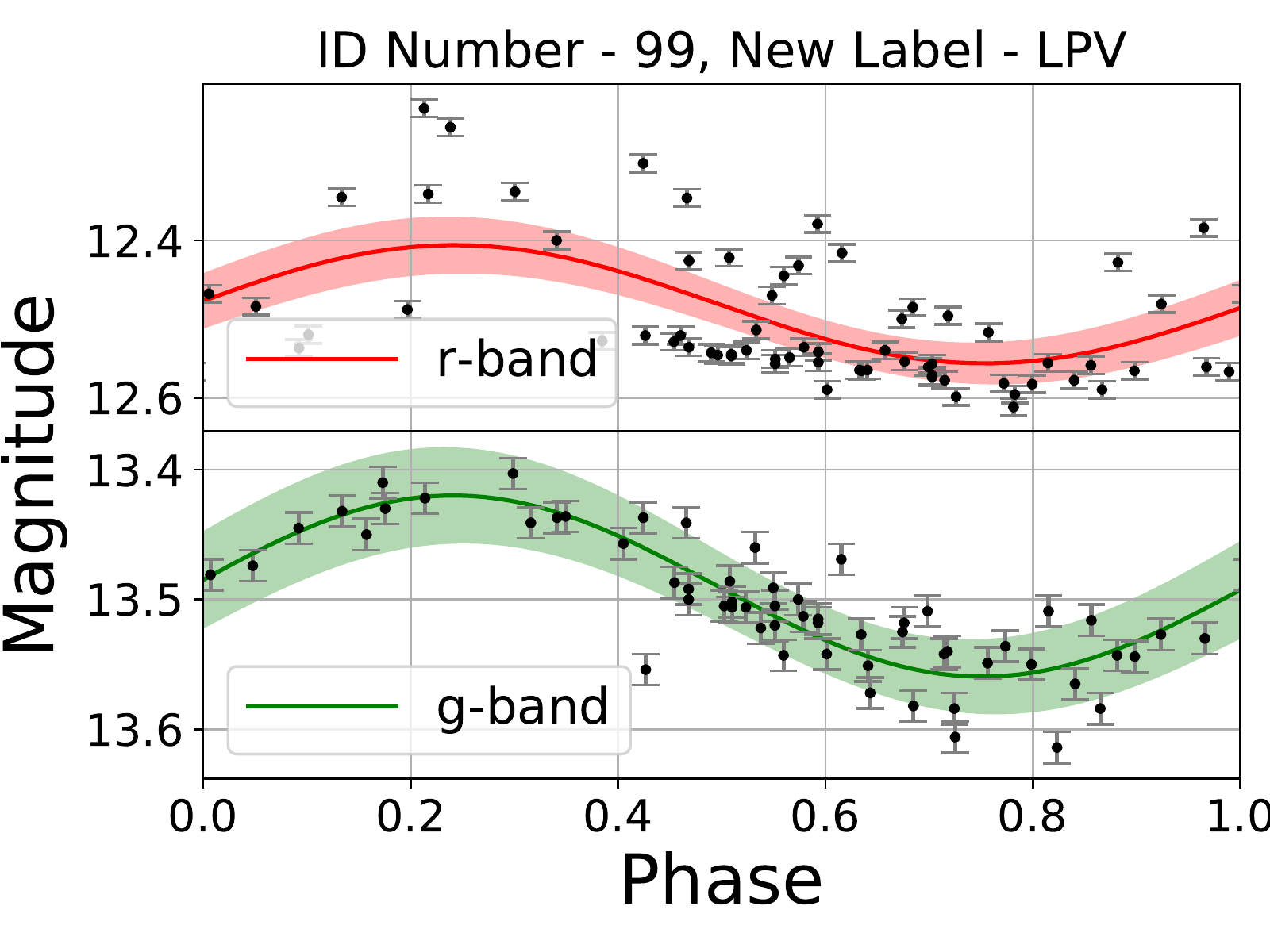}{0.3\textwidth}{(c) Semi-Regular (SR)}
          }
\gridline{\fig{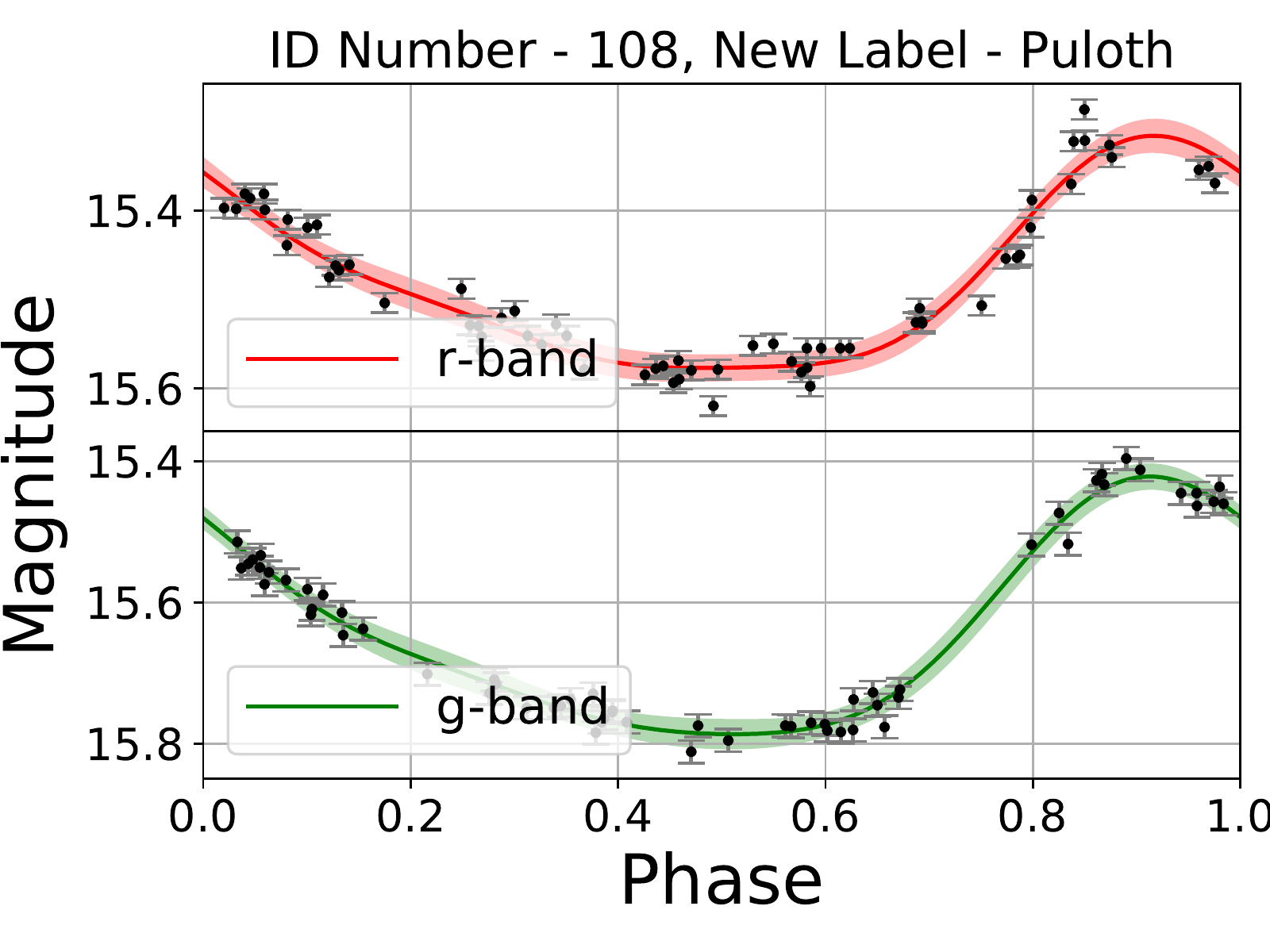}{0.3\textwidth}{(d) Delta Scuti (DSCT)}
          \fig{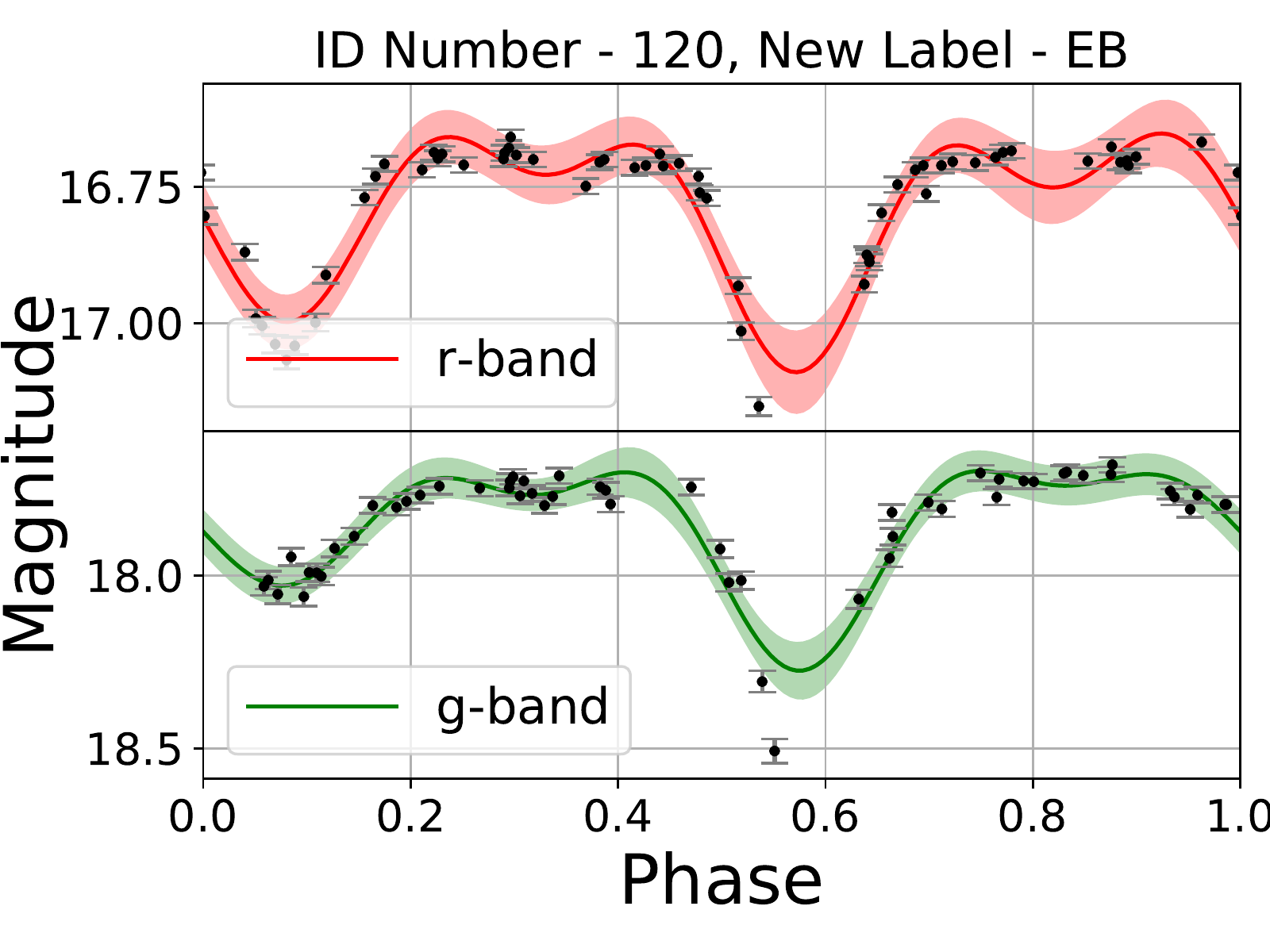}{0.3\textwidth}{(e) Eclipsing Algol (EA)}
          \fig{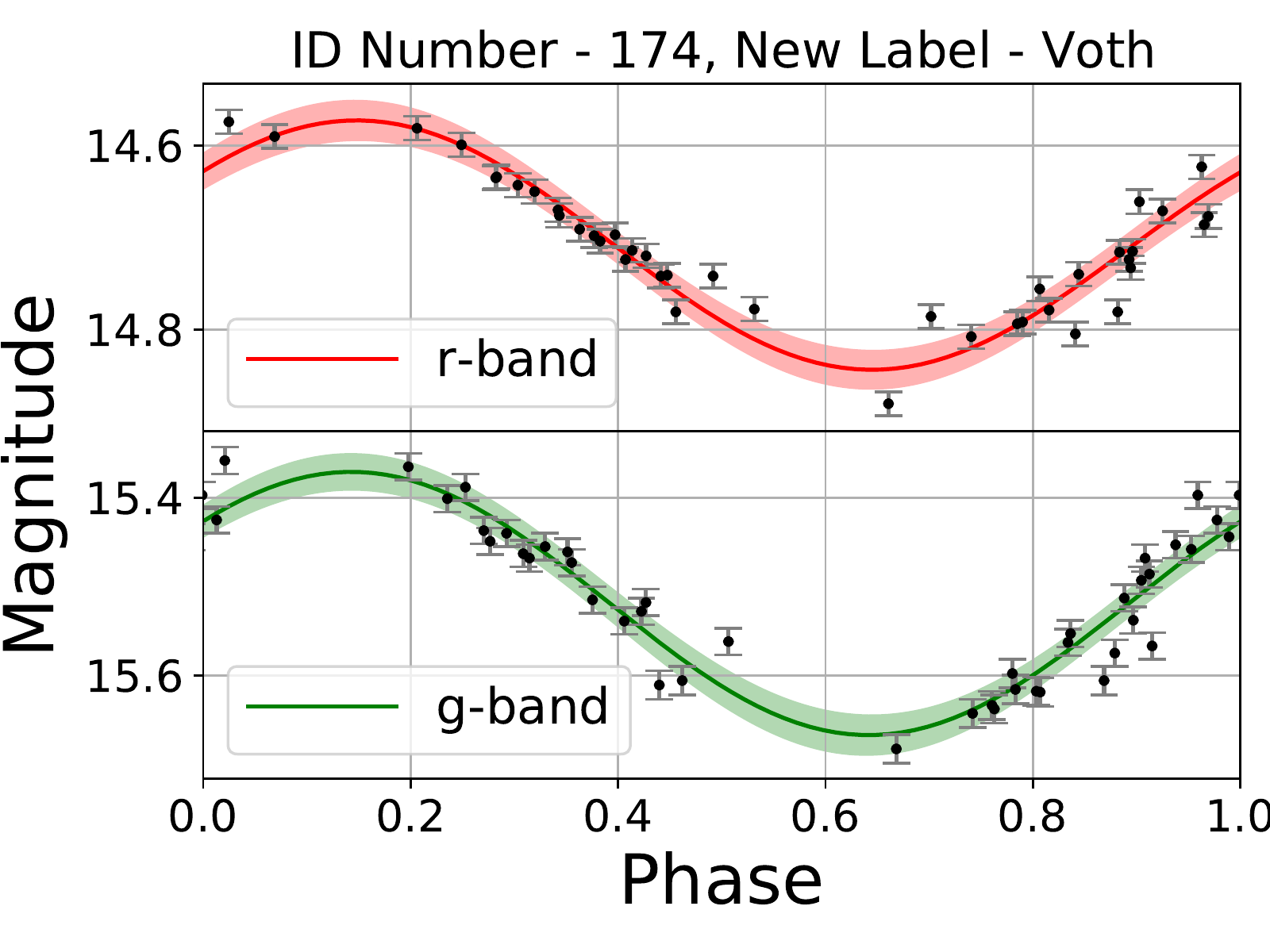}{0.3\textwidth}{(f) RS Canum Venaticorum (RSCVN)}
          }
\gridline{\fig{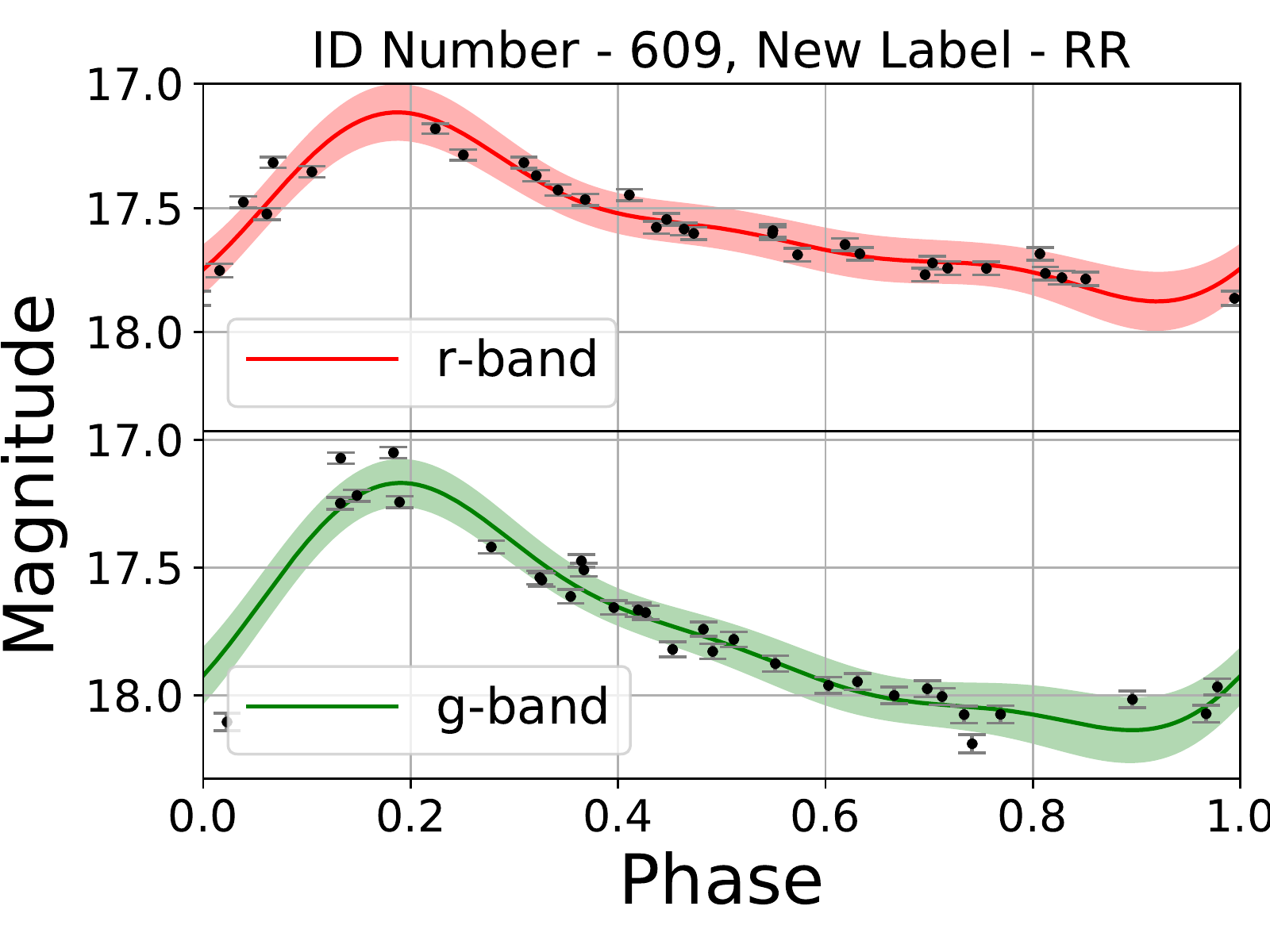}{0.3\textwidth}{(g) RR Lyrae (RR)}
          \fig{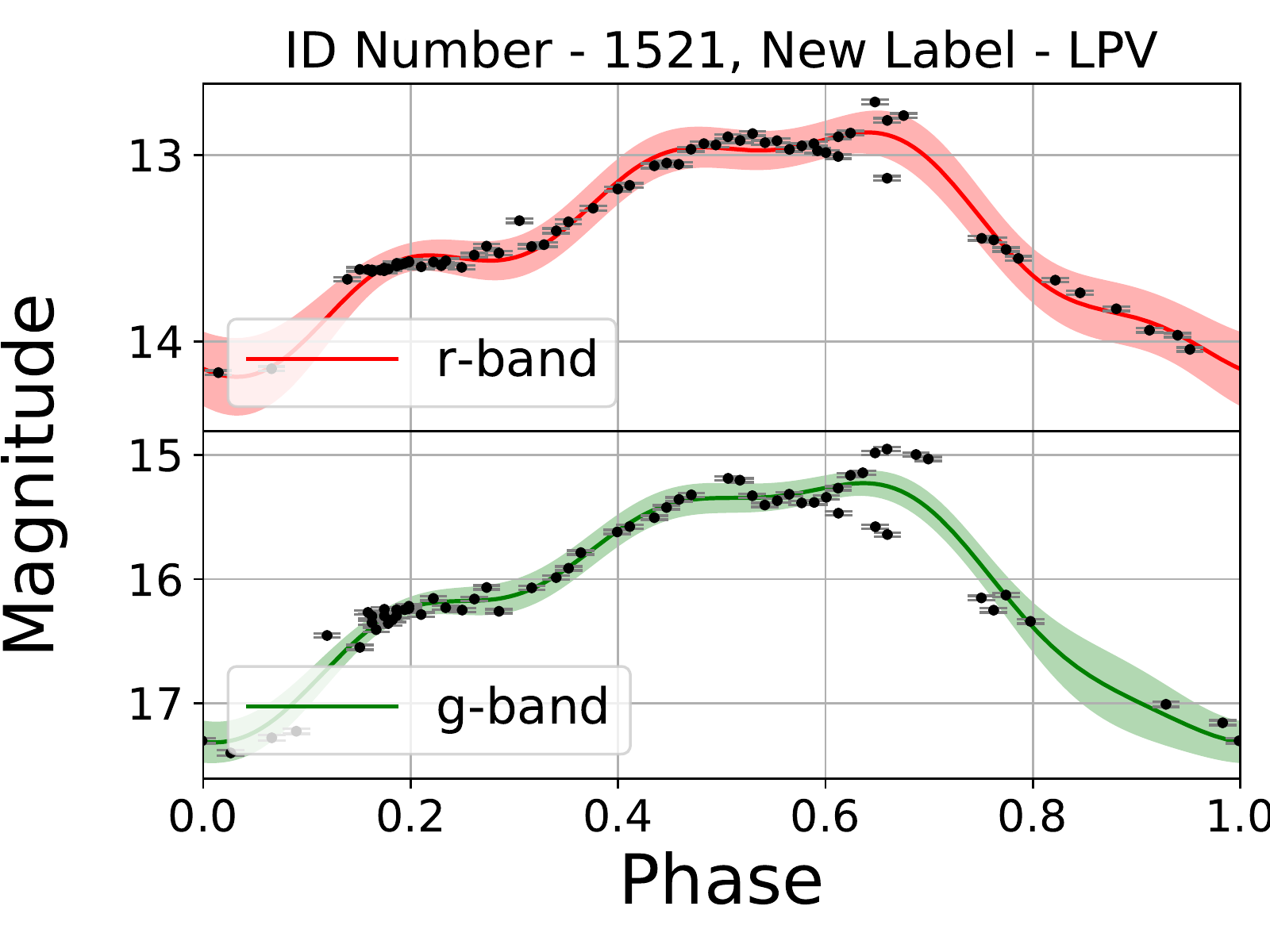}{0.3\textwidth}{(h) Mira}
          \fig{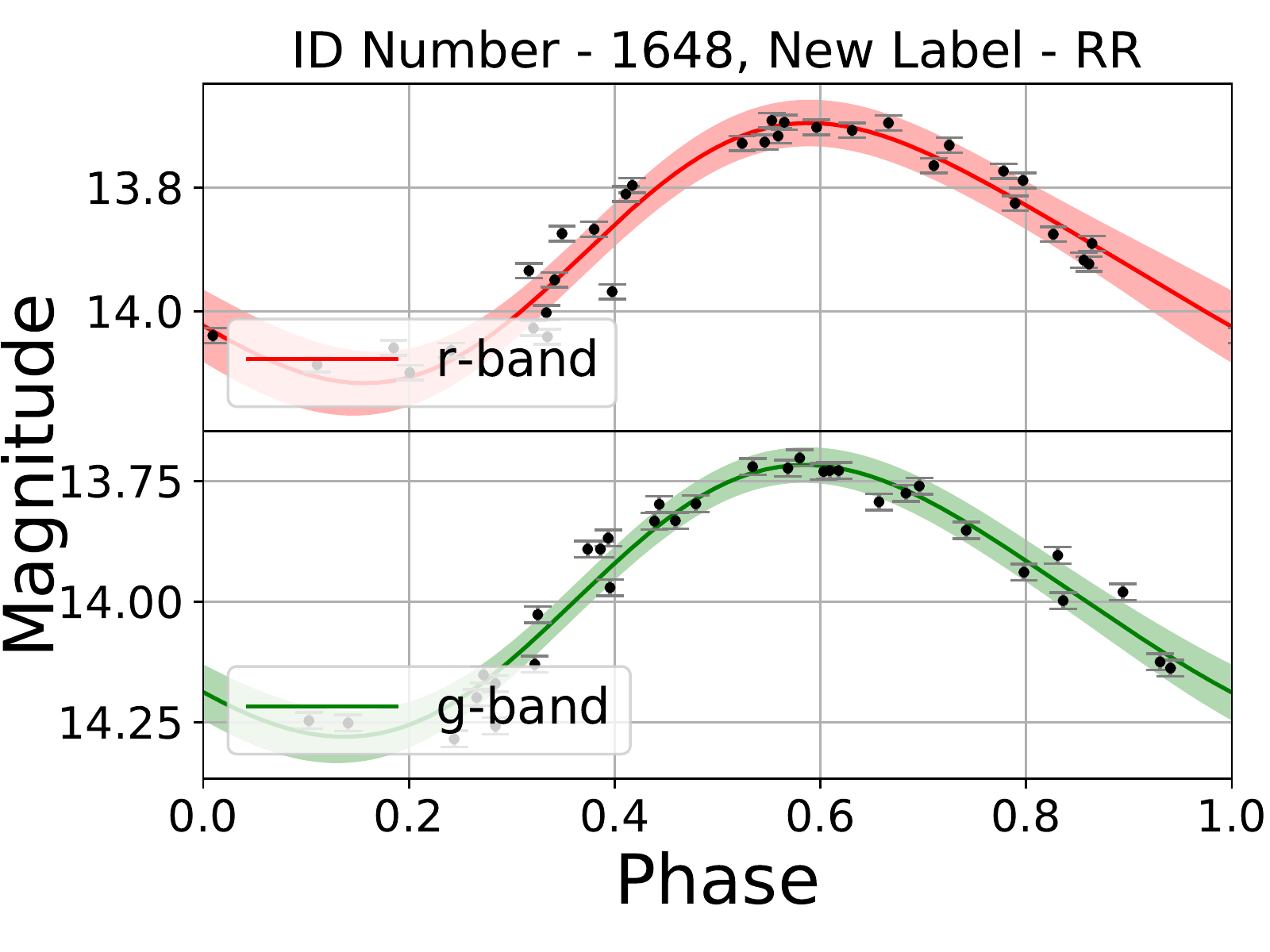}{0.3\textwidth}{(i) RR Lyrae Type C (RRc)}
          }
\gridline{\fig{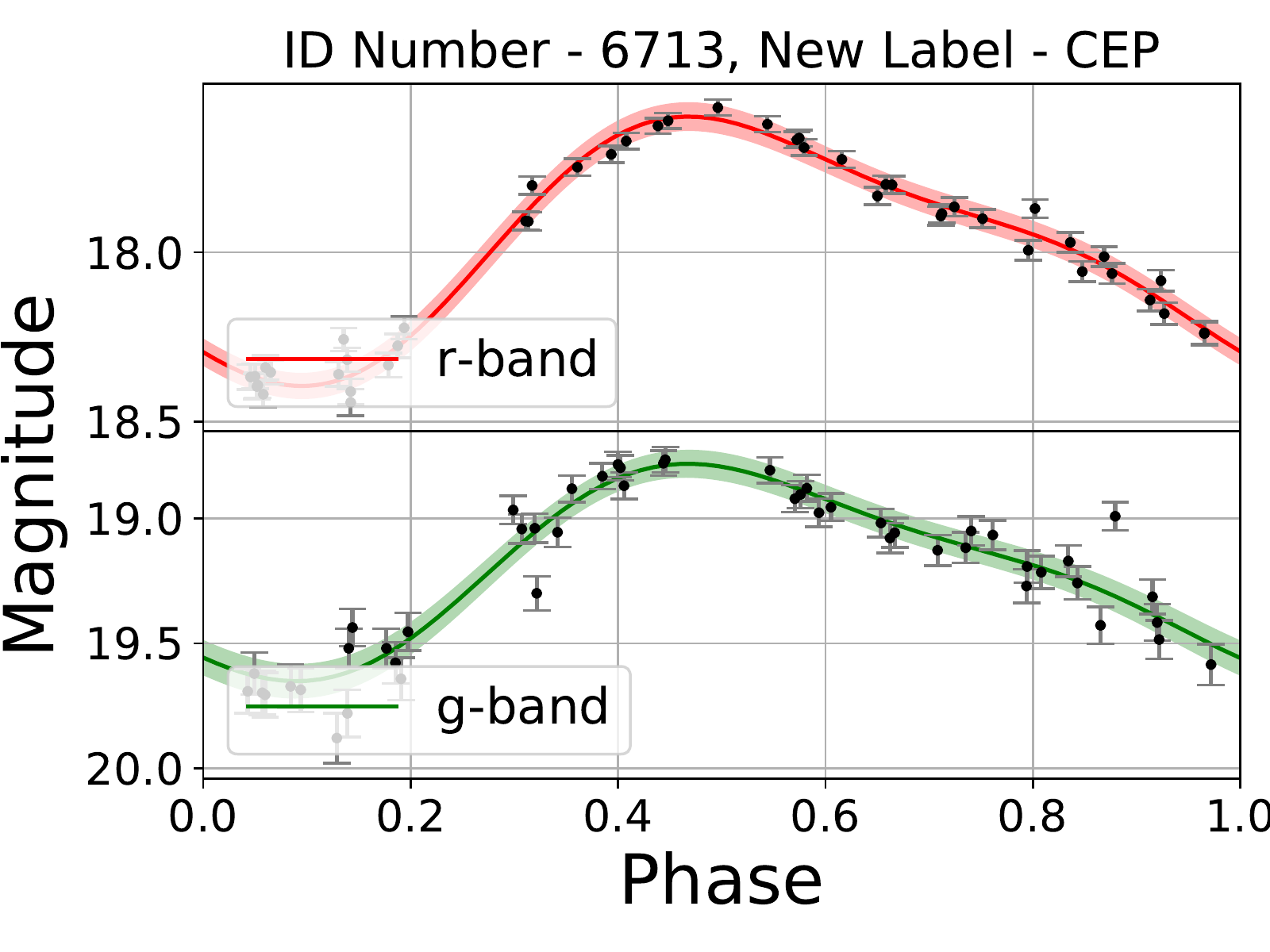}{0.3\textwidth}{(j) Type II Cepheid (CEPII)}
          \fig{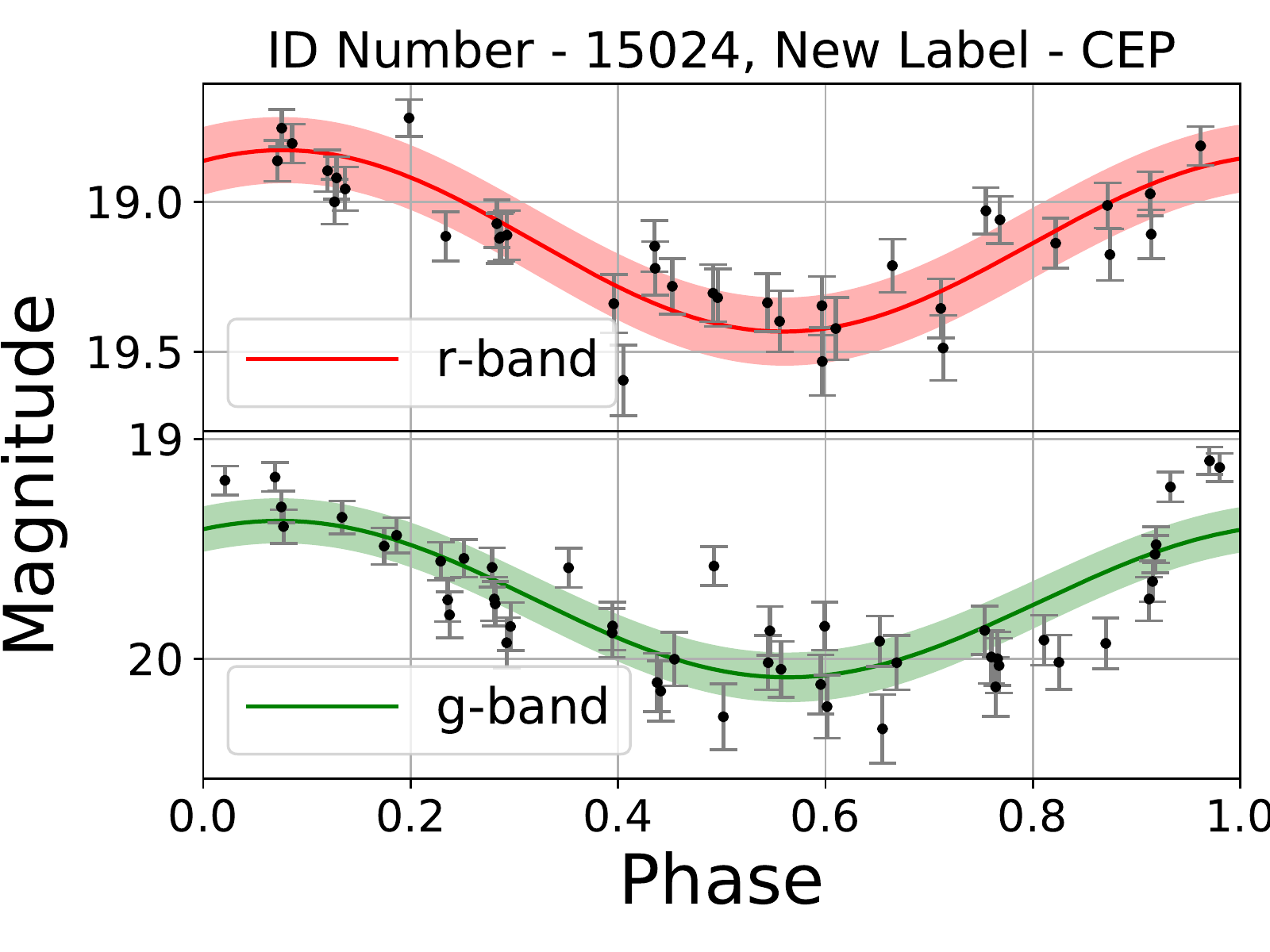}{0.3\textwidth}{(k) Cepheid (CEP)}
          }
\caption{Example plots of the MGPR interpolation for different types of periodic variable stars. The shaded region represents $2\sigma$ from the mean function. We provide the ID numbers and classes from the ZTF CPVS \citep{Chen_2020}, as well as class labels (see Appendix \ref{app:newlabels} for more details) from Cheung et al. (2021, in prep). \label{fig:grbandgpr}}
\end{figure*}

\subsection{Data Pre-processing} \label{subsec:data}
\begin{figure}[ht!]
	\centering
	\includegraphics[width=1.0\linewidth]{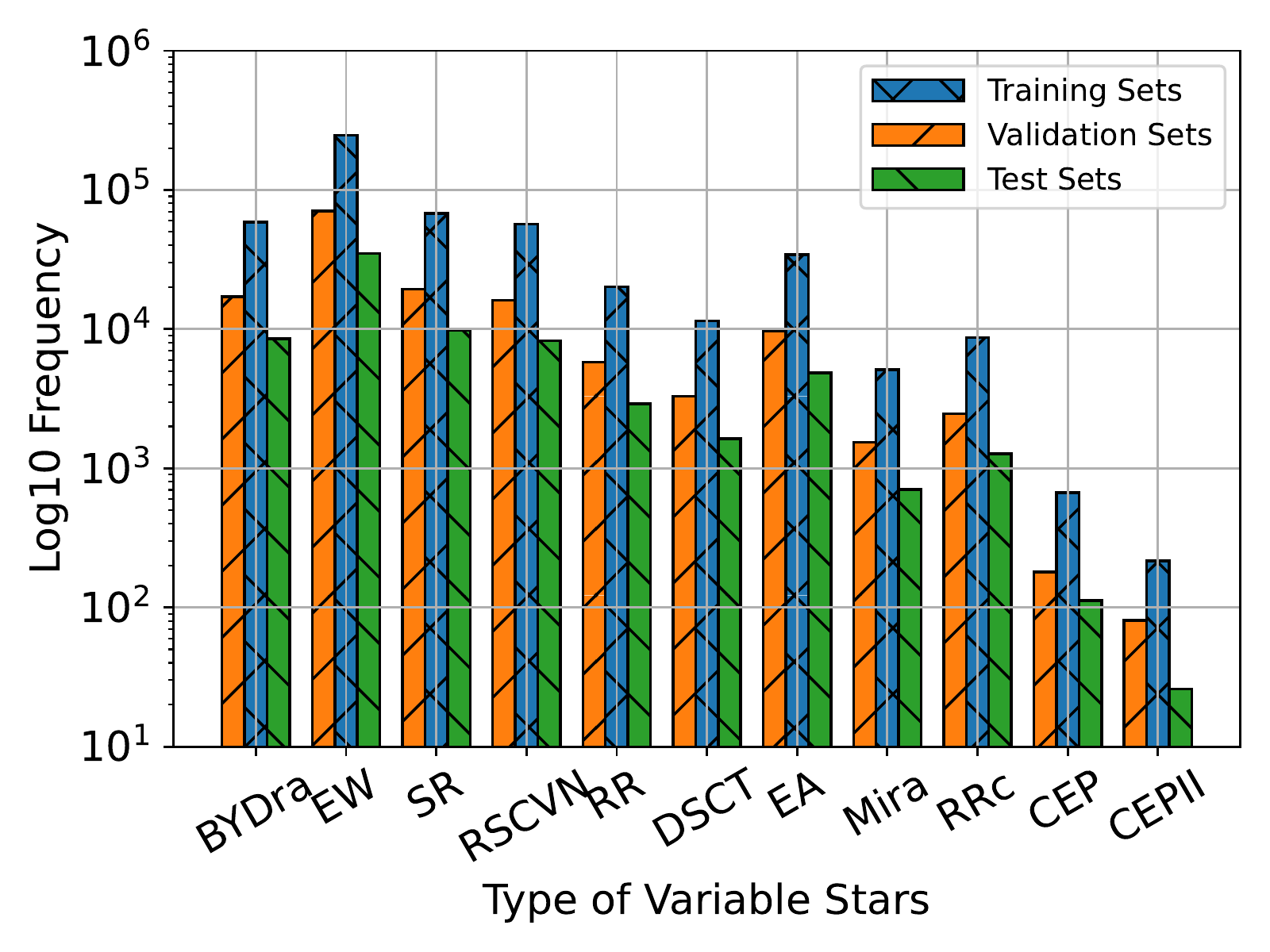}
	\caption{Distribution of the training, validation, and test sets for different categories of periodic variable stars. The classes are not uniformly distributed. However, their distributions are similar among the training, validation, and test sets. See Appendix \ref{app:briefreview} for the description of these class labels and a brief review of periodic variable stars. \label{fig:statistic}}
\end{figure}
We extract $730,184$ out of $781,602$ available periodic variable star light curves in the ZTF CPVS \citep{Chen_2020} which contain at least one $2.5\sigma$ and $3\sigma$ detection in $g$- and $r$-band, respectively. We phase-fold their light curves in both bands using their common period provided by Table 2 in \citet{Chen_2020} using the following
\begin{equation}
    \phi = \frac{t}{\tau} - \Big[\frac{t}{\tau}\Big]
\end{equation}
Here, $\phi$ is the phase, $t$ is the time of measurement in Modified Julian Date (MJD), and $\tau$ is the common period in both filters. The big brackets in the second term of the equation represent the floor function. \\

We use the open-source software \texttt{george} \citep{dan_foreman_mackey_2014_11989} to interpolate the data points using multivariate Gaussian process regression (MGPR) \citep{10.1093/mnras/stz2362, 2021arXiv210604370Q} in a normalized wavelength space and temporal phase space. Our MGPR uses a mean function $\eta$ and a covariance function $K$ (also called the kernel). The most general kernel is the radial basis function
\begin{equation}
    K(\vec{r}, \vec{r}') = \text{exp}\Big(-\frac{|\vec{\phi} - \vec{\phi}'|^{2}}{l_{\phi}^{2}}\Big)\text{exp}\Big(-\frac{|\vec{\lambda} - \vec{\lambda}'|^{2}}{l_{\lambda}^{2}}\Big)
\end{equation}
Here, $\vec{r} = (\vec{\phi}, \vec{\lambda})$ is a high dimensional position vector and $\vec{\lambda}$ is the normalized wavelength. We normalize the wavelengths using
\begin{equation}
    \lambda \rightarrow \frac{\lambda - \lambda_{l}}{\lambda_{h} - \lambda_{l}}
\end{equation}
Here, $\lambda_{h}$ and $\lambda_{l}$ represent the upper and lower limit of wavelengths for the visual band. $l_{\phi}$ and $l_{\lambda}$ measure the correlation along the phase and wavelength direction, respectively. We also multiply the radial basis function by the constant kernel
\begin{equation}
    K(\vec{r}, \vec{r}') = C
\end{equation}
in which we assume a constant variance over all data points. Additionally, we add the white noise kernel
\begin{equation}
K(\vec{r}, \vec{r}') =
\begin{cases}
      \delta & \text{if $\vec{r} = \vec{r}'$}, \\
      0 & \text{otherwise}
   \end{cases}
\end{equation}
Here, $\delta$ measures the white noise level of the observed fluxes; in developing this methodology, we find that the addition of the white noise kernel greatly improves the MGPR results. \\

We minimize hyperparameters $l_{\phi}$, $l_{\lambda}$, $C$, and $\delta$ with respect to the (negative) log-likelihood function \citep{books/lib/RasmussenW06}. We adopt the optimizing algorithm based on the Limited-Memory Broyden–Fletcher–Goldfarb–Shanno method \citep{53712fe04a3448cfb8598b14afab59b3} provided by the \texttt{Scipy} package to minimize the objective function. We show representative interpolated light curve in $g$- and $r$-band in Figure \ref{fig:grbandgpr} for reference. The mean function is set to be $\eta = 0$. Our empirical findings suggest that either assuming a non-zero $\eta$ or fitting $\eta$ as a hyper-parameter will yield unrealistic results, such as the divergence of the correlation length $l_{\phi}$\footnote{Occasionally, $l_{\phi}$ may diverge or approach zero. In these cases, we fine-tune the initial guess for $l_{\phi}$ until we obtain a reasonable interpolation.}. To enforce periodic boundary conditions, we repeat and stack each band of light curves by $3$ times. The interpolation results of the middle stack will be taken as the output. Upon fitting the hyper-parameters, we generate light curves with a total of $160$ evenly spaced data points along the phase direction for both bands. This number is close to the average number of detection for the ZTF CPVS.\\

Finally, we transform each band of the light curves using a standard-scaler transformation
\begin{equation}
    \begin{aligned}
    m \rightarrow \frac{m - \bar{m}}{\sigma} \\
    \delta m \rightarrow \frac{\delta m}{\sigma} \\
    \end{aligned}
\end{equation}
Here, $m$ represents the $g$-/$r$- band (apparent) magnitudes, $\delta m$ represents the magnitude uncertainties, $\bar{m}$ represents the average magnitude, and $\sigma$ is the standard deviation of the $g$-/$r$- band magnitudes. This normalization is a common form of data pre-processing that speeds up training. We split the data into training to validation to test ratio of $7:2:1$, and we have shown the distribution of these three sets of data in a histogram in Figure \ref{fig:statistic}.

\subsection{The Convolutional Variational Autoencoder} \label{subsec:vae}
\begin{figure*}[htb!]
	\centering
	\includegraphics[width=1.0\linewidth]{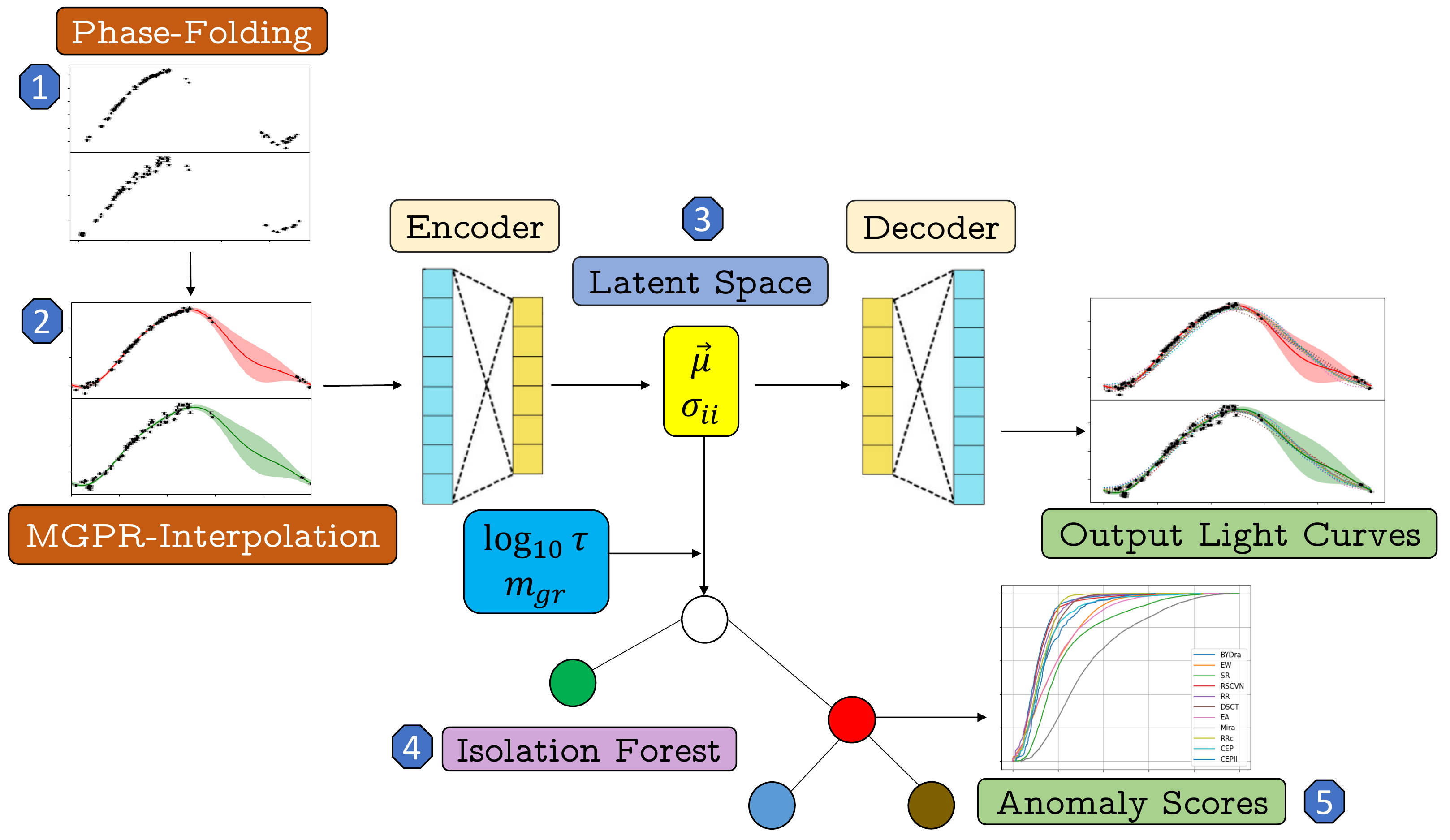}
	\caption{The anomaly detection pipeline, which we enumerate here. (1) We first phase-fold the raw detection data in $g$- and $r$- bands using the common period from ZTF PVSC. (2) The phased-folded data is then interpolated using the MGPR to generate light curves in both bands. (3) The interpolated light curves are then passed to the encoder to generate latent vectors $\vec{\mu}$ and (diagonal) covariance matrix $\sigma^{2}I_{n}$ elements using a variational, convolutional architecture. (4) Next, log$_{10}\tau$ and $m_{\text{gr}}$ are appended to $\vec{\mu}$ and all are passed to the isolation forest. (5) We rank the events according to the anomaly scores from the isolation forest, with the highest-scoring events being the most anomalous. \label{fig:pipeline}}
\end{figure*}
The ZTF CPVS contains over $700$ thousand light curves with approximately 100 observations per event. To conduct an anomaly search, a set of \textit{features} must be extracted from each light curve. Here, we will use a neural network to learn a latent feature space. Since the light curves are the results of physical processes in the source, it is reasonable to expect that they could be characterized by a small number of hidden variables (latent features) that determine the physics sampled by observations (e.g., temperature, mass, or the age of the stars). To learn this low-dimensional latent space, we adopt the variational autoencoder \citep[VAE,][]{2013arXiv1312.6114K}. A VAE is a generative, probabilistic model in which the neural network learns a distribution rather than a discrete value of the latent features for each event. The distribution is assumed to be a multivariate Gaussian with a mean $\vec{\mu}$ and a diagonal covariance matrix $\sigma^{2}I_{n}$ \citep{MAL-056}. VAEs are shown to be promising in generalizing features and generating new data from learned factors \citep{boutin2020iterative, 2020arXiv200812595G}. In addition, VAEs are also capable of learning a smooth latent representation, in which physically similar events naturally cluster in the latent space \citep{7803340, 2020arXiv200801487O}, making the identification of anomalies easier than traditional autoencoders (which can have discontinuous latent spaces). \\

The VAE consists of an encoder and decoder. We use convolutional layers to construct the encoder/decoder because we wish to capture the temporal and wavelength correlations of the light curves. Following the approach of \citet{Villar_2020} and \citet{2021arXiv210312102V}, we stack both bands of interpolated light curves horizontally to form an ``image" of size $2 \times 160$. Inspired by the work of \citet{10.1093/mnras/stab1248}, we choose to apply periodic padding to the ``images" during the encoding process to enforce periodic boundary conditions. Finally, we choose a latent space of size $2 \times 10$ after a rough parameter search where we vary the latent vector size from $2 \times 1$ to $2 \times 10$. We choose the LeNet structure for the other encoding/decoding layers \citep{10.1162/neco.1989.1.4.541}, and we summarize the hyper-parameters and network structures in Appendix \ref{app:structure}. We train the autoencoder by minimizing the mean-square error function which is given as
\begin{equation}
    \mathcal{L} = \sum_{\text{All Samples}}\frac{D}{N}\sum_{N}(Y_{\text{Input}} - Y_{\text{Output}})^{2}
\end{equation}
Here, $D$ is the size of the ``image", and $N$ is the total number of data points. In addition, we add the standard Kullback-Leibler (KL) divergence \citep{2019arXiv190708956O}
\begin{equation}
    \mathcal{L_{
    \text{KL}}} = -\frac{1}{2}\sum_{\text{All Samples}}[1 + \text{log}(\sum_{i}\sigma_{ii}^{2}) - \sum_{i}\sigma_{ii}^{2} - |\vec{\mu}|^{2}]
\end{equation}
Here, $\sigma_{ii}^{2} = [\sigma^{2}I_{n}]_{ii}$. We minimize the loss function using the ADAM optimizer \citep{2014arXiv1412.6980K} implemented in the \texttt{TensorFlow} package \citep{tensorflow2015-whitepaper} with default learning parameters.

\subsection{Anomaly Detection via Isolation Forest} \label{subsec:search}
Here we describe the process of searching for anomalies via an isolation forest. The isolation forest works by building an ensemble of binary isolation trees in the learned latent space. The anomalies are identified as those having short average path lengths to reach isolation \citep{4781136}. In addition to the latent vectors $\vec{\mu}$, the log of common period log$_{10}\tau$, and the difference between the average $g$- and $r$-band magnitude ($m_{\text{gr}} := \langle m_{\text{g}} \rangle - \langle m_{\text{r}} \rangle$) are passed to the isolation forest \citep{10.1145/2133360.2133363}. We choose to include these two additional features because the information about the intrinsic period and temperature are lost during the pre-processing and normalization procedure. We use the isolation forest implemented in the \texttt{scikit-learn} package with $100000$ base estimators \footnote{We have done a convergence test and found that a value of $5000$ is enough to ensure convergence in the anomaly ranking.}. The full pipeline is shown in Figure \ref{fig:pipeline}.

\subsection{Coordinate Transformation} \label{subsec:transform}
As we shall see in Section \ref{sec:results}, the presence of the annular structure in the latent space inspired us to look for the anomalies in a spherical coordinate transformation of our learned latent space. We transform the latent space using the following \citep{10.2307/2308932}
\begin{equation}
\begin{aligned}
    r = \sum_{j}^{10} x_{j}^{2} \\
    \vartheta_{i} = \text{cos}^{-1} \left(\frac{x_{i}}{\sum_{j = i}^{10} x_{j}^{2}}\right) , & 1 < i < 9 \\
    \vartheta_{9} =
    \begin{cases}
    \text{cos}^{-1}\left(\frac{x_{8}}{\sqrt{x_{8}^{2} + x_{9}^{2}}}\right), & x_{9} \leq 0 \\
    2\pi - \text{cos}^{-1}\left(\frac{x_{8}}{\sqrt{x_{8}^{2} + x_{9}^{2}}}\right), & \text{otherwise}
    \end{cases}
\end{aligned}
\end{equation}
We report top anomalies found in both coordinate systems.

\section{Results and Discussion} \label{sec:results}
\begin{deluxetable}{cc}
\caption{Features from \citet{Chen_2020} which have been used to calculate the correlation matrix in addition to our learned latent variables. See Figure \ref{fig:correlation} for the correlation matrix. \label{tab:physicalpara}}
\tablewidth{0pt}
\tablehead{
\colhead{Feature} & \colhead{Description}
}
\startdata
$\tau$ & Common $g$- and $r$-band Period \\
$R_{21}$ & Amplitude Ratio $a_{2}/a_{1}$ \\
$\phi_{21}$ & Phase Difference $\phi_{2} - 2\phi_{1}$ \\
$\langle m_{\text{g}} \rangle$ & Mean $g$-band Apparent Magnitude \\
$\langle m_{\text{r}} \rangle$ & Mean $r$-band Apparent Magnitude  \\
$\tau_{\text{g}}$ & $g$-band Period \\
$\tau_{\text{r}}$ & $r$-band Period \\
$N_{\text{g}}$ & Number of Detection in $g$-band \\
$N_{\text{r}}$ & Number of Detection in $r$-band \\
$R_{21\text{g}}$ & $g$-band Amplitude Ratio \\
$R_{21\text{r}}$ & $r$-band Amplitude Ratio \\
$\phi_{21\text{g}}$ & $g$-band Phase Difference \\
$\phi_{21\text{r}}$ & $r$-band Phase Difference
\enddata
\tablecomments{In \citet{Chen_2020}, the light curves are fitted with a fourth-order Fourier function $f = a_{0} + \Sigma_{n = 1}^{4}a_{n}\text{cos}(2\pi nt/\tau + \phi_{n})$}
\end{deluxetable}
\begin{deluxetable}{c|cc}
\caption{Distribution matrix of class labels for the top 100 anomalies in the Cartesian and Spherical latent space. \label{tab:anomaliesratio}}
\tablewidth{0pt}
\tablehead{
\colhead{Type} & \colhead{Cartesian} & \colhead{Spherical}
}
\startdata
SR & $72$ & $85$ \\
Mira & $28$ & $14$ \\
RSCVN & - & $1$
\enddata
\end{deluxetable}
\begin{figure*}[ht!]
\gridline{\fig{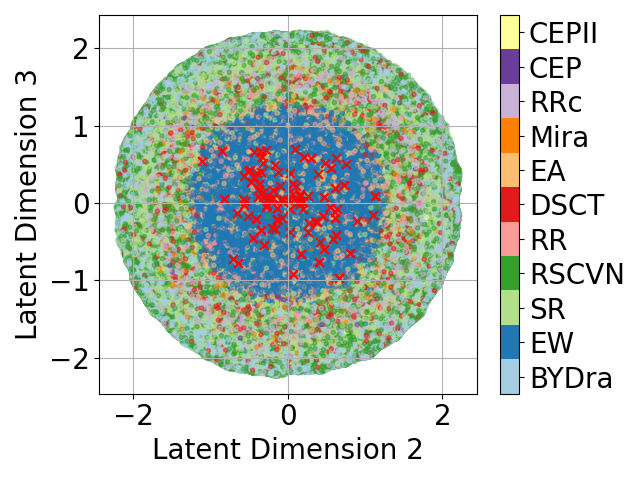}{0.3\textwidth}{(a)}
          \fig{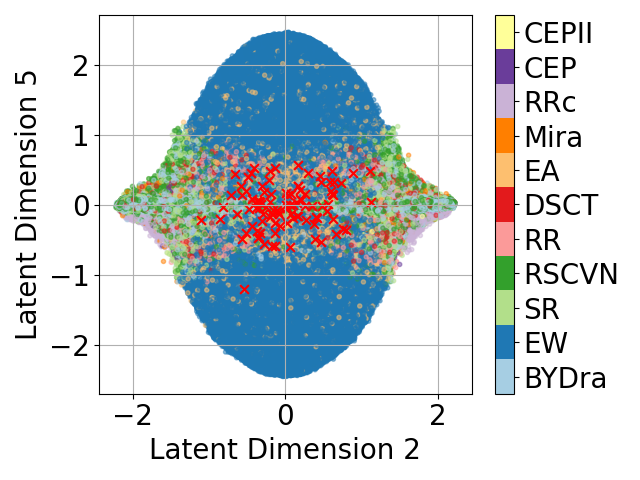}{0.3\textwidth}{(b)}
          \fig{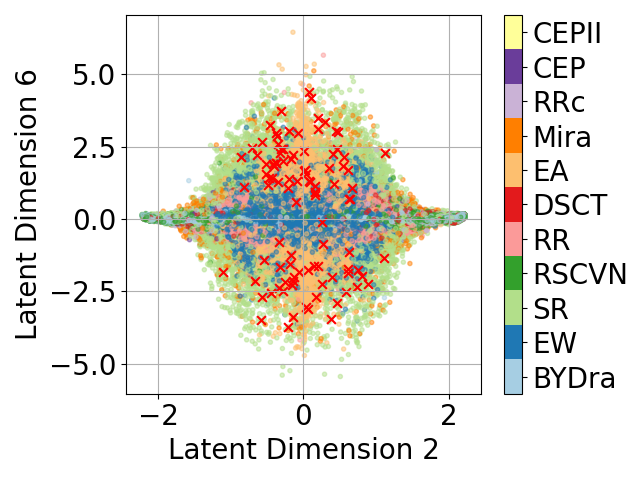}{0.3\textwidth}{(c)}
         }
\gridline{\fig{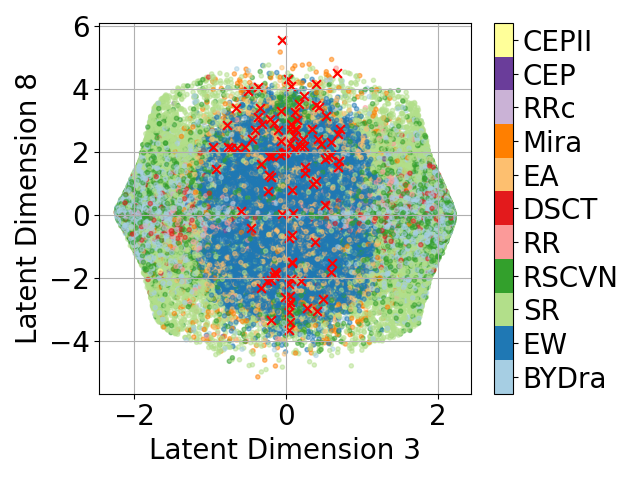}{0.3\textwidth}{(d)}
          \fig{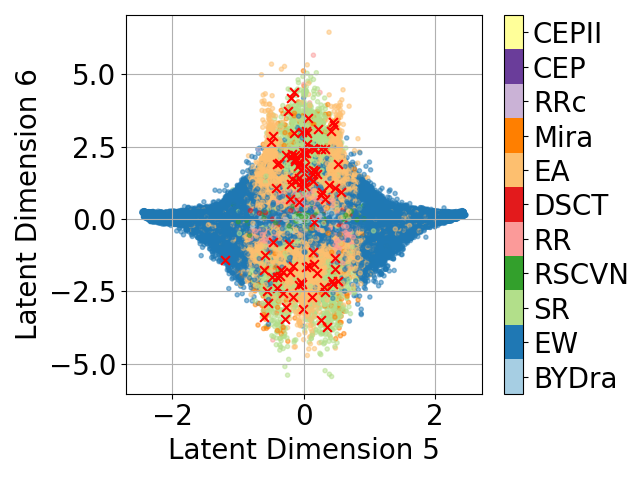}{0.3\textwidth}{(e)}
          \fig{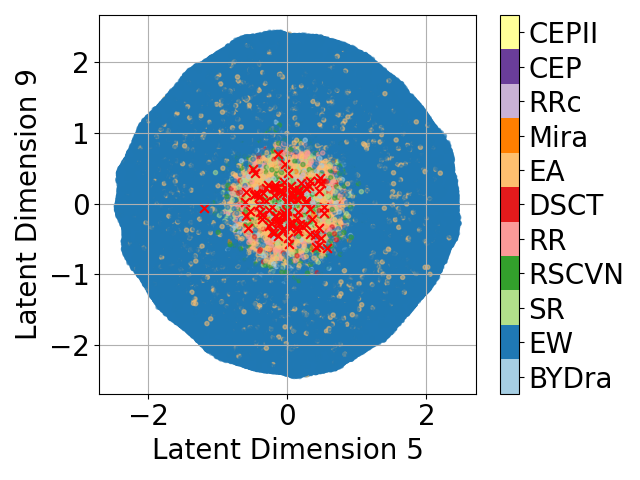}{0.3\textwidth}{(f)}
         }
\caption{Representative latent distributions for different periodic variable stars, labeled by distinct colors. The anomalies are marked as red-crosses. Note that we have also included $m_{\text{gr}}$ and log$_{10}\tau$ as extra features in the isolation forest, but their distributions are omitted from this plot. \label{fig:latent}}
\end{figure*}
\begin{figure*}[ht!]
	\centering
	\includegraphics[width=0.9\linewidth]{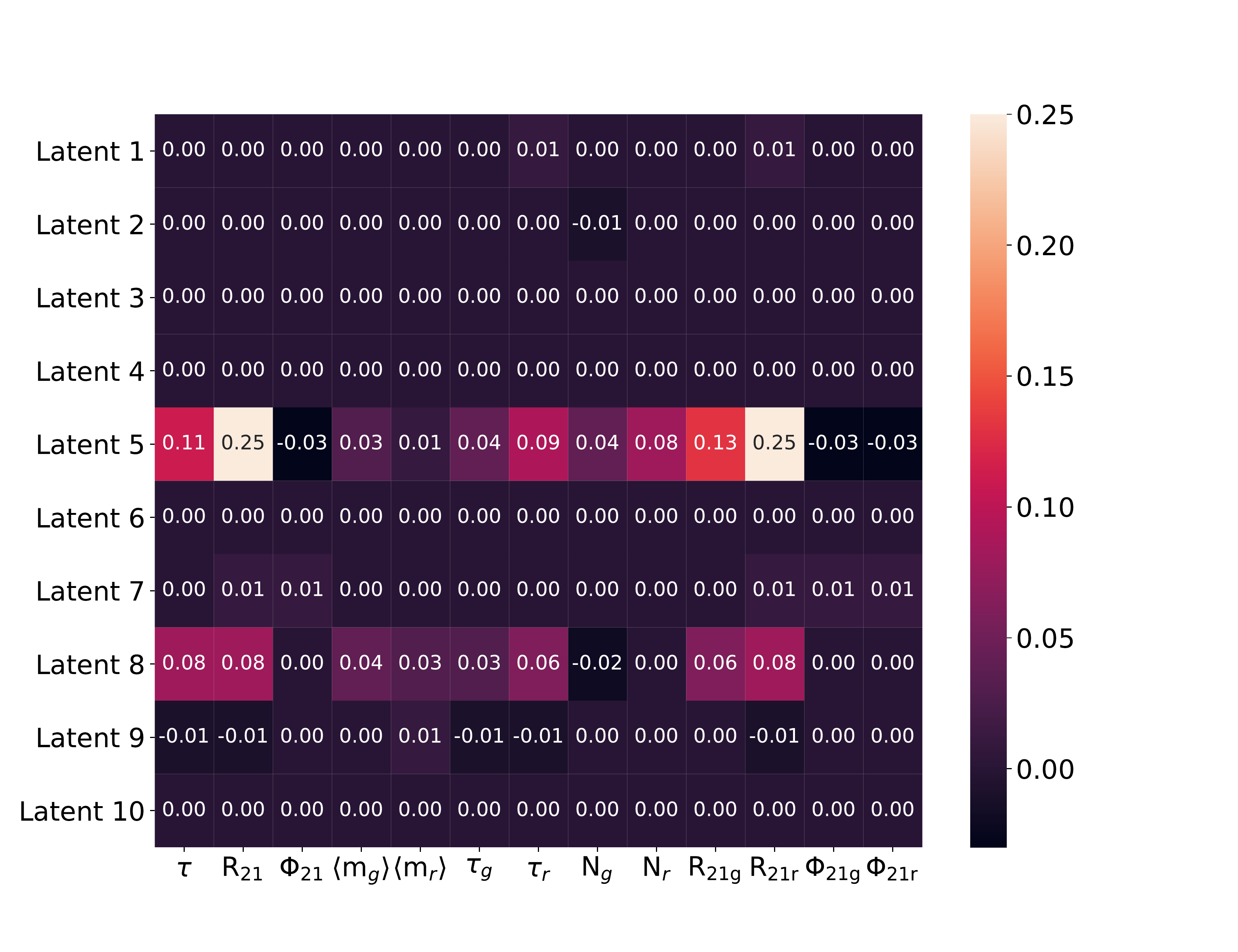}
	\caption{Correlation matrix between the latent variables and the derived quantities extracted from \citet{Chen_2020}. The correlation matrix ranges from $-1$ to $1$. The positive ends represent a positive correlation between two variables and vice versa. Some of the latent features (e.g., Latent variables 5 and 8) show medium correlation with $R_{21}$, which is a measurement of higher-order modes in the Fourier transform. \label{fig:correlation}}
\end{figure*}
\begin{figure*}[ht!]
\gridline{\fig{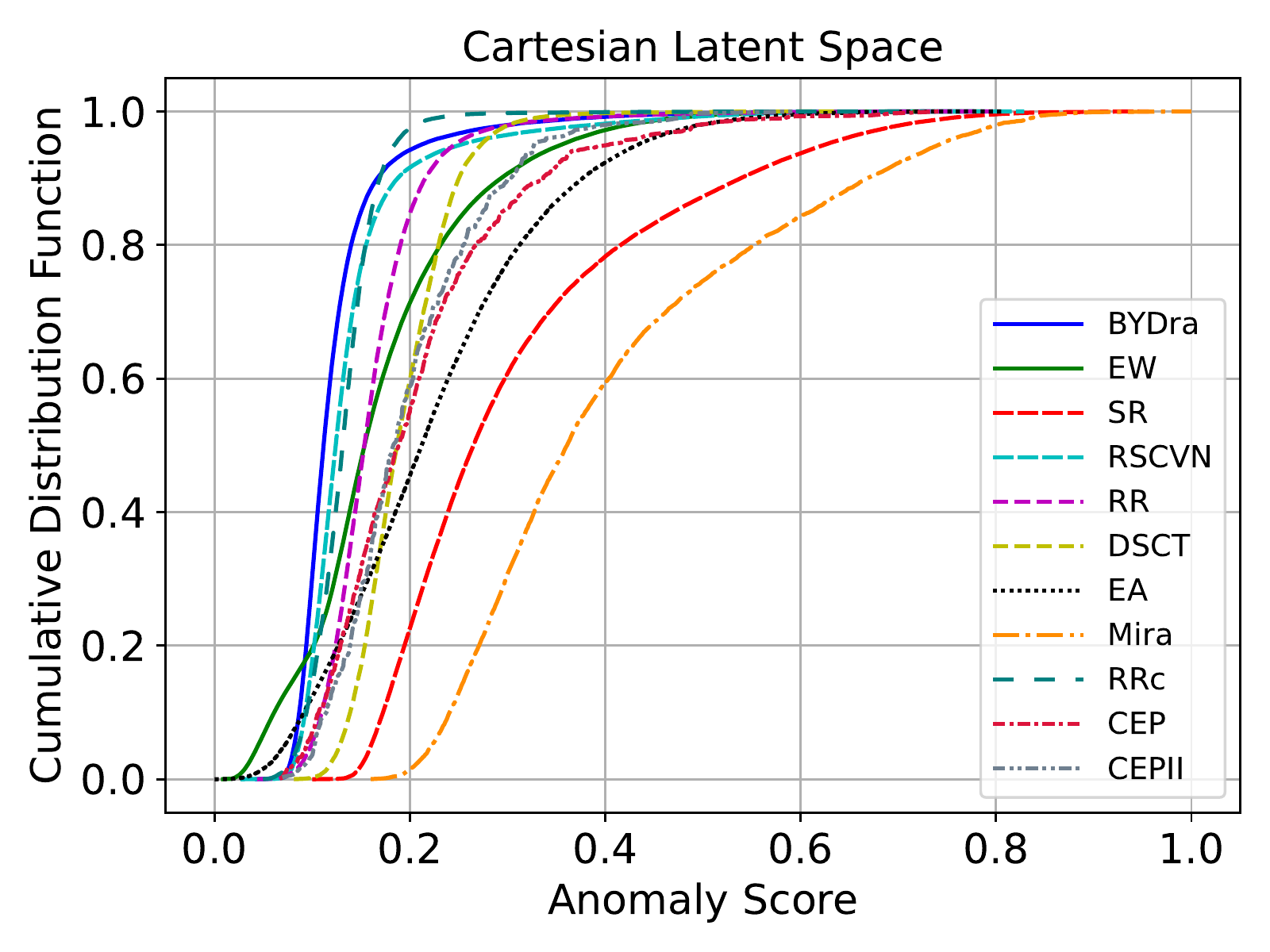}{0.5\textwidth}{(a)}
          \fig{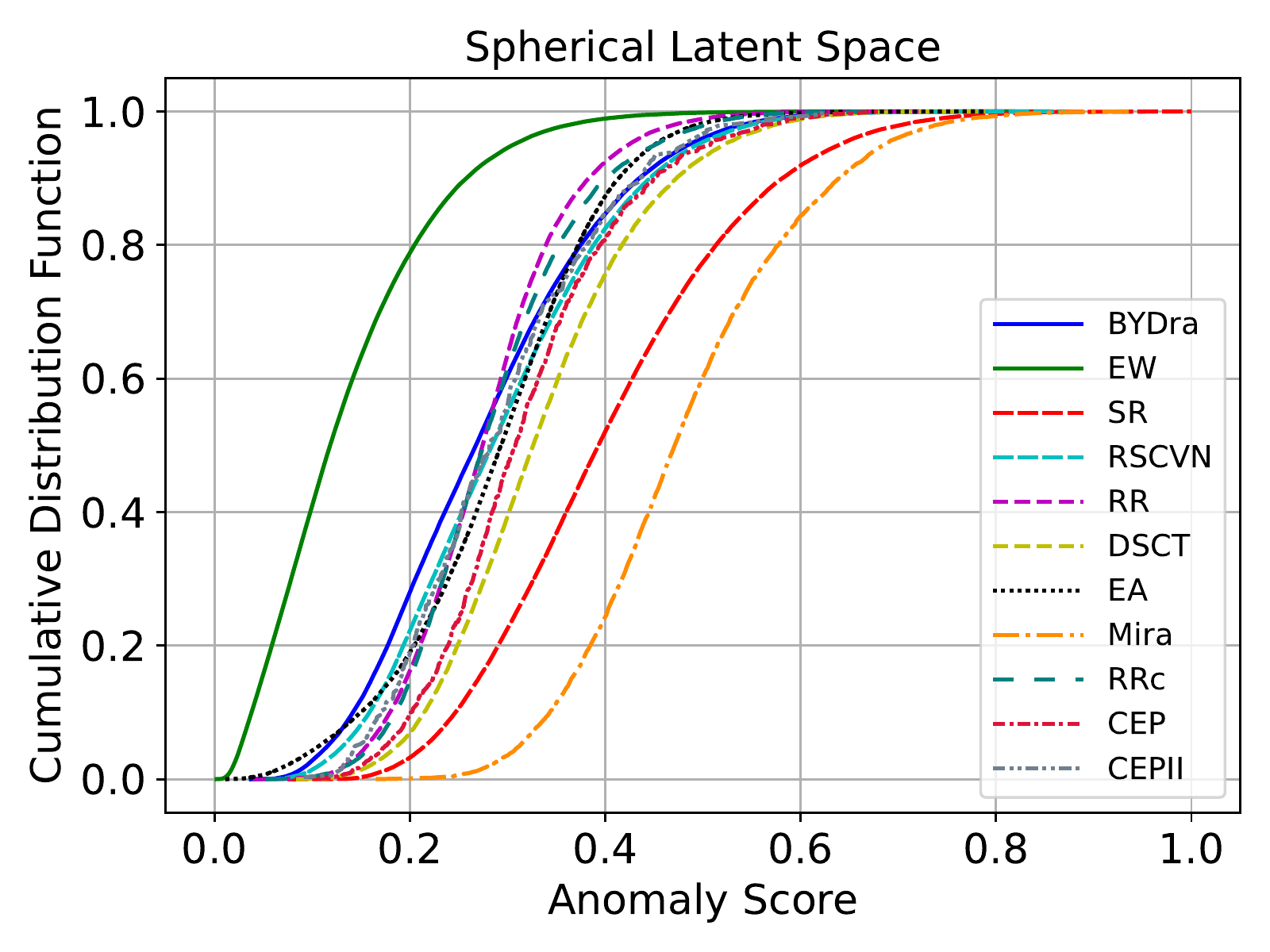}{0.5\textwidth}{(b)}
         }
\caption{The cumulative distribution function (CDF) of the ranked anomaly scores for different categories of periodic variables in the (a) Cartesian, and (b) Spherical latent space. Note that $M_{\text{gr}}$ and log$_{10}\tau$ is added to the isolation forest \textit{after} the transformation has done. In both cases, the periodic variable stars classified by \citet{Chen_2020} as Mira and SR variables are relatively clustered towards the higher anomaly scores, indicating they are more anomalous than other types of periodic variables. \label{fig:cdfcarsph}}
\end{figure*}
\begin{figure*}[ht!]
\gridline{\fig{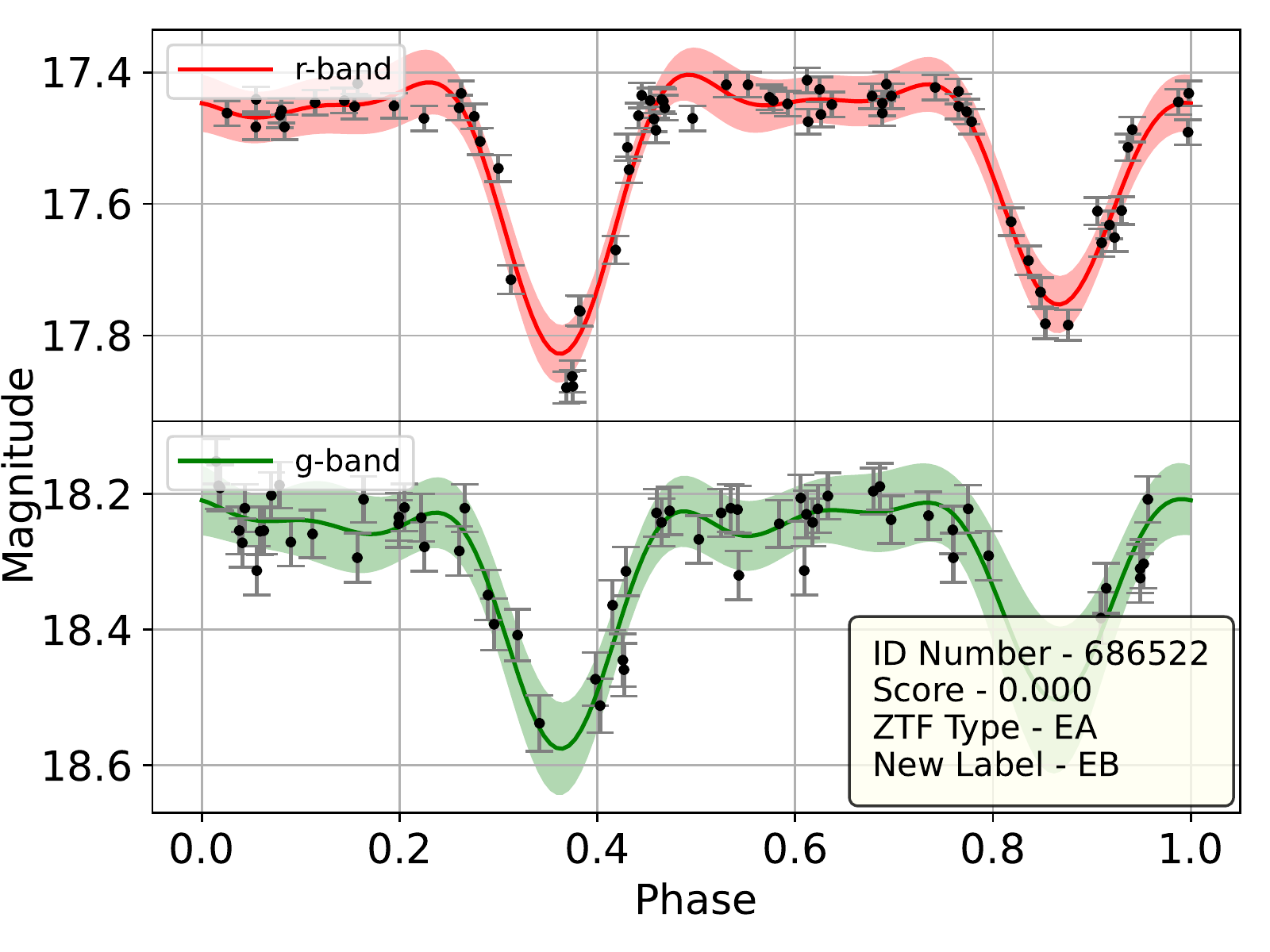}{0.5\textwidth}{(a)}
          \fig{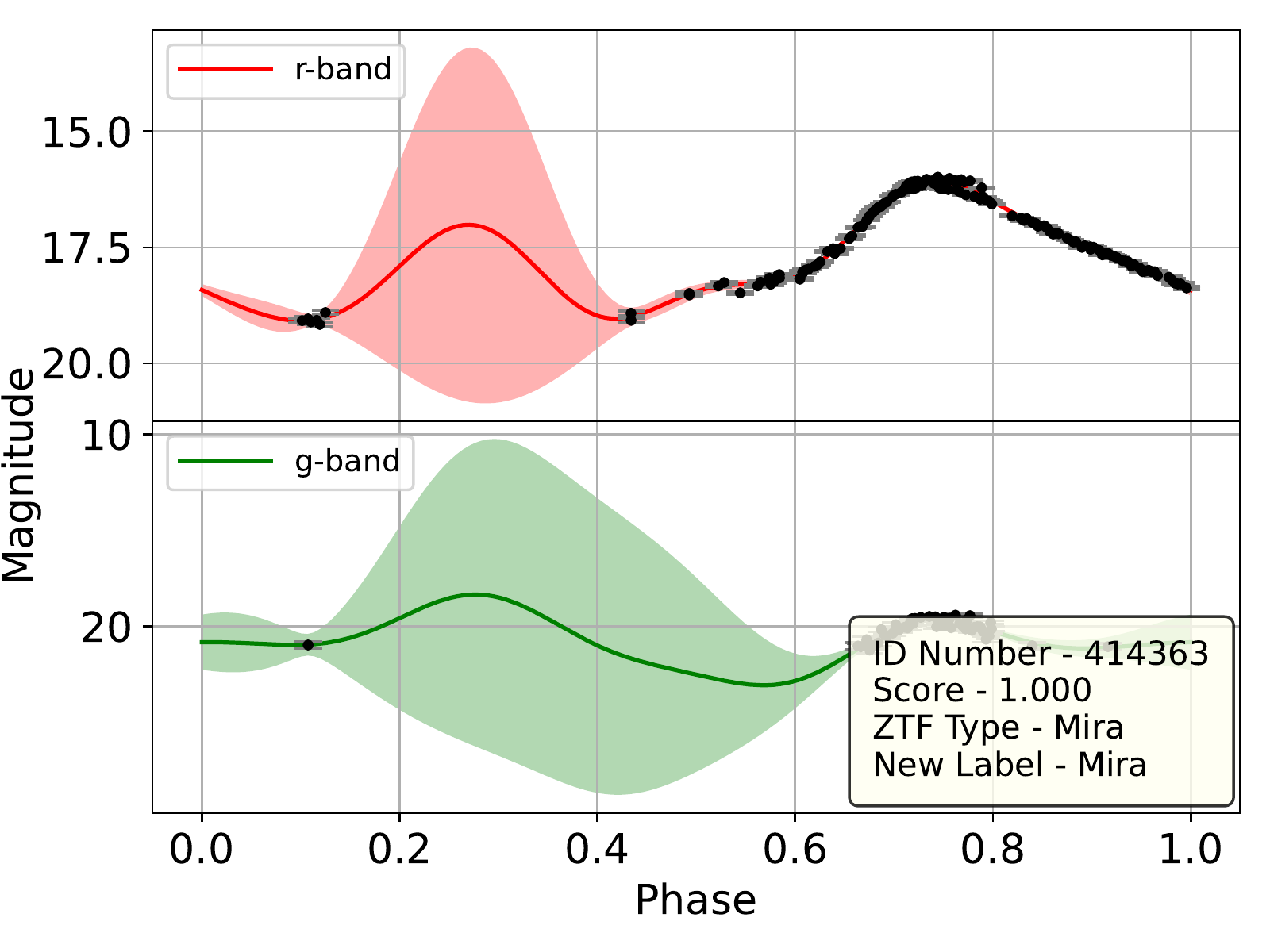}{0.5\textwidth}{(b)}
         }
\caption{Examples of periodic variable stars in (\textit{left}) the least anomalous, and (\textit{right}) the most anomalous group of objects in the Cartesian latent space. In the left-hand plot, there is a clear primary period, corresponding to the primary/secondary eclipses. Furthermore, the dips are of similar magnitudes both bands  (of order $0.2$ mag) and are in-phase. In the right-hand plot, oscillations are seemingly more irregular with fluctuations spanning over several magnitudes. Note that the shaded region represents $2\sigma$ from the mean function.  \label{fig:anomaliescartesian}}
\end{figure*}
\begin{figure*}[ht!]
\gridline{\fig{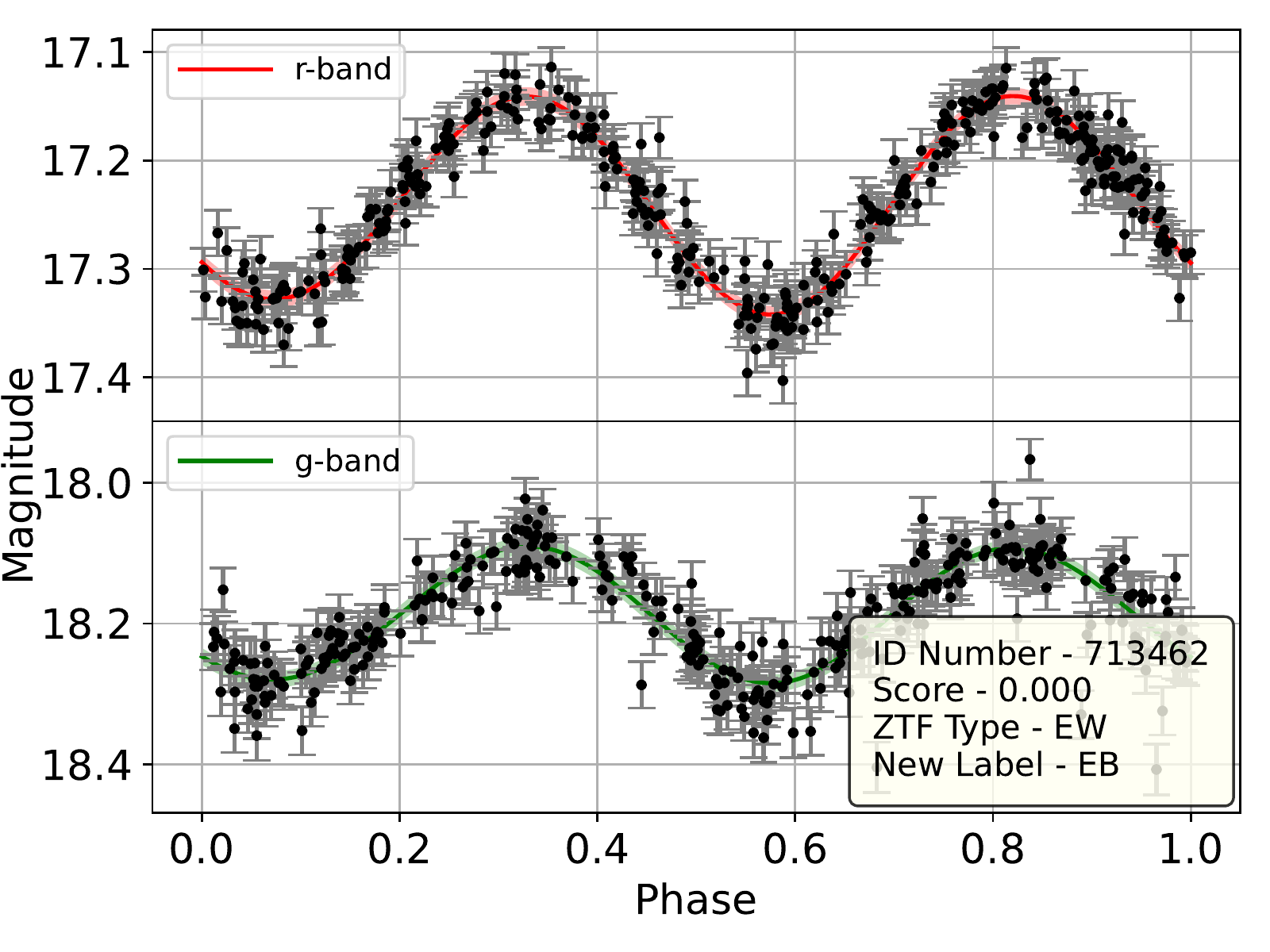}{0.5\textwidth}{(a)}
          \fig{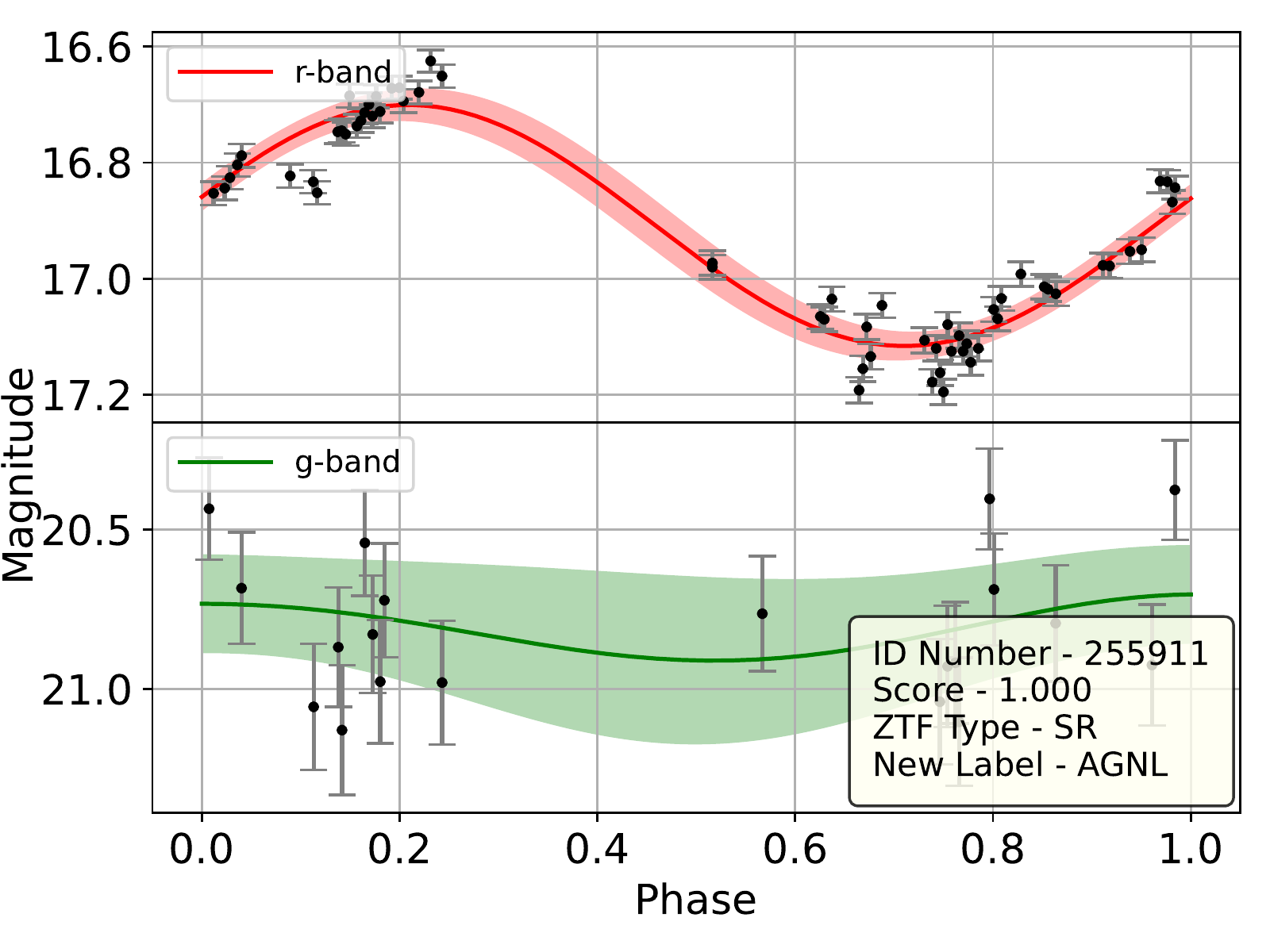}{0.5\textwidth}{(b)}
         }
\caption{Same as Figure \ref{fig:anomaliescartesian}, but for the Spherical latent space. On the left-hand side, the shape of the $g$-band and $r$-band light curves are positively correlated (the maxima and minima matched with each other). However, on the right-hand side, there is seemingly a $\sim \frac{\pi}{2}$ phase lag. We additionally note, however, the poor data quality of the $g$-band magnitude. \label{fig:anomaliesspherical}}
\end{figure*}
\begin{figure*}[ht!]
\gridline{\fig{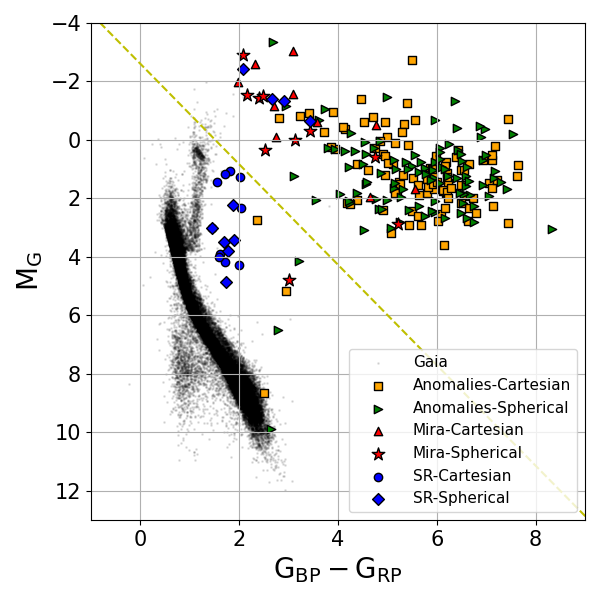}{0.5\textwidth}{(a)}
          \fig{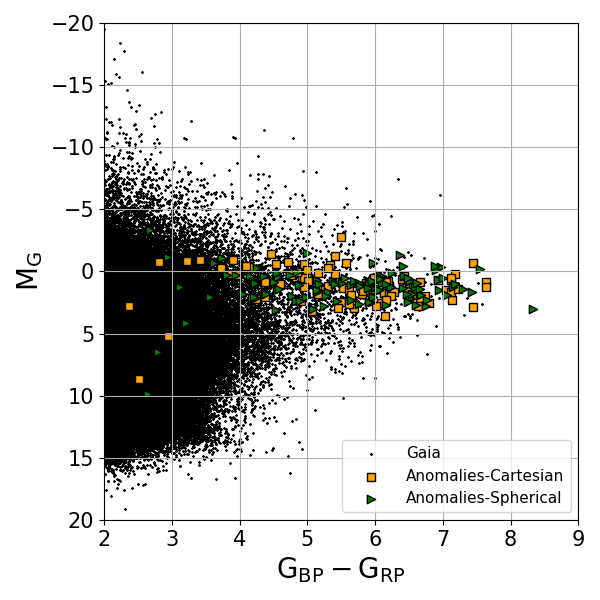}{0.5\textwidth}{(b)}
         }
\caption{Distribution of the top 100 anomalous events, and the top 10 normal Mira  (in red) and SR (in blue) variables on the Gaia Hertzsprung-Russell diagram. The black dots are Gaia data taken from \citet{2018A&A...616A...2L}. The vertical axis is the absolute magnitude in the Gaia $g$-band filter, while the horizontal axis represents the difference between the Gaia Optical $B-$band and $r$-Band (apparent) magnitudes. In computing the absolute $g$-band magnitudes, we choose the geometric distance computed by \citet{2021AJ....161..147B}, and we choose the $g$-band apparent magnitudes given by \citet{2021A&A...649A...1G}. The data are available in the Vizier Catalog \citep{2014yCat....1.2023S}. The searches are facilitated through the Python package \texttt{Astroquery} \citep{2019AJ....157...98G}. In (a), the clustered black dots represent the main-sequence stars, while the dashed straight line represents a linear cut given by \citet{2018A&A...618A..58M} that separate young stellar objects (YSOs) and long-period variables. Our anomalies are seemingly evolved, bright, cool, and red stars, with some exceptions being YSOs. In (b), the clustered data are cold stars with $G_{\text{BP}} - G_{\text{BP}} > 2$. Our anomalies could not be obtained by a simple trivial cut in the Gaia Hertzsprung-Russell Diagram. Note that the Gaia $g$-band absolute magnitudes shown in (a) and (b) have not been corrected for interstellar extinction. \label{fig:hrdiagram}}
\end{figure*}
\begin{figure*}[ht!]
\gridline{\fig{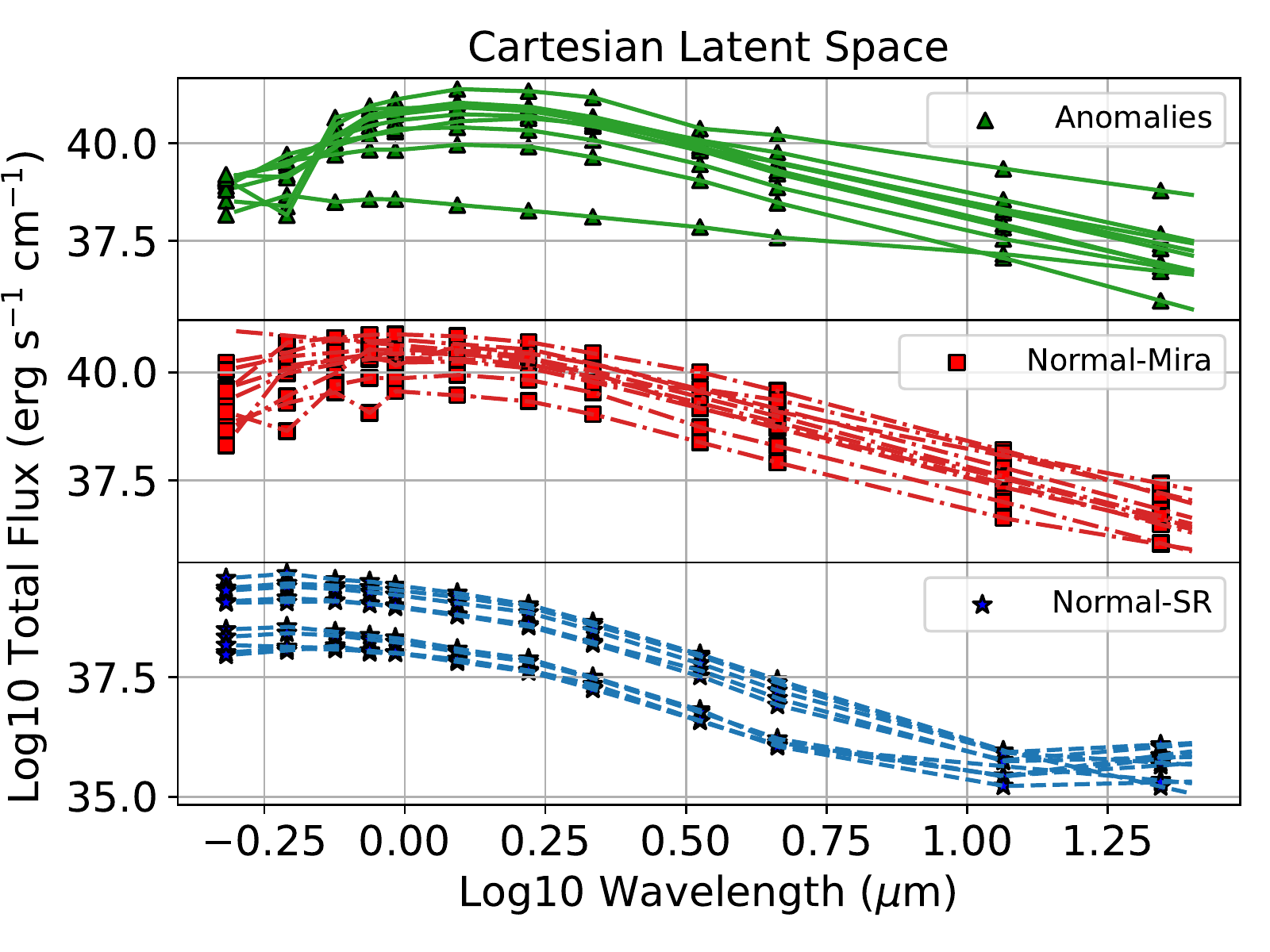}{0.5\textwidth}{(a)}
          \fig{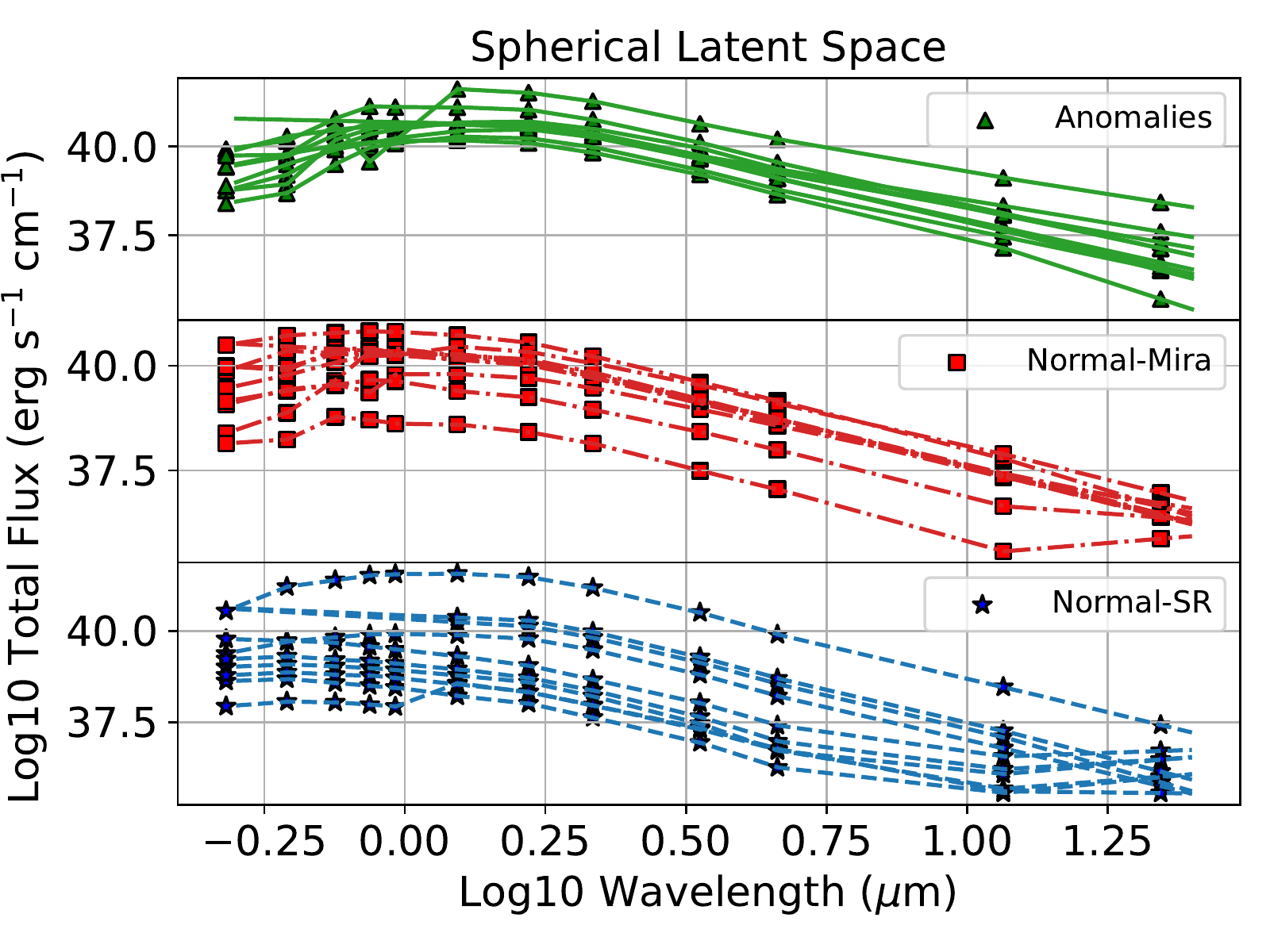}{0.5\textwidth}{(b)}
         }
\caption{SEDs for the selected events in (a) Cartesian, and (b) Spherical latent space. In each plot, we select the top 10 anomalies for illustration. Since the majority of the anomalies are Mira and SR variables, we include the top 10 normal Mira and SR variables in both latent spaces for comparison. For each plot, the top panel is for the anomalies, the middle panel is for the normal Mira variables, and the bottom panel is for the normal SR variables. Lines between data points are linearly interpolated. In these plots, we show spectral data of 12 band filters: the W1, W2, W3, and W4 bands provided by the WISE \citep{2014yCat.2328....0C} survey; The 2MASS \citep{2006AJ....131.1163S} J, H, and K$_{\text{s}}$ bands from the URAT \citep{2015AJ....150..101Z} survey, and the g, r, i. z, y bands from the Pan-STARRS1 \citep{2016arXiv161205560C} survey. Effective wavelengths of these band filters are extracted from \citet{2019ApJ...877..116W}. Note that we have corrected the spectral flux for interstellar extinction. We find the anomalies cooler than the most `normal' Mira and SR variables. \label{fig:spectralanomalies}}
\end{figure*}
\begin{figure}[ht!]
	\centering
	\includegraphics[width=1.0\linewidth]{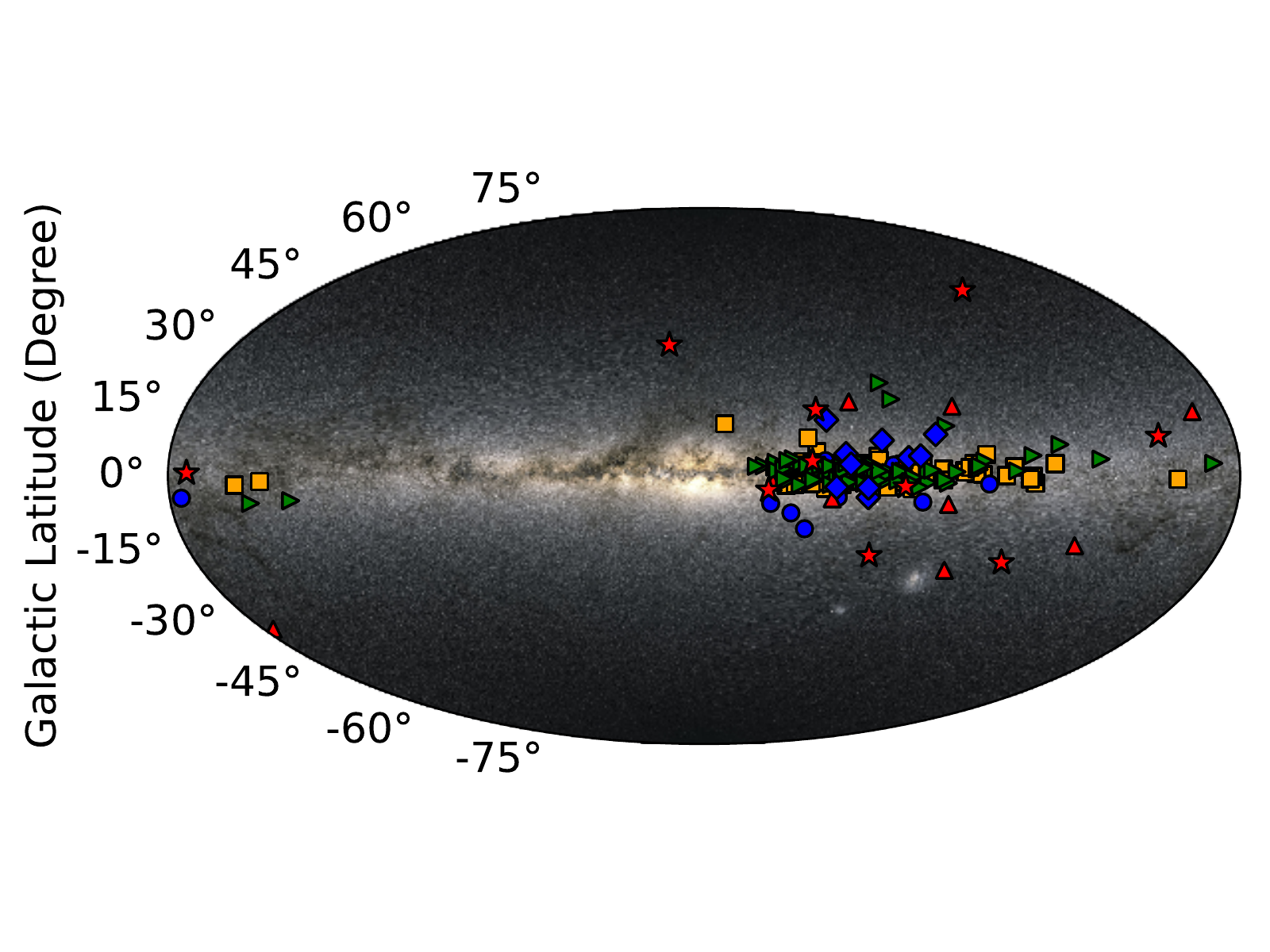}
	\caption{Distribution of anomalous variables and ``normal" SR/Mira variables in Milky Way galactic coordinates. Markers are the same as in Figure \ref{fig:hrdiagram} (a). Almost all the anomalies in both latent spaces cluster in the Milky Way galactic disk, indicating they might be consistent with young and massive stars. The plotting of the background milky way sky-map is facilitated through the python package \texttt{mw-plot}. Image credit: ESA/Gaia/DPAC. \label{fig:milky}}
\end{figure}
\begin{figure}[ht!]
	\centering
	\includegraphics[width=1.0\linewidth]{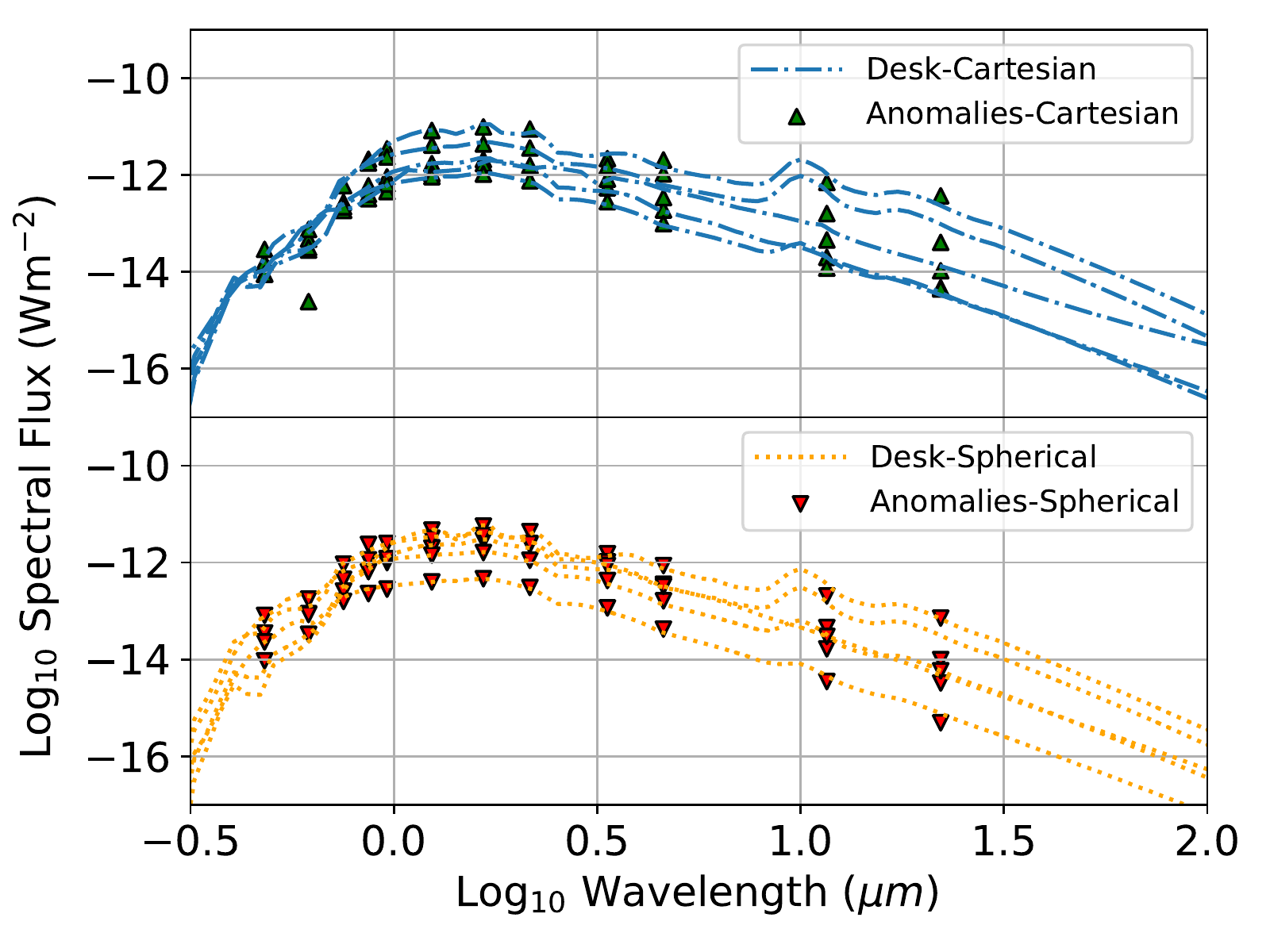}
	\caption{SEDs of 5 examples of anomalous periodic variables, plotted with best fit models fit with \texttt{DESK}. Upper panel: Cartesian latent space. Lower panel: Spherical latent space. \label{fig:dusty}}
\end{figure}
\begin{figure}[ht!]
	\centering
	\includegraphics[width=1.0\linewidth]{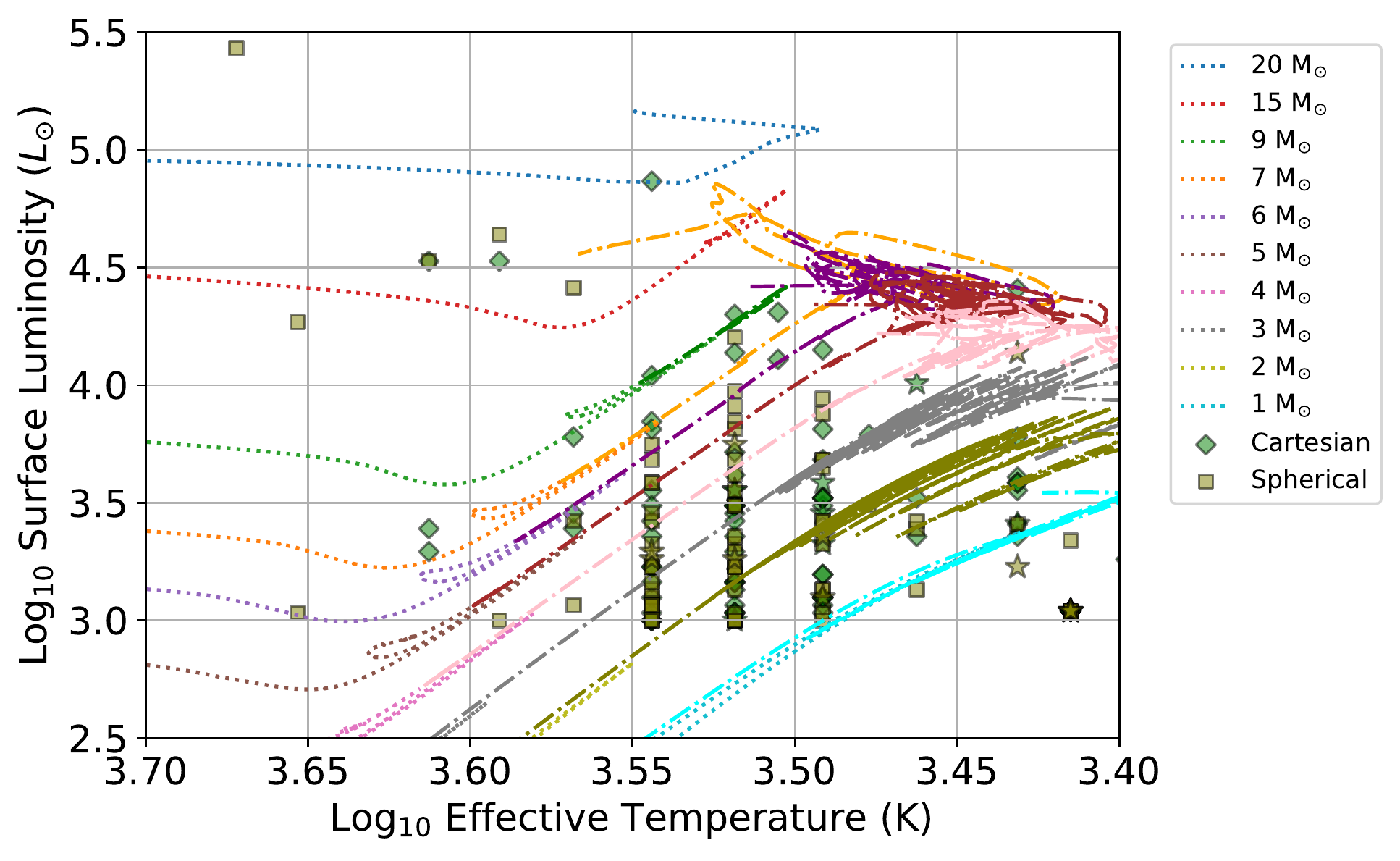}
	\caption{Luminosity against temperature relations for the top 100 anomalies in both Cartesian and Spherical latent spaces. The top 10 anomalies in both latent spaces are marked as stars. We appended also theoretical stellar evolution paths for different amounts of zero-age main-sequence masses. For each stellar evolution path, the dotted lines are tracks that correspond to the RGB phase, and the dash-dotted lines are tracks that correspond to the AGB phase. \label{fig:lmt}}
\end{figure}
\begin{figure}[ht!]
	\centering
	\includegraphics[width=1.0\linewidth]{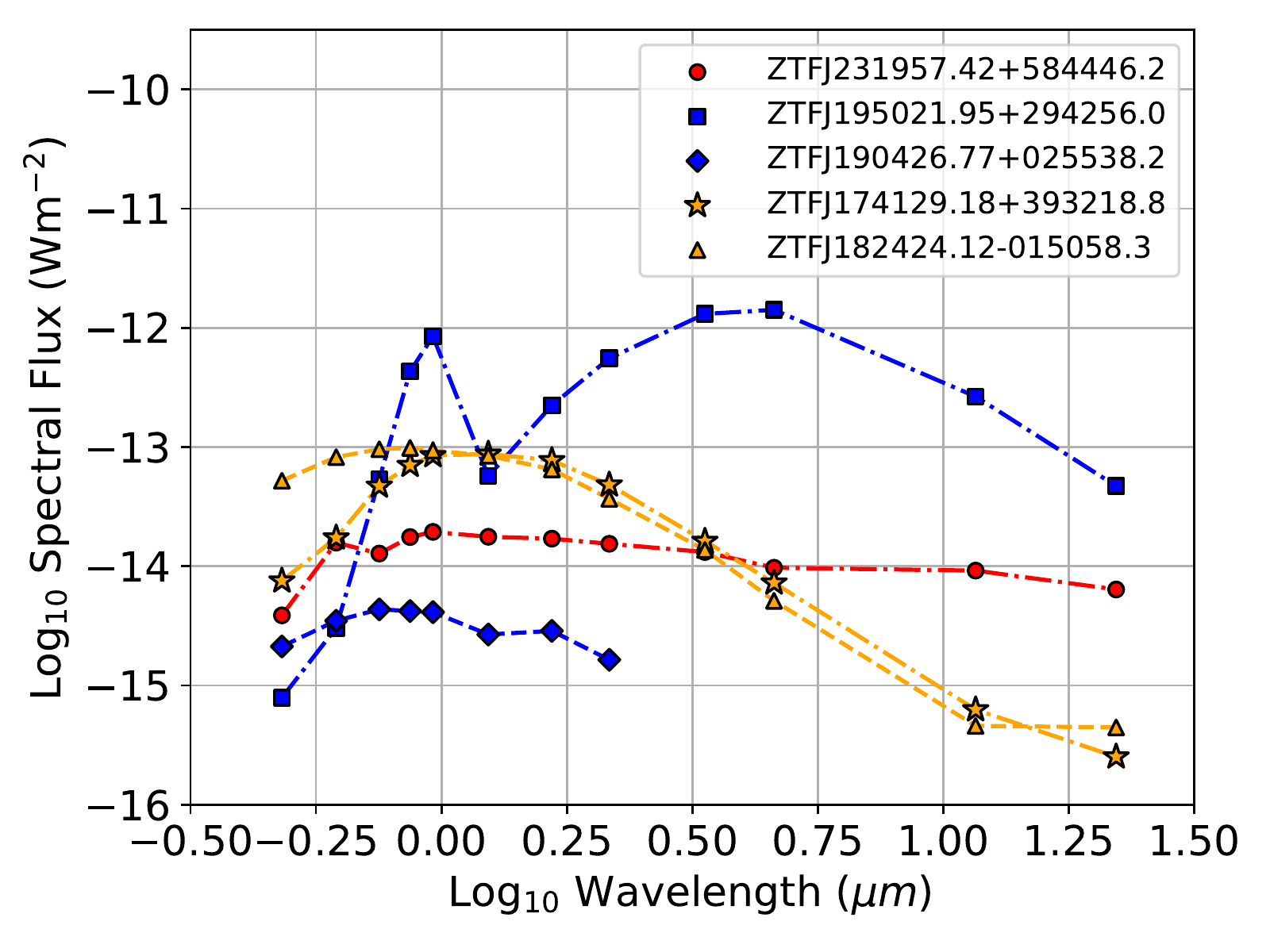}
	\caption{Representative plots of SEDs for periodic variables that are inconsistent with models of evolved stars. Lines are linear interpolation for better visualization. Their likely nature is represented by different colors: red for class II YSO, orange for YSOs in between class II/class III, and blue for stars that appear elongated in images (i.e., may contain contamination). See Section \ref{subsec:individual} for more details. \label{fig:ysofit}}
\end{figure}

\subsection{Latent Variable Distributions} \label{subsec:latent}
Representative 2-D projections of our learned $10-$dimensional latent space are shown in Figure \ref{fig:latent}, color-coded based on their ZTF CPVS classifications. The latent variables exhibit annular structure (see, for example, Figures \ref{fig:latent} (a), (b), (c), and (e)), including rings and circles of different sizes that are formed by the clustering of different categories of data. Figures \ref{fig:latent} (d) and (f) show clustering of periodic variables of Type EW, EA, SR, RR and Mira. Our representative plots suggest that our autoencoder is capable of identifying separating periodic variables that belong to different classes; the use of our autoencoder as a classifier is explored in Cheung et al. (in prep). \\

We next explore if the latent space variables are correlated to hand-engineered, observable parameters of the systems. We extract a series of derived features from \citet{Chen_2020}; we have briefly described their physical meanings in Table~\ref{tab:physicalpara}. We compute the correlation matrix (using the Pearson correlation coefficient) between the latent variables and the above-mentioned physical features, and we show their results in Figure \ref{fig:correlation}. We find that most of the latent features are uncorrelated with the derived features, hinting that the neural network is learning something beyond these observable features. One of the latent variables shows a mild correlation of $r_p \sim 0.25$ with $R_{21}$, which is the ratio of coefficient between high-order modes to the fundamental mode in a Fourier series expansion. One interpretation of this correlation is that the neural network is capturing periodic variables that oscillate with higher-order Fourier modes. We note that the latent space has dimensions equally ($r_p\approx0.25$) correlated with $R_{21\text{r}}$ (the same ratio measured only with the $r$-band photometry), while it is in weaker ($r_p\approx0.13$) correlation with $R_{21\text{g}}$. This makes sense, as in most cases, $\tau \simeq \tau_{\text{r}}$ due to noiser $g$-band light curves.

\subsection{Overall Anomaly Score Distributions} \label{subsec:anomalyscore}
We rank all $730,184$ periodic variables according to the normalized anomaly scores described in Section \ref{sec:method}. In this section, we explore the overall results of our anomaly detection scores, as well as the top anomalies within the ZTF CPVS. \\

We show the resulting cumulative distribution function for the anomaly scores within each photometrically classified class in Figure \ref{fig:cdfcarsph} (a). The BYDra, RSCVN, and RRc variables have similar distributions which are clustered towards less anomalous scores. Note that the BYDra and RSCVN variables are some of the most common classes in our training set (see Figure \ref{fig:statistic}). Also scattered at the lower end of anomaly scores are RR, EW, EA, DSCT, CEP, and CEPII variables. We note in particular that there is a small bump of EW variables at the low anomaly end of the plot, indicating that the majority of the least anomalous events should be the EW variables. Finally, the SR and Mira variables are clustered towards the high anomaly end of the plot, indicating that they should constitute the majority of the top anomalies. \\

We train the isolation forest again on the spherical transformation of the latent space, and we show the resulting cumulative distribution function of the anomaly scores in Figure \ref{fig:cdfcarsph} (b). The transformation to the spherical coordinates changes the distribution for the anomalous sources. In this case, BYDra, RSCVN, RR, DSCT, EA, RRc, CEP, and CEPII variables have similar anomaly score distributions. The EW variables are now obviously less anomalous than the remaining periodic variables and still constitute the majority of the least anomalous event. Although the categorical-wise ranking of the overall anomalies changes, the SR and Mira variables remain clustered towards the high anomaly end.

\subsection{The Most and Least Anomalous Variables} \label{subsec:theanomalies}
We next turn our attention to the most and least anomalous variable stars as identified by our pipeline, focusing on the top/bottom 100 events in both the Cartesian and Spherical coordinate systems. They represent $\sim 0.01$\% of the total population. The \textit{least} anomalous events in the Cartesian space are mostly EW variables, consistent with the fact that they (1) generally have a regularly shaped light curve that can be represented in limited numbers of Fourier modes and (2) are the most common variable type in our sample. We show example light curves in Figure \ref{fig:anomaliescartesian} (a). The EW variables typically have a highly regular primary and secondary eclipse, with typical eclipse depths $\lesssim 1$ mag. \\

On the other hand, the \textit{most} anomalous events are mostly photometrically classified as Mira and SR variables. We show one example light curve in Figure \ref{fig:anomaliescartesian} (b) for reference. These show highly irregular variability. Moreover, some of the anomalies exhibit larger fluctuations that span over several magnitudes. We note that these anomalies typically have red colors ($m_{gr} > 2$), and we revisit this observation in Section \ref{subsec:discuss}. We list our anomaly scores (as well as relevant features from \citealt{Chen_2020}) in Appendix \ref{app:top100}. \\

We perform a similar analysis on the top and least anomalies in the Spherical latent space, finding that the least anomalous events are similar to those in the Cartesian latent space. We show one example light curve in Figure \ref{fig:anomaliesspherical} (a). We also find that the topmost anomalous events are also mostly photometrically classified as SR and Mira variables, again consistent with the search in Cartesian space. We show one example light curve in Figure \ref{fig:anomaliesspherical} (b). We find that some of the anomalies show weak or even anti-correlation between the $g$- and $r$-band light curves. Furthermore, they have red colors ($m_{gr} > 2$), and we also revisit this observation in Section \ref{subsec:discuss}. \\

We remark that unphysical MGPR results (e.g., unphysically bright/dim interpolations) could not be completely eliminated, and they could be misjudged as outliers in the latent spaces. However, we have manually checked that the ratio of total unphysical MGPR results to the total of samples is less than $0.01 \%$, and the ratio of contributions of the unphysical MGPR results to the top anomalies are $1/100$ and $3/100$ for the Cartesian and Spherical latent spaces respectively. In short, our anomaly detection pipeline is only mildly affected by the limited quality of the input data set.

\subsection{Understanding the Physical Origins of the Anomalies} \label{subsec:discuss}
In this section, we further investigate the physical nature of our identified anomalies. We have extracted Gaia\footnote{The Gaia band filters are different from the ZTF band filters. To avoid ambiguity, we would use a small letter for ZTF while a capital letter for Gaia.} $g$-band absolute magnitudes $M_{\text{G}}$, and the difference between the Gaia $B-$band and $r$-band magnitudes $G_{\text{BP}} - G_{\text{RP}}$ of the top 100 anomalies in both the Cartesian and Spherical latent space. We plot their distribution against $10,000,000$ randomly sampled stars from Gaia (making up the main sequence) in Figure \ref{fig:hrdiagram} (a). We show that the majority of the anomalies clustered around $-2 < M_{\text{G}} < 4$ and $4 < G_{\text{BP}} - G_{\text{RP}} < 8$, indicating that they are mostly bright stars with extremely cool surface temperatures. This is consistent with the fact that the majority of the anomalies are classified Mira and SR variables from the ZTF light curves. We also show in Figure \ref{fig:hrdiagram} (b) that our anomalies are not a trivial cut in the Gaia catalog, spanning a wide range of colors. We show the distribution of the anomalies and the least anomalous Mira and SR variables for both latent space in the Milky Way galactic coordinates in Figure \ref{fig:milky}. The top anomalies are more tightly clustered in the Milky Way galactic disk compared to their ``normal" counterparts. Their distribution suggested that they are consistent with young and massive stars. Moreover, there is likely significant interstellar reddening that should be taken into account. \\

To analyze their more complete spectral energy distributions (SEDs), we correct for interstellar reddening by extracting the extinction coefficients computed by the 3-D Milky Way dust map Bayestar19 \citep{2019ApJ...887...93G}. The extinction coefficients $A(\lambda)$ are wavelength dependent and can be computed by
\begin{equation}
    A(\lambda) = E(B-V)R(\lambda)
\end{equation}
Here, $E(B-V)$ is the difference in B-V color due to reddening, and $R(\lambda)$ is the extinction vector given in Table 1 of \citet{2019ApJ...887...93G}. They assumed $R_{V} = 3.3$. The extinction is specified by the object right ascension (RA), declination (Dec), and distance. We use the RA and Dec given by the ZTF CPVS, and we use the Gaia geometric distances. The dust map is probabilistic, and we adopt the mean value for the extinction. Note that Table 1 of \citet{2019ApJ...887...93G} does not include extinction coefficients for band filters of W1, W2, W3, and W4 of the WISE surveys. We have made use of scaling relations given by \citet{2016ApJS..224...23X} to approximate the corresponding extinction corrections. The extinction coefficient $A_{i}$ for $i = $ W1, W2, W3, and W4 is given by $A_{i} = \alpha_{i}A_{K_{s}}$, and $\alpha = 0.591$, $0.463$, $0.537$, and $0.364$ for $i = $ W1, W2, W3, and W4. We use the \texttt{DUSTMAP} \citep{2018JOSS....3..695M} package for these extinction corrections. \\

We download the SEDs of the top 100 anomalies and the top 10 ``normal" Mira/SR variables in both latent spaces from the Vizier catalog \citep{2014yCat....1.2023S}. We show these SEDs in Figure \ref{fig:spectralanomalies} (a) and (b). Note that normal Mira/SR variable SEDs typically peak around or just below $1$ $\mu$m, while the anomalous SEDs peak just above $1$ $\mu$m. This translates to an oversimplified black body temperature (using Wien's law) of $\gtrsim 3000$ K for the normal type stars, while the anomalies have cooler temperatures of $\lesssim 2500$ K. Again, this is consistent with the Gaia colors shown in Figure \ref{fig:hrdiagram} (a). There are two potential reasons why these stars are redder: these anomalous stars may be \textit{intrinsically} cooler or they are significantly reddened by local dust. The oversimplified temperature estimates, assuming no dust, are unrealistically cold, so we assume that these stars likely are enshrouded by unaccounted for dust. \\

To model the local dust and underlying SED, we use the open-source code \texttt{DESK} \citep{2020JOSS....5.2554G}. This software package wraps a simple minimizing mean-squared error fitter around the radiative transfer \texttt{DUSTY} code \citep{2000ASPC..196...77N, 2012A&A...543A..36G}. We choose an oxygen \citep{2011ApJ...728...93S} or carbon \citep{2011A&A...532A..54S} rich stellar envelope model with a mixture of amorphous carbon and silicate dust grains, whose spectra are calculated based on the two-dimensional radiative transfer code \texttt{2DUST} \citep{2003ApJ...586.1338U}. Optical constants are given by \citet{1996MNRAS.282.1321Z} and \citet{1992A&A...261..567O}. They are shown to be robust in modeling the spectra for dusty Asymptotic Giant Branch (AGB) and Red Giants Branch (RGB) stars. We assume spherical geometry and a black body SED. Furthermore, we assume that the gas and dust have no relative velocity (Goldman et al., private communication). There are a number of free parameters to be fit \citep[see, for example, Table 1 of][]{2011ApJ...728...93S, 2011A&A...532A..54S}. Here we focus on the surface temperature $T$, luminosity $L$, optical depth $\tau_{10\mu m}$ (at 10 $\mu$m), and the inner radius of the dust shell $R_{\text{in}}$. The \textit{dust} mass-loss rate $\dot{M_{\text{D}}}$ can be inferred from these quantities by \citep{2010A&A...524A..49S}
\begin{equation}
    \dot{M_{\text{D}}} = 4\pi \frac{\tau_{10\mu m}}{\kappa_{10\mu m}}R_{\text{in}}v_{\text{exp}}
\end{equation}
Here, $\kappa_{10\mu m}$ is the opacity at 10 $\mu$m, and $v_{\text{exp}} = 10$ km/s is the expansion velocity. Note that spherical symmetry is implied. The \textit{total} mass-loss rate $\dot{M}$ can be obtained by assuming certain gas to dust ratio, which we take to be $200$ \citep{1994MNRAS.267..711W, 2011ApJ...728...93S, 2011A&A...532A..54S}; however, we remark that the exact value of the gas to dust ratio is poorly known \citep{2008MNRAS.383..399L}. We fit the top 100 anomalies in both the Cartesian and Spherical latent space. We illustrate the performance of the fitting by plotting the SEDs together with the best fit \texttt{DESK} models for five examples of anomalies in both latent spaces in Figure \ref{fig:dusty}. Our fitting results give an average (taken over all samples) of $\tau_{10\mu m} \approx 0.14$ and $0.15$ for the Cartesian and Spherical latent spaces respectively\footnote{The 2DUST carbon-rich stellar envelope model provides only $\tau_{11.3\mu m}$. However, $\tau_{10\mu m}$ should be fairly similar to $\tau_{11.3\mu m}$, and therefore should have no significant impact in terms of order of magnitude estimations.}, although some extend to large optical depths ($\tau_{10\mu m} > 1$). Their mass-loss rates span a large range of values from $10^{-10}$ to $10^{-5}$ $M_{\odot}$ year$^{-1}$. The range of values is also qualitatively consistent with some estimated mass-loss rates of evolved stars \citep{1988A&AS...72..259D, 1990A&A...231..134N, 2005A&A...438..273V, 2009ApJ...698.1136G, 2009A&A...506.1277G, 2015MNRAS.448..502M, Hofner2018,2019ApJ...885..113M}. As expected, we observe greater opacities accompanied by greater mass-loss rates. Taken together, the results indicate that these anomalous periodic variables are typically dusty. \\

To better understand the nature of these anomalies, we plot their luminosities $L$ versus surface effective temperatures $T_{\text{eff}}$ obtained from the SED fitting in Figure \ref{fig:lmt}. Our anomalies span a range of temperature of $3.4 < \text{log}_{10}T_{\text{eff}} (\text{K}) < 3.65$. Most have a luminosity of $3 < \text{log}_{10}L (L_{\odot}) < 4$, with some more luminous individuals having $\text{log}_{10}L (L_{\odot}) > 4$. These results indicate that they are cool and luminous, which is consistent with evolved stars. In the same Figure, we tabulated the MIST model of theoretical stellar evolution paths computed by \citet{2016ApJS..222....8D} and \citet{2016ApJ...823..102C} using the \texttt{MESA} code \citep{2011ApJS..192....3P, 2013ApJS..208....4P, 2015ApJS..220...15P}. For the \texttt{MESA} models, we have assumed a metallicity of $Z = 0.045$, and a rigid rotational velocity of $0.4$ of the Kepler velocity; these parameters have been slightly tuned to match the sample. The position of our anomalous stars on the HR diagram spans the range for single stars from roughly $1$ $M_{\odot}$ to $20$ $M_{\odot}$, although many of the anomalous stars seems consistent with having a lower mass ($\lesssim 5$ $M_\odot$). They are largely consistent with RGB (data points lying on top of dotted lines) and AGB (data points lying on top of dash-dotted lines) stars. \\

Assuming that these anomalies are indeed RGB/AGB stars, we consider possible explanations for why these events were selected as anomalous by our algorithm. Some anomalous stars that are consistent with RGB models (as shown in Figure \ref{fig:lmt}) could be very evolved, past the RGB tip. The increase of the luminosity is consistent with the evolution of very evolved RGB stars that have considerable mass-loss and produce significant dust \citep{1981ApJ...247..607C, 2003ApJ...582L..43C}. Some RGB stars are also observed to pulsate irregularly \citep{2020JAVSO..48...50P}. These are consistent with our findings in their highly irregular light curves and large mass-loss rates. \\

For anomalies consistent with the AGB, the variability may be explained by thermal pulsations, where significant mass-losses are expected \citep{1999IAUS..191..591I, 2005AJ....130..776T, 2014ApJ...790...22R, 2016ApJ...822...73R}. Thermal pulsating AGB stars have been observed with changing periods, which could lead to a semi-regular light curve \citep{2016AN....337..293U, 2016JAVSO..44..179N}. We remark that spectroscopic follow-up, specifically to measure carbon and s-process elemental abundances \citep{2004MNRAS.352..984S}, would be beneficial to confirm whether these anomalies are actually in the thermal pulsating AGB phase. However, we also note that their mass-loss rates are lower than the typical value for the AGB superwind phase ($\sim 10^{-4}$ $M_{\odot}$ year$^{-1}$, see, for instance, \citet{2008MNRAS.390L..59L}). Last but not least, the stellar object itself could be pulsating with secondary or higher-order period \citep{2004ApJ...604..800W, 2008BaltA..17..223M, 2009MNRAS.399.2063N, 2015MNRAS.452.3863S, 2019ApJ...879...62M} that makes their light curves highly irregular. Nonetheless, we do not rule out the possibilities that some of these anomalies are low mass post-AGB stars, given that they could be located on top of the post-AGB track. Indeed, irregular post-AGB pulsators have been observed \citep{2007MNRAS.375.1338K}. However, such a scenario would be highly unlikely. Their Galactic scale height as shown in Figure \ref{fig:milky} post loose constraints on their ages ($\lesssim 10$ Gyr), which is approximately the time for low-mass stars to take to evolve into the post-AGB phase. \\

Why are these highly variable events seemingly all dusty? Mass-loss events which can drive dust production include thermal pulsations \citep{1984ApJ...283..313C, 2012Natur.490..232M}, and dust-driven stellar winds \citep{2003A&A...409..715W, 2008A&A...486..497W}. The variable and highly irregular light curves could be related to the distribution and dynamics of the dust (e.g., asymmetric and in-homogeneous local dust distribution, as well as the motion of dust and their interaction with the electromagnetic signals). In addition, since the Galactic positions of most of the anomalous stars are consistent with them being young and massive, another possibility to consider is dust production in the outflows of a non-conservative binary mass transfer. This conjecture is justified by the multiplicity fraction of stars, which is known to be increasing with mass \citep[e.g.][]{2017ApJS..230...15M}, and by observations of massive mass transferring systems which show copious amounts of dust production \citep[e.g.,][]{2011MNRAS.418.1959S, 2015MNRAS.450.2551M}. However, our study has only taken a limited amount of spectral information, and therefore we suggest detailed spectroscopic follow-up studies to confirm or rule out some of our enumerated possibilities. \\

Finally, we remark that the \textit{masses} of these anomalies are poorly constrained. Although their Galactic scale heights indicate that they are compatible with young and massive ($\gtrsim 5$ $M_\odot$) evolved stars, having low-mass evolved stars in the Galactic plane is also compatible with state-of-the-art Milky Way stellar initial mass functions\footnote{Since the Milky Way galactic disk is a star factory, while stellar initial mass function favors the production of low-mass stars.} \citep{2021MNRAS.507..398H, 2021MNRAS.504.2557D}. Indeed, low-mass evolved stars have been observed around the Milky Way galactic disk \citep{2009A&A...503L..21M, 2009A&A...500L..25D, 2016A&A...593A.125S, 2019ApJ...883..177N}. Spectroscopic studies are essential in understanding the \textit{metallicities} and thus the \textit{age} of these anomalous stars. In addition, it is possible that the seemingly low mass objects are extincted by additional dust. Although we have accounted for the interstellar extinction, there is always a possibility that dust substructures in the Milky Way are unaccounted for. Furthermore, systematic errors in stellar evolution models that we have assumed might contribute significantly to their estimated temperature and luminosity. Detailed stellar modeling is needed to better constrain their masses.  \\

\subsubsection{Anomalies Consistent With Super-AGB Stars} \label{subsec:particular}
In this section, we highlight two particularly interesting examples of anomalous periodic variable stars with exceptionally large luminosities and unusual light-curve morphologies. First, ZTFJ200229.70+295140.8 is classified as a SR variable in the ZTF CPVS. Our SED fitting result yields $T_{eff} = 3200$ K and $\text{log}_{10}L (L_{\odot}) = 4.31$. It shows double-peaked features in the unfolded light curve taken from the ASAS-SN catalog \citep{2014ApJ...788...48S, 2017PASP..129j4502K, 2018MNRAS.477.3145J}. We note that this object is classified as a Carbon star in the SIMBAD \citep{simbad} catalog, indicating a carbon-rich atmosphere. Second, ZTFJ191700.77+100441.0 is classified as a Mira variable, with $T_{eff} = 4100$ K and $\text{log}_{10}L (L_{\odot}) = 4.5$. Its unfolded light curve as captured by ZTF also shows weak double-peak features. These two stars share characteristics that are consistent with super-AGB stars discovered by \citet{2021AAS...23812103O}, which are stars that could explode as electron-capture supernovae \citep{2015ApJ...807..184F, 2019PASA...36....6L, 2019ApJ...886...22Z, 2020ApJ...889...34L}. \\

\subsubsection{Anomalies Inconsistent With Evolved Stars} \label{subsec:individual}
In this section, we highlight some of the anomalies that are not consistent with an evolved star model, due to either lower luminosities or bluer colors. \\

Periodic variables ZTFJ231957.42+584446.2, ZTFJ174129.18+393218.8, and ZTFJ182424.12-015058.3 have absolute $g$-band magnitudes of $5.2$, $9.9$, and $4.1$ respectively. They are classified as SR variables in the ZTF CPVS; however, they are dimmer than typical SR variables. Their SEDs could not be well-fitted by the \texttt{DESK} code. Their positions on the Gaia HR-Diagram (c.f. Figure \ref{fig:hrdiagram} (a)) indicate that they may instead be YSOs, objects which are evolving from a protostar to the zero-age main sequence. Some YSOs oscillate periodically \citep{2019A&A...627A.135B}, while the others undergo highly irregular accretion-disk-driven variations \citep{2015SciA....1E0686S, 2018JAVSO..46...83H, 2021AJ....162..101V}. YSOs are characterized by the width and slope $\alpha_{IR}$ of their SEDs at $\lambda = 2.2$ $\mu$m
\begin{equation}
    \alpha_{IR} = \frac{\partial\text{log}(\lambda F_{\lambda})}{\partial \text{log}(\lambda)}\bigg|_{\lambda = 2.2 \mu m}
\end{equation}
To approximate the slope at $\lambda = 2.2$ $\mu$m, we linearly interpolate the spectral flux between the $K_{s}$ and W1 band. \\

The SED of ZTFJ231957.42+584446.2 (shown as red circles in Figure \ref{fig:ysofit}) is broad, with a slope of $\alpha_{IR} \sim -0.37$ at $\lambda = 2.2$ $\mu$m. It is consistent with a YSO that is surrounded by hot circumstellar materials, in which the SED of the latter is superposed on that of the stellar black body \citep{Andre2011, Castro2013}, known as a Class II YSO \citep{Andre2011}. \\

The SEDs of ZTFJ174129.18+393218.8 (shown as orange stars in Figure \ref{fig:ysofit}) and ZTFJ182424.12-015058.3 (shown as orange triangles in Figure \ref{fig:ysofit}), do \textit{not} show significant infrared excess. Their SEDs are also narrower when compared to that of ZTFJ231957.42+584446.2, and have measured slopes of $\alpha_{IR} \sim -2.48$ and $-2.25$ at $\lambda = 2.2$ $\mu$m, respectively. One possible explanation for the irregularity and lack of IR excess could be that these are YSOs evolving in between class II and class III \citep{Andre2011}, in which the latter is characterized by having a narrow SED and a slope of $\alpha_{IR} < -1.5$ at $\lambda = 2.2$ $\mu$m that is believed to be caused by optically thin debris. We remark that in addition to spectroscopic follow-up, X-ray observations could also constrain the YSO origin of these objects \citep{2007A&A...463..275G}, as Class II/III YSOs should have strong X-ray emission. \\

Periodic variables ZTFJ195021.95+294256.0 and ZTFJ190426.77+025538.2 were classified as SR and RSCVN variables, respectively. Their SEDs (shown as blue squares and diamonds in Figure \ref{fig:ysofit}) show infrared excesses superposed on a black body SED. However, we note that they are both highly elongated when viewed in archival PanSTARRS DR1 \citep{2017yCat.2349....0C} images \citep[image extraction facilitated by the Aladin Sky Atlas,][]{2000A&AS..143...33B, 2014ASPC..485..277B}, hinting that they could be blended by another stellar object, i.e. either it is a binary system or there is another star in the line of sight. Higher-resolution, follow-up observations could better resolve these potential binaries.

\section{Conclusion} \label{sec:conclusion}
In this study, we presented an unsupervised learning approach of searching for anomalous periodic variables. Light curves of periodic variable stars as given by the ZTF Catalog of Periodic Variable Stars were encoded using a convolutional variational autoencoder to create a low dimensional latent representation. We searched for anomalous periodic variables in this learned latent space via an isolation forest. Our pipeline generates a list of these anomalous periodic variables, which can be sorted by class label.\\

We identify anomalies that are mostly semi-regular and Mira variables having high-variability, irregular light curves. Through modeling the spectral energy distributions, we hypothesize that most of the anomalous events are evolved stars (Red Giant Branch or Asymptotic Giant Branch stars) surrounded by dusty envelopes and concentrated in the Milky Way galactic disk. A fraction of our anomalous events seems consistent with young stellar objects. We discussed their peculiarities and enumerate several possible explanations to account for their anomalous light curves. Detailed spectroscopic follow-up observations and light curve modeling are \textit{essential} in order to fully understand the anomalies detected in our data-driven pipeline and how these anomalies fit into our current understanding of late-stage stellar processes. \\

Astrophysical science has entered the big-data era. The number of periodic variables will only continue to grow exponentially as new surveys, such as LSST, come online. We show that our pipeline is a robust and automatic way to search for anomalous periodic variables within a sea of light curves. We anticipate such a method can be applied to upcoming surveys with increasingly large data sets.

\acknowledgments
We thank the Flatiron institute for providing computer cluster access. We thank Prof. Ming-Chung, Chu in the Chinese University of Hong Kong for providing the NVIDIA RTX2080-Ti GPU as the computational resources for the neural network training. We also thank Mathieu Renzo and Adam Jermyn at the Center of Computational Astrophysics, Flatiron Institute for their valuable discussion on the anomalous periodic variables stars. We additionally acknowledge the SNAD team, whose database was frequently utilized throughout this work. VAV acknowledges support by the Simons Foundation through a Simons Junior Fellowship (\#718240) during the early phases of this project, as well as support in part by the NSF through grant AST-2108676.

\appendix
\section{Network Structure} \label{app:structure}
\begin{deluxetable}{lc}[ht!]
\caption{Hyper-parameters for the autoencoder \label{tab:hyperpara}}
\tablewidth{0pt}
\tablehead{
\colhead{Hyper-parameters} & \colhead{Size}
}
\startdata
Epoch & $895$ \\
Batch Size & $1024$ \\
Convolution Layers $N_{L}$ & $3$ \\
Neurons In The First Layer $N_{1}$ & $32$ \\
Neurons In The Second Layer $N_{2}$ & $64$ \\
Neurons In The Third Layer $N_{3}$ & $128$ \\
Neurons In After The Flatten Layer $N_{D}$ & $256$ \\
Latent Dimension & $10$ \\
Kernel Size & $(3, 4)$ \\
Strides & $(1, 2)$ \\
Dropout Fraction & $0.1$
\enddata
\tablecomments{Rough-gridded, parameter searches were conducted over several of the hyperparameters. Final hyperparameters are listed here.}
\end{deluxetable}
We choose the convolutional neural network as the backbone for the encoder and the decoder, as described in Section \ref{sec:method}. The encoder consists of $N_{L} = 3$ convolutional layers, $1$ dense layer, and a $2\times 10$ latent dimension. The structure for the encoder is described below
\begin{enumerate}
    \item \textbf{Input layer} of size $2\times 160$
    \item \textbf{Convolutional layer} of filter size $N_{1}$ with the \textit{ReLu activation} and \textit{Dropout}.
    \item \textbf{Convolutional layer} of filter size $N_{2}$ with the \textit{ReLu activation} and \textit{Dropout}.
    \item \textbf{Convolutional layer} of filter size $N_{3}$ with the \textit{ReLu activation} and \textit{Dropout}.
    \item \textbf{Dense layer} with $N_{D}$ neurons, \textit{Linear Activation}
    \item \textbf{Latent space} of size $2\times 10$
\end{enumerate}
For simplicity, we describe the network architecture for the encoder only. The decoder is the symmetric counterpart of the encoder, without periodic padding and dropout. We summarize the hyperparameters in Table \ref{tab:hyperpara}. To optimize the chosen hyperparameters, we perform rough grid searches on $N_{1}$, $N_{2}$, $N_{3}$, $N_{D}$, the size of the latent space, and dropout fractions. We applied early-stopping with a patience of $500$ epochs. We train our final model for $895$ epochs, which took roughly $6$ hours on the NVIDIA RTX2080-Ti GPU.

\section{Additional Labels} \label{app:newlabels}
\begin{figure*}[ht!]
\gridline{\fig{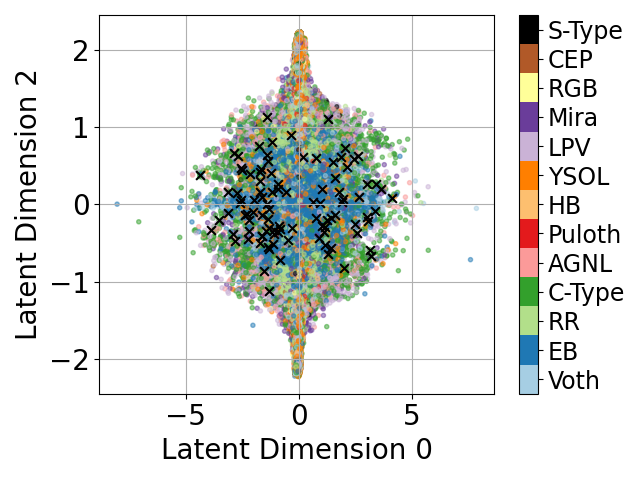}{0.3\textwidth}{(a)}
          \fig{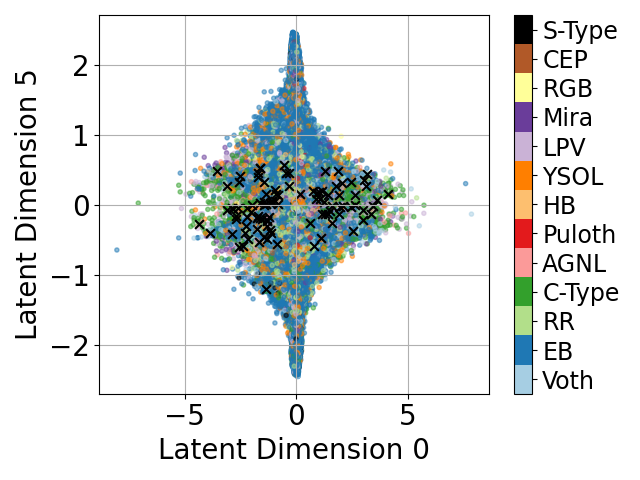}{0.3\textwidth}{(b)}
          \fig{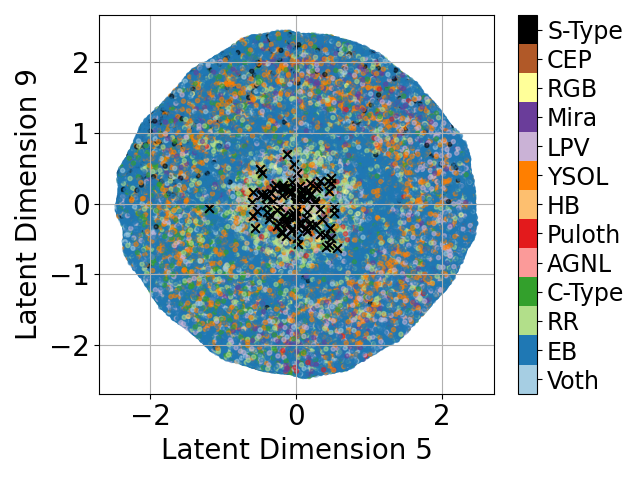}{0.3\textwidth}{(c)}
         }
\gridline{\fig{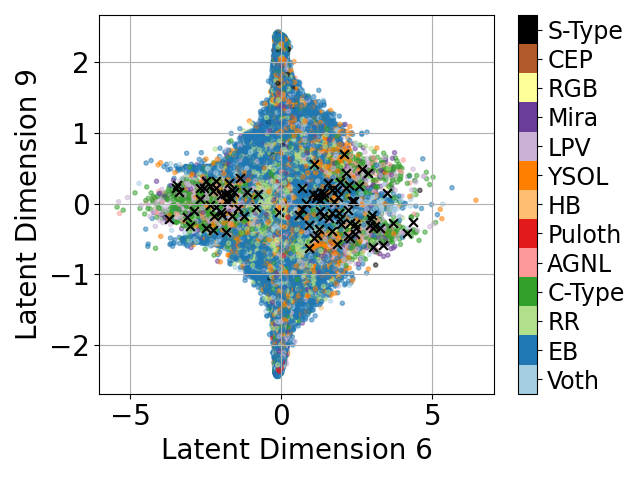}{0.3\textwidth}{(d)}
          \fig{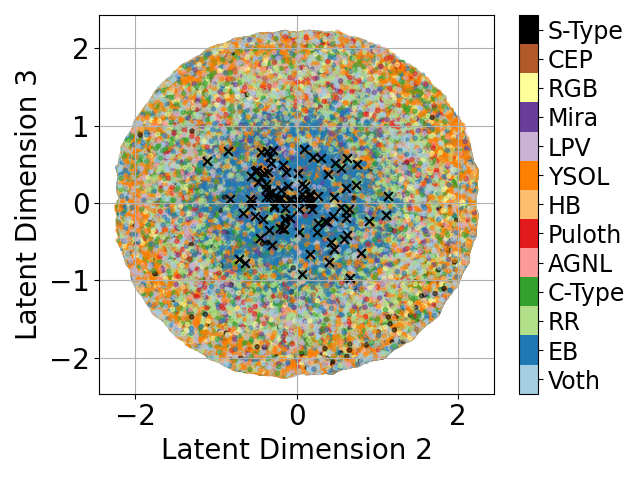}{0.3\textwidth}{(e)}
          \fig{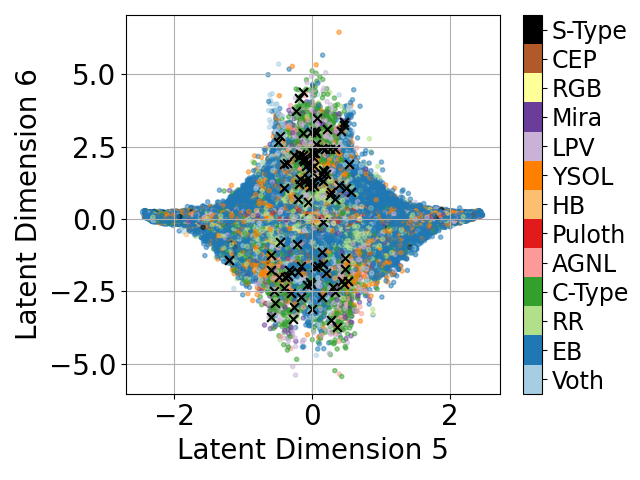}{0.3\textwidth}{(f)}
         }
\caption{Same as Figure \ref{fig:latent}, but using class labels obtained from Cheung et al. (2021, in prep), and the anomalies are marked as dark crosses. These example latent plots show clustering of different classes of periodic variables, and they are seemingly separating from each other. \label{fig:peterslabel}}
\end{figure*}

In addition to the class labels provided by \citet{Chen_2020}, we have acquired class labels for periodic variable stars generated by Cheung et al. (2021, in prep) using a hierarchical isolation forest, where the labels were cross-matched with the SIMBAD database \citep{simbad}. They include YSO like (YSOL), Eclipsing Binary (EB), RR Lyrae variables (RR), Other Variables (V$_\mathrm{oth}$), Mira variables (Mira), AGN like (AGNL), other Pulsating Variables (Pul$_\mathrm{oth}$), Horizontal Branch stars (HB), Long-Period Variable stars (LPV), S-Type stars (S-Type), Carbon stars (C-Type), Red Giant Branch stars (RGB), and Cepheid variables (CEP). The classification is done by training the random forest with the latent vector $\vec{\mu}$ plus extra hand-engineered features, including the joint period, and the amplitude and mean magnitudes of both the $g$- and $r$- band light curves. We have shown the latent distribution with these new class labels in Figure \ref{fig:peterslabel}. As with the original ZTF CPVS labels, the latent space shows separation between the various physical classes.

\section{Error Evaluation} \label{app:error}
\begin{figure*}[ht!]
\gridline{\fig{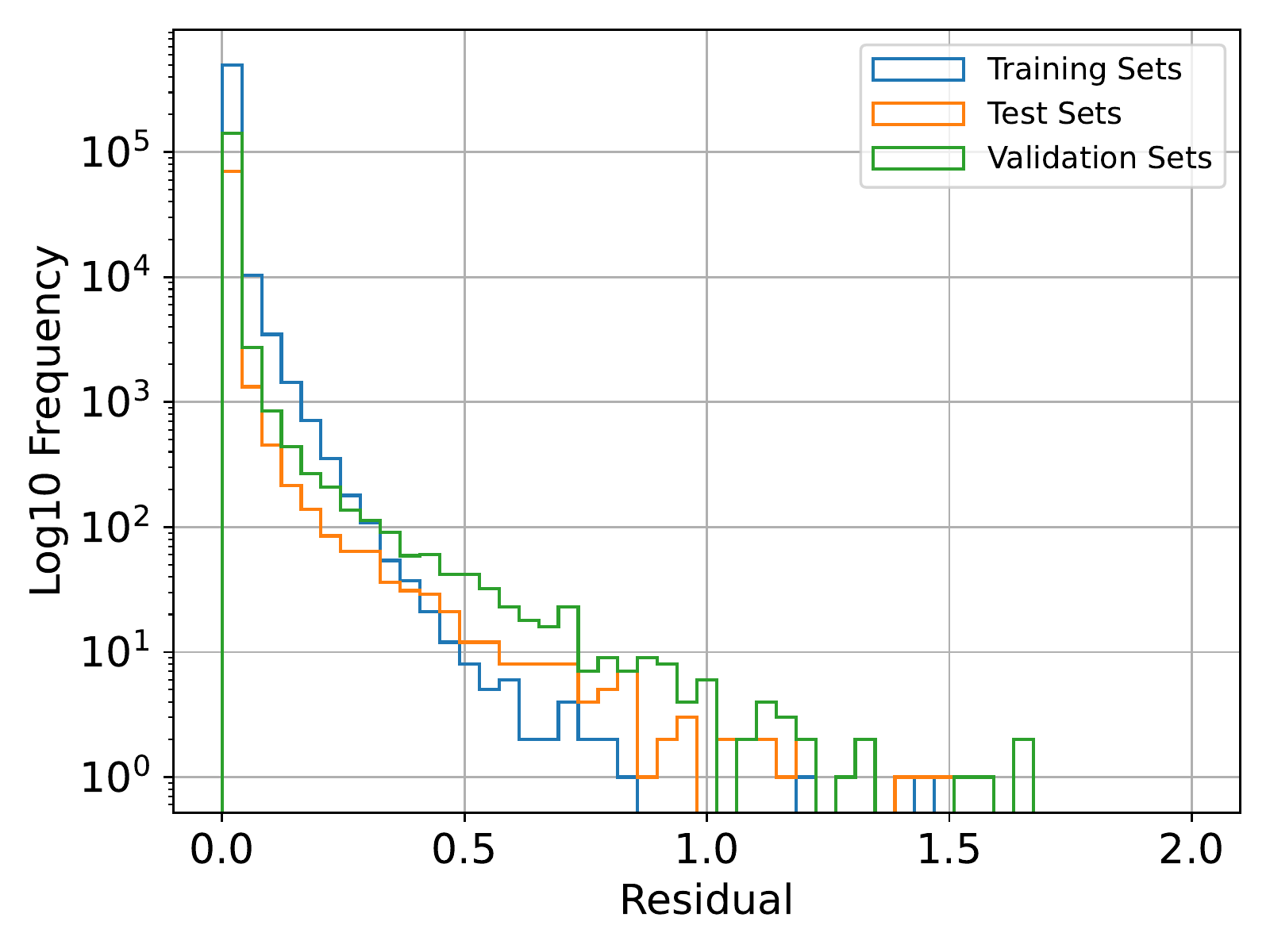}{0.5\textwidth}{(a)}
          \fig{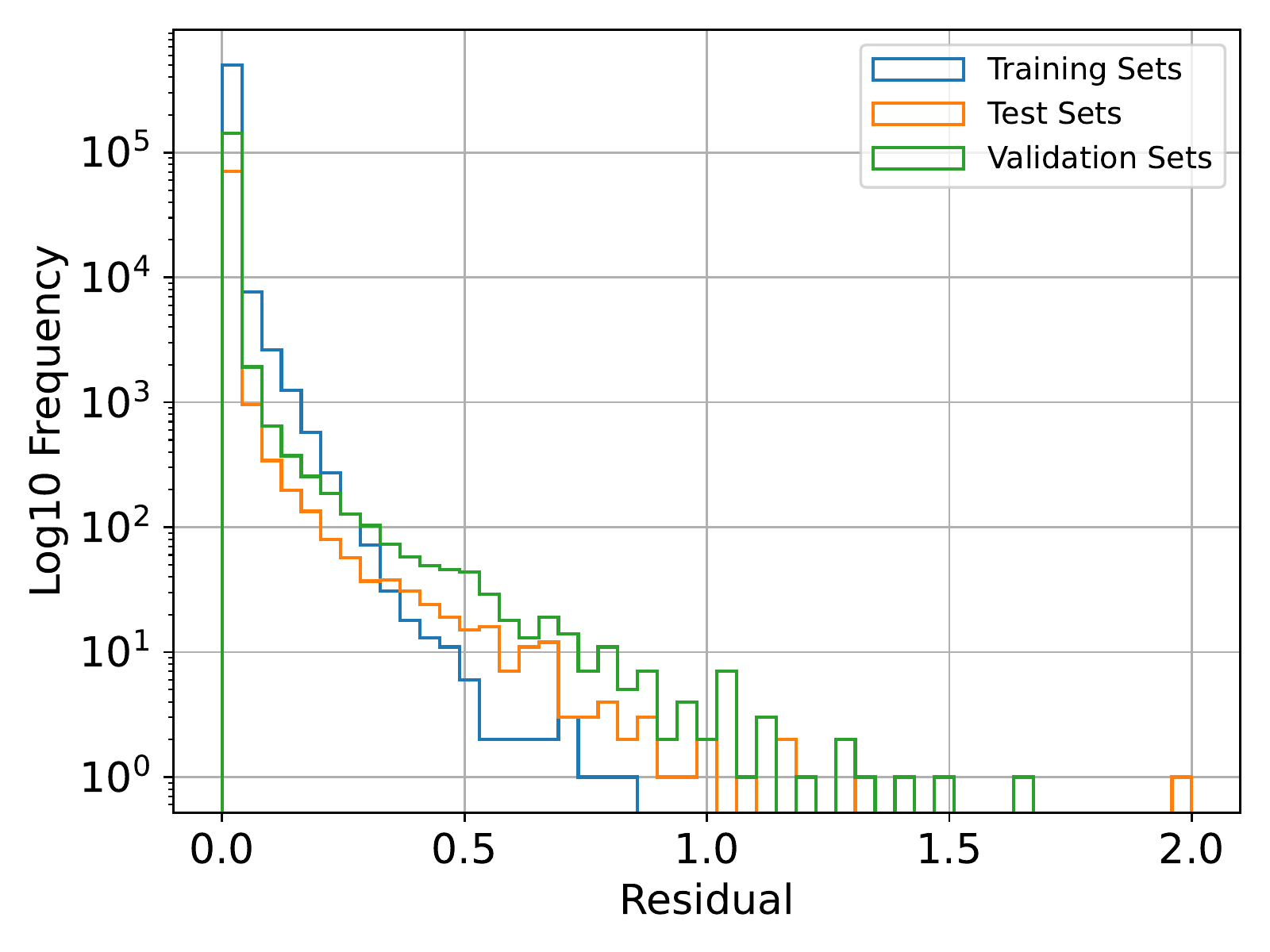}{0.5\textwidth}{(b)}
          }
\caption{Histogram of the residual function for (a) $g$-band, and (b) $r$-band respectively. The majority of the samples from the training, validation, and test sets are having residual $< 1$, indicating that our neural network is reconstructing the light curves with reasonable accuracy. \label{fig:histogramerror}}
\end{figure*}
\begin{figure*}[ht!]
\gridline{\fig{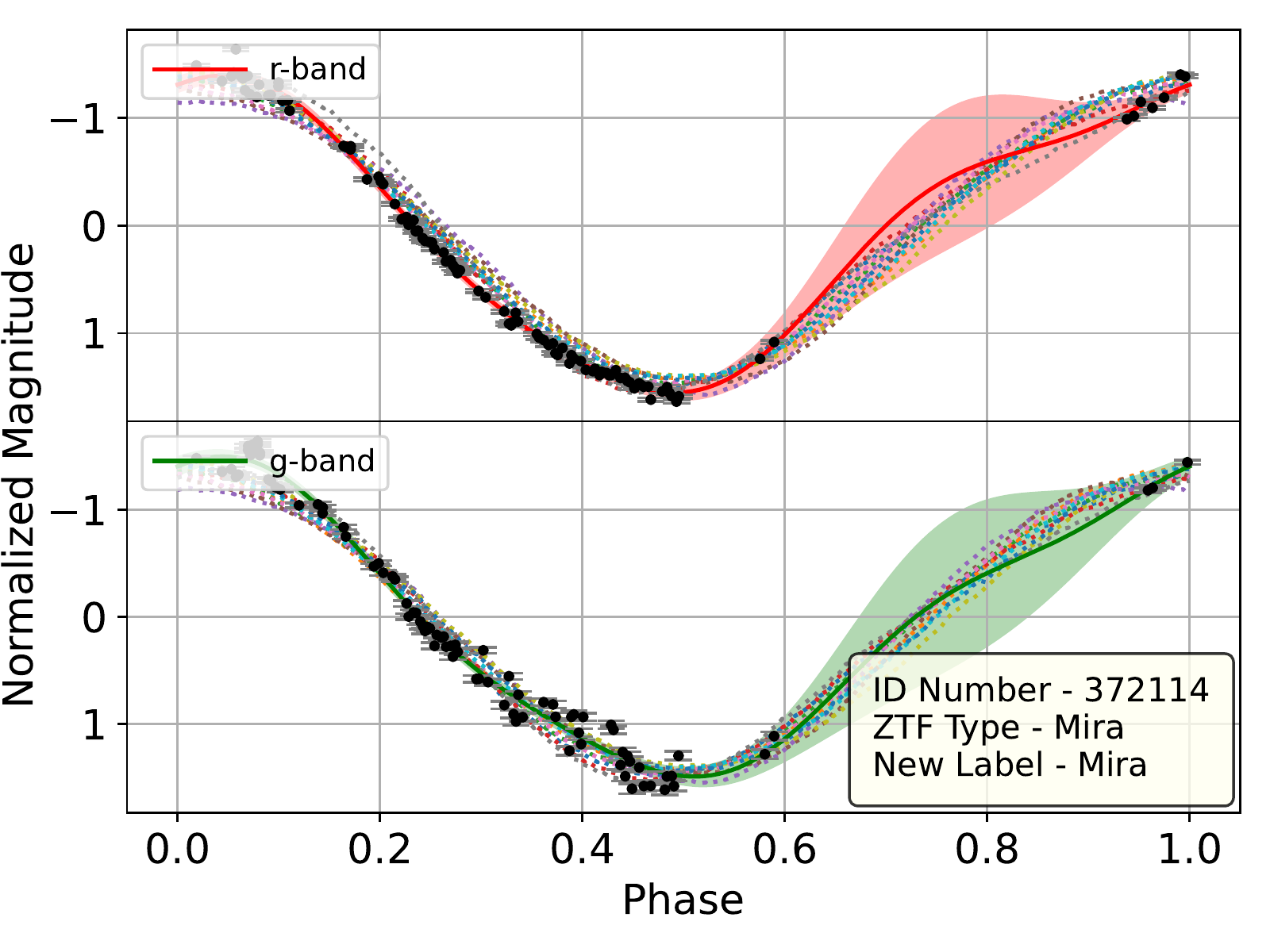}{0.5\textwidth}{(a)}
          \fig{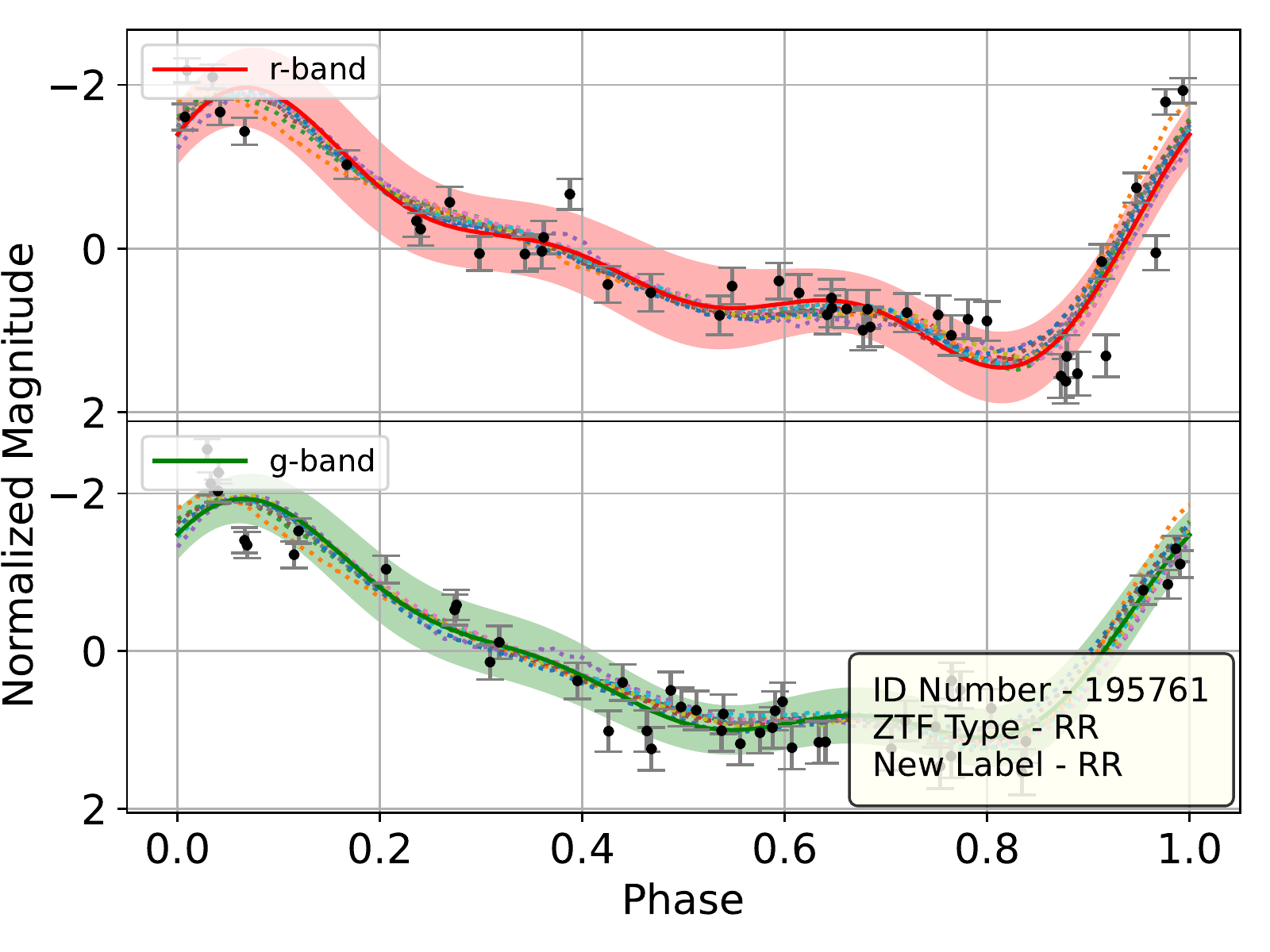}{0.5\textwidth}{(b)}
          }
\caption{Example plots of the MGPR versus the neural network reconstruction results for the (a) Mira, and (b) RR variables extracted from the test sets. The light curves sampled from the latent spaces are shown in dotted lines. Note that the magnitudes are standard-scalar normalized. These example plots show that our neural network is capable of reproducing the light curves, although our convolutional variational autoencoder seems to underestimate the GP-estimated error. \label{fig:testaccuracy}}
\end{figure*}
We estimate the error of our neural network reconstruction using the following residual function
\begin{equation}
    \text{Res} = \frac{1}{N}\sum_{N} \frac{(Y_{\text{NN}} - Y_{\text{MGPR}})^{2}}{\sigma_{GPR}^{2}}
\end{equation}
Here, $N$ is the number of data points \textit{per light curves for each band}. To compute $Y_{\text{NN}}$, we randomly sampled $10$ light curves (as a multivariate Gaussian) from the latent space and compute the average of normalized magnitude for each band of the light curve. The results for the $g$- and $r$- band for the training, testing, and validation sets are shown as histograms in Figure \ref{fig:histogramerror}. We observe that most of the events have residual values of less than $1$, which indicates that our neural network is encoding and decoding the light curves with reasonable accuracy. We have also shown two examples in the test sets of the MGPR versus neural network reconstructions in Figure \ref{fig:testaccuracy}.

\section{Brief Description Of The Periodic Variable Stars} \label{app:briefreview}
\textbf{BY Draconis (BY Dra)} are periodic variable stars of the K and M spectral type, mainly found in the solar neighborhood. They are characterized by quasiperiodic photometric oscillations lasting less than a day to months and low amplitude variations ranging from a few hundredths to 0.5 mag \citep{1978A&A....64..153B, 2006PASP..118.1506L}.\\

\textbf{Eclipsing W UMa (EW)} variables are strongly interacting, eclipsing binaries systems in which both stars usually fill their critical Roche lobes and share a common envelope. They typically have orbital periods between about 5 and 20 hours. Their light curves do not show sharp `dips' as typical eclipsing binaries do, and therefore it is impossible to identify the exact times of onset and end of eclipses. \citep{selam, Qian_2017}. \\

\textbf{Eclipsing Algol (EA)} variables consist of a pair of stars that are comparable in size but with different masses. The more massive, hotter star is typically a main sequence mid-B to mid-F star, which has not filled its Roche lobe. The less massive, cooler, fainter, and larger star is believed to evolve beyond the main sequence \citep{2004A&A...417..263B, 2006ApJ...649..973C}. \\

\textbf{Semi-Regular (SR)} variables are pulsating K or M type giants or supergiants located on the AGB of the Hertzsprung-Russell diagram. They show quasi-periodic or even multi-periodic oscillation with periods between 30 and several thousand days. They are also characterized by their small variation in magnitude of $< 0.2$ mag \citep{1999A&A...346..542K, sr1999, Alard_2001, 2014AN....335.1072F}. \\

\textbf{RS Canum Venaticorum (RSCVN)} variables are binaries with chromospherically active late-type stars. They show out-of-eclipse variability likely caused by starspots that rotate into and out of the field of view. Depending on the location of the star spot, the period of variability may or may not match the period of rotation \citep{1990ASPC....9..195R, 2001JAVSO..29...82P, 2017RAA....17...55C}. \\

\textbf{RR Lyrae (RR)} variable stars common in globular clusters. They follow standardizable period-luminosity relations in the near-infrared. Moreover, they have been used to measure the age of globular clusters, as well as the chemical and dynamical properties of nearby galaxies \citep{2003LNP...635..105C, 2004rrls.book.....S, braga_2019, 2021ApJ...909..200B, 2021ApJ...915...50H}. \\

\textbf{RR Lyrae Type C (RRc)} variable stars are a sub-class of the RR variable stars. Unlike the majority of RRa/RRb variable stars which radially oscillate in their fundamental mode, RRc stars are pulsating in their first overtone. RRc stars also tend to lie on the hot side of the instability strip in the Hertzsprung-Russell diagram \citep{2012JAVSO..40..327S, 2013JAVSO..41...75P}. \\

\textbf{Delta Scuti (DSCT)} variables are stars located on the lower edge of the instability strip or are evolving from the main-sequence to the giant branch. They are of spectral type A or F with a period of less than $0.3$ days. They mostly oscillate with nonradial p-modes. They follow period-luminosity relation, which also makes them a useful standardizable candle to the Milky Way galactic Bulge and the Magellanic Clouds \citep{1979PASP...91....5B, 1999A&A...352..547P, 2000ASPC..210....3B}. \\

\textbf{Mira} variables are late-type, long-period variable stars located on the coolest and most luminous part of the AGB in the Hertzsprung-Russell diagram. They exhibit large luminosity amplitudes with periods of more than 80 days. They are also shown to follow a period-luminosity relation in the infrared region, which have been used as distance indicators \citep{2002MNRAS.329L...7F, 2005AJ....130..776T, 2017ApJS..232...16Y}. \\

\textbf{Cepheid (CEP)} variables are young supergiants. They are bright and follow a period-luminosity relation, so they are widely used as distance indicators. They have also been used to determine the Hubble constant, as well as to confirm the accelerated expansion of the universe \citep{Madore1998, 2007AJ....133.1810B, 2013MNRAS.430..546D, 2019Sci...365..478S}. \\

\textbf{Type II Cepheid (CEPII)} variables are located in the Cepheid instability strip. They are low-mass, old, He-burning stars that are either in the post-horizontal branch, AGB, or post-AGB phase. They are fainter than classical Cepheids. They also follow period-luminosity relations in the optical and near-infrared bands and have been utilized as distance indicators \citep{2018A&A...619A..51B, 2018MNRAS.477.2276K, 2019ApJ...883...58D, 2020A&A...644A..95B}.

\newpage
\section{Truncated Table For All Periodic Variables} \label{app:top100}
\begin{deluxetable}{cccccc}[ht!]
\caption{Truncated table for all periodic variables space ranked by their anomaly scores computed in the Cartesian latent space. Here, we list only the top 15 anomalous periodic variables together with their attribute. A full machine-readable table is available in the electronic version of the manuscript. \label{tab:topanomalycartesian}}
\tablewidth{0pt}
\tablehead{
\colhead{ZTF Identifier} & \colhead{Catalog Internal Identifier} & \colhead{Class Label} & \colhead{RA (Deg)} & \colhead{Dec (Deg)} & \colhead{Anomaly Score}
}
\startdata
ZTFJ192024.63+051215.3 & 414363 & Mira & 290.10266 & 5.20426 & 1.00000 \\
ZTFJ200658.04+290821.0 & 575168 & SR & 301.74186 & 29.13917 & 0.95635 \\
ZTFJ190726.45+110744.0 & 366883 & SR & 286.86024 & 11.1289 & 0.95476 \\
ZTFJ185635.63+094400.8 & 331786 & Mira & 284.14848 & 9.73358 & 0.93694 \\
ZTFJ193506.54+171844.1 & 466871 & SR & 293.77729 & 17.31226 & 0.93541 \\
ZTFJ192659.95+191124.4 & 438180 & SR & 291.74983 & 19.19012 & 0.93135 \\
ZTFJ194827.66+264926.9 & 518456 & SR & 297.11528 & 26.82415 & 0.92922 \\
ZTFJ231957.42+584446.2 & 764950 & SR & 349.98927 & 58.74618 & 0.92879 \\
ZTFJ193956.80+202637.6 & 486646 & SR & 294.98669 & 20.44378 & 0.92510 \\
ZTFJ201228.51+304322.6 & 587449 & Mira & 303.1188 & 30.72296 & 0.92477 \\
ZTFJ191200.21+130256.2 & 385339 & SR & 288.0009 & 13.04897 & 0.92455 \\
ZTFJ193108.60+222622.7 & 452461 & SR & 292.78585 & 22.43965 & 0.92393 \\
ZTFJ201141.08+335655.8 & 585656 & SR & 302.92119 & 33.94884 & 0.92137 \\
ZTFJ184248.01+031052.7 & 297439 & SR & 280.70005 & 3.18133 & 0.92079 \\
ZTFJ184326.14+031005.3 & 298955 & SR & 280.85892 & 3.16816 & 0.91883 
\enddata
\end{deluxetable}
\begin{deluxetable}{cccccc}[ht!]
\caption{Same as Table \ref{tab:topanomalycartesian},  but for the anomaly scores computed in the Spherical latent space. \label{tab:topanomalyspherical}}
\tablewidth{0pt}
\tablehead{
\colhead{ZTF Identifier} & \colhead{Catalog Internal Identifier} & \colhead{Class Label} & \colhead{RA (Deg)} & \colhead{Dec (Deg)} & \colhead{Anomaly Score}
}
\startdata
ZTFJ181542.63-094000.5 & 255911 & SR & 273.92763 & -9.66683 & 1.00000 \\
ZTFJ200425.27+313412.9 & 568824 & SR & 301.1053 & 31.57027 & 0.98176 \\
ZTFJ191410.29+125023.8 & 392650 & SR & 288.54289 & 12.83997 & 0.98037 \\
ZTFJ204714.25+452208.0 & 637533 & SR & 311.80938 & 45.36891 & 0.97722 \\
ZTFJ195243.06+245556.1 & 533756 & SR & 298.17943 & 24.93227 & 0.97436 \\
ZTFJ193737.09+244839.4 & 476451 & SR & 294.40457 & 24.81095 & 0.96000 \\
ZTFJ211036.37+543858.3 & 664688 & SR & 317.65156 & 54.64955 & 0.95911 \\
ZTFJ183750.68-042050.4 & 286982 & SR & 279.46118 & -4.34736 & 0.95854 \\
ZTFJ185007.52+045822.0 & 314960 & SR & 282.53137 & 4.9728 & 0.95718 \\ 
ZTFJ201229.74+374537.7 & 587495 & SR & 303.12392 & 37.76049 & 0.95595 \\
ZTFJ194238.06+244155.5 & 497353 & SR & 295.6586 & 24.69875 & 0.95250 \\
ZTFJ190143.24+052613.6 & 346249 & SR & 285.4302 & 5.43712 & 0.95114 \\
ZTFJ185447.33+053955.4 & 326476 & SR & 283.69722 & 5.66541 & 0.94981 \\
ZTFJ184808.18-041601.8 & 310240 & SR & 282.03409 & -4.26717 & 0.94732 \\
ZTFJ190559.43+122112.9 & 361526 & Mira & 286.49764 & 12.3536 & 0.93624 \\
\enddata
\end{deluxetable}

\bibliography{main}{}

\begin{thebibliography}{}
\expandafter\ifx\csname natexlab\endcsname\relax\def\natexlab#1{#1}\fi
\providecommand{\url}[1]{\href{#1}{#1}}
\providecommand{\dodoi}[1]{doi:~\href{http://doi.org/#1}{\nolinkurl{#1}}}
\providecommand{\doeprint}[1]{\href{http://ascl.net/#1}{\nolinkurl{http://ascl.net/#1}}}
\providecommand{\doarXiv}[1]{\href{https://arxiv.org/abs/#1}{\nolinkurl{https://arxiv.org/abs/#1}}}

\bibitem[{Abadi {et~al.}(2015)Abadi, Agarwal, Barham, Brevdo, Chen, Citro,
  Corrado, Davis, Dean, Devin, Ghemawat, Goodfellow, Harp, Irving, Isard, Jia,
  Jozefowicz, Kaiser, Kudlur, Levenberg, Man\'{e}, Monga, Moore, Murray, Olah,
  Schuster, Shlens, Steiner, Sutskever, Talwar, Tucker, Vanhoucke, Vasudevan,
  Vi\'{e}gas, Vinyals, Warden, Wattenberg, Wicke, Yu, \&
  Zheng}]{tensorflow2015-whitepaper}
Abadi, M., Agarwal, A., Barham, P., {et~al.} 2015, {TensorFlow}: Large-Scale
  Machine Learning on Heterogeneous Systems.
\newblock \url{https://www.tensorflow.org/}

\bibitem[{Alard {et~al.}(2001)Alard, Blommaert, Cesarsky, Epchtein, Felli,
  Fouque, Ganesh, Genzel, Gilmore, Glass, \& et~al.}]{Alard_2001}
Alard, C., Blommaert, J. A. D.~L., Cesarsky, C., {et~al.} 2001, The
  Astrophysical Journal, 552, 289–308, \dodoi{10.1086/320440}

\bibitem[{Andr{\'e}(2011)}]{Andre2011}
Andr{\'e}, P. 2011, Spectral Classification of Embedded Stars, ed. M.~Gargaud,
  R.~Amils, J.~C. Quintanilla, H.~J.~J. Cleaves, W.~M. Irvine, D.~L. Pinti, \&
  M.~Viso (Berlin, Heidelberg: Springer Berlin Heidelberg), 1549--1553,
  \dodoi{10.1007/978-3-642-11274-4_504}

\bibitem[{{Bailer-Jones} {et~al.}(2021){Bailer-Jones}, {Rybizki}, {Fouesneau},
  {Demleitner}, \& {Andrae}}]{2021AJ....161..147B}
{Bailer-Jones}, C.~A.~L., {Rybizki}, J., {Fouesneau}, M., {Demleitner}, M., \&
  {Andrae}, R. 2021, \aj, 161, 147, \dodoi{10.3847/1538-3881/abd806}

\bibitem[{{Banerjee} {et~al.}(2021){Banerjee}, {Garain}, {Paul}, {Shaikh}, \&
  {Sarkar}}]{2021ApJ...910...23B}
{Banerjee}, P., {Garain}, D., {Paul}, S., {Shaikh}, R., \& {Sarkar}, T. 2021,
  \apj, 910, 23, \dodoi{10.3847/1538-4357/abded3}

\bibitem[{Baron \& Poznanski(2016)}]{10.1093/mnras/stw3021}
Baron, D., \& Poznanski, D. 2016, Monthly Notices of the Royal Astronomical
  Society, 465, 4530, \dodoi{10.1093/mnras/stw3021}

\bibitem[{{Bellm} {et~al.}(2019){Bellm}, {Kulkarni}, {Graham}, {Dekany},
  {Smith}, {Riddle}, {Masci}, {Helou}, {Prince}, {Adams}, {Barbarino},
  {Barlow}, {Bauer}, {Beck}, {Belicki}, {Biswas}, {Blagorodnova}, {Bodewits},
  {Bolin}, {Brinnel}, {Brooke}, {Bue}, {Bulla}, {Burruss}, {Cenko}, {Chang},
  {Connolly}, {Coughlin}, {Cromer}, {Cunningham}, {De}, {Delacroix}, {Desai},
  {Duev}, {Eadie}, {Farnham}, {Feeney}, {Feindt}, {Flynn}, {Franckowiak},
  {Frederick}, {Fremling}, {Gal-Yam}, {Gezari}, {Giomi}, {Goldstein},
  {Golkhou}, {Goobar}, {Groom}, {Hacopians}, {Hale}, {Henning}, {Ho}, {Hover},
  {Howell}, {Hung}, {Huppenkothen}, {Imel}, {Ip}, {Ivezi{\'c}}, {Jackson},
  {Jones}, {Juric}, {Kasliwal}, {Kaspi}, {Kaye}, {Kelley}, {Kowalski},
  {Kramer}, {Kupfer}, {Landry}, {Laher}, {Lee}, {Lin}, {Lin}, {Lunnan},
  {Giomi}, {Mahabal}, {Mao}, {Miller}, {Monkewitz}, {Murphy}, {Ngeow},
  {Nordin}, {Nugent}, {Ofek}, {Patterson}, {Penprase}, {Porter}, {Rauch},
  {Rebbapragada}, {Reiley}, {Rigault}, {Rodriguez}, {van Roestel}, {Rusholme},
  {van Santen}, {Schulze}, {Shupe}, {Singer}, {Soumagnac}, {Stein}, {Surace},
  {Sollerman}, {Szkody}, {Taddia}, {Terek}, {Van Sistine}, {van Velzen},
  {Vestrand}, {Walters}, {Ward}, {Ye}, {Yu}, {Yan}, \&
  {Zolkower}}]{2019PASP..131a8002B}
{Bellm}, E.~C., {Kulkarni}, S.~R., {Graham}, M.~J., {et~al.} 2019, \pasp, 131,
  018002, \dodoi{10.1088/1538-3873/aaecbe}

\bibitem[{{Benedict} {et~al.}(2007){Benedict}, {McArthur}, {Feast}, {Barnes},
  {Harrison}, {Patterson}, {Menzies}, {Bean}, \&
  {Freedman}}]{2007AJ....133.1810B}
{Benedict}, G.~F., {McArthur}, B.~E., {Feast}, M.~W., {et~al.} 2007, \aj, 133,
  1810, \dodoi{10.1086/511980}

\bibitem[{Bernhard {et~al.}(2021)Bernhard, Hümmerich, Paunzen, \&
  Supíková}]{10.1093/mnras/stab2065}
Bernhard, K., Hümmerich, S., Paunzen, E., \& Supíková, J. 2021, Monthly
  Notices of the Royal Astronomical Society, \dodoi{10.1093/mnras/stab2065}

\bibitem[{{Bhardwaj} {et~al.}(2019){Bhardwaj}, {Panwar}, {Herczeg}, {Chen}, \&
  {Singh}}]{2019A&A...627A.135B}
{Bhardwaj}, A., {Panwar}, N., {Herczeg}, G.~J., {Chen}, W.~P., \& {Singh},
  H.~P. 2019, \aap, 627, A135, \dodoi{10.1051/0004-6361/201935418}

\bibitem[{{Bhardwaj} {et~al.}(2021){Bhardwaj}, {Rejkuba}, {de Grijs}, {Yang},
  {Herczeg}, {Marconi}, {Singh}, {Kanbur}, \& {Ngeow}}]{2021ApJ...909..200B}
{Bhardwaj}, A., {Rejkuba}, M., {de Grijs}, R., {et~al.} 2021, \apj, 909, 200,
  \dodoi{10.3847/1538-4357/abdf48}

\bibitem[{Blumenson(1960)}]{10.2307/2308932}
Blumenson, L.~E. 1960, The American Mathematical Monthly, 67, 63.
\newblock \url{http://www.jstor.org/stable/2308932}

\bibitem[{{Boch} \& {Fernique}(2014)}]{2014ASPC..485..277B}
{Boch}, T., \& {Fernique}, P. 2014, in Astronomical Society of the Pacific
  Conference Series, Vol. 485, Astronomical Data Analysis Software and Systems
  XXIII, ed. N.~{Manset} \& P.~{Forshay}, 277

\bibitem[{{Bonnarel} {et~al.}(2000){Bonnarel}, {Fernique}, {Bienaym{\'e}},
  {Egret}, {Genova}, {Louys}, {Ochsenbein}, {Wenger}, \&
  {Bartlett}}]{2000A&AS..143...33B}
{Bonnarel}, F., {Fernique}, P., {Bienaym{\'e}}, O., {et~al.} 2000, \aaps, 143,
  33, \dodoi{10.1051/aas:2000331}

\bibitem[{Boutin {et~al.}(2020)Boutin, Zerroug, Jung, \&
  Serre}]{boutin2020iterative}
Boutin, V., Zerroug, A., Jung, M., \& Serre, T. 2020, in NeurIPS 2020 Workshop
  SVRHM.
\newblock \url{https://openreview.net/forum?id=jE6SlVTOFPV}

\bibitem[{{Braga} {et~al.}(2018){Braga}, {Bhardwaj}, {Contreras Ramos},
  {Minniti}, {Bono}, {de Grijs}, {Minniti}, \& {Rejkuba}}]{2018A&A...619A..51B}
{Braga}, V.~F., {Bhardwaj}, A., {Contreras Ramos}, R., {et~al.} 2018, \aap,
  619, A51, \dodoi{10.1051/0004-6361/201833538}

\bibitem[{{Braga} {et~al.}(2020){Braga}, {Bono}, {Fiorentino}, {Stetson},
  {Dall'Ora}, {Salaris}, {da Silva}, {Fabrizio}, {Marinoni}, {Marrese},
  {Mateo}, {Matsunaga}, {Monelli}, \& {Wallerstein}}]{2020A&A...644A..95B}
{Braga}, V.~F., {Bono}, G., {Fiorentino}, G., {et~al.} 2020, \aap, 644, A95,
  \dodoi{10.1051/0004-6361/202039145}

\bibitem[{{Braga, V. F.} {et~al.}(2019){Braga, V. F.}, {Stetson, P. B.}, {Bono,
  G.}, {Dall\'{}Ora, M.}, {Ferraro, I.}, {Fiorentino, G.}, {Iannicola, G.},
  {Inno, L.}, {Marengo, M.}, {Neeley, J.}, {Beaton, R. L.}, {Buonanno, R.},
  {Calamida, A.}, {Contreras Ramos, R.}, {Chaboyer, B.}, {Fabrizio, M.},
  {Freedman, W. L.}, {Gilligan, C. K.}, {Johnston, K. V.}, {Lub, J.}, {Madore,
  B. F.}, {Magurno, D.}, {Marconi, M.}, {Marinoni, S.}, {Marrese, P. M.},
  {Mateo, M.}, {Matsunaga, N.}, {Minniti, D.}, {Monson, A. J.}, {Monelli, M.},
  {Nonino, M.}, {Persson, S. E.}, {Pietrinferni, A.}, {Sneden, C.}, {Storm,
  J.}, {Walker, A. R.}, {Valenti, E.}, \& {Zoccali, M.}}]{braga_2019}
{Braga, V. F.}, {Stetson, P. B.}, {Bono, G.}, {et~al.} 2019, A\&A, 625, A1,
  \dodoi{10.1051/0004-6361/201834893}

\bibitem[{{Breger}(1979)}]{1979PASP...91....5B}
{Breger}, M. 1979, \pasp, 91, 5, \dodoi{10.1086/130433}

\bibitem[{{Breger}(2000)}]{2000ASPC..210....3B}
{Breger}, M. 2000, in Astronomical Society of the Pacific Conference Series,
  Vol. 210, Delta Scuti and Related Stars, ed. M.~{Breger} \& M.~{Montgomery},
  3

\bibitem[{{Budding} {et~al.}(2004){Budding}, {Erdem}, {{\c{C}}i{\c{c}}ek},
  {Bulut}, {Soydugan}, {Soydugan}, {Baki{\c{s}}}, \&
  {Demircan}}]{2004A&A...417..263B}
{Budding}, E., {Erdem}, A., {{\c{C}}i{\c{c}}ek}, C., {et~al.} 2004, \aap, 417,
  263, \dodoi{10.1051/0004-6361:20034135}

\bibitem[{{Busko} \& {Torres}(1978)}]{1978A&A....64..153B}
{Busko}, I.~C., \& {Torres}, C.~A.~O. 1978, \aap, 64, 153

\bibitem[{Byrd {et~al.}(1995)Byrd, Lu, Nocedal, \&
  Zhu}]{53712fe04a3448cfb8598b14afab59b3}
Byrd, R., Lu, P., Nocedal, J., \& Zhu, C. 1995, SIAM Journal of Scientific
  Computing, 16, 1190, \dodoi{10.1137/0916069}

\bibitem[{{Cacciari} \& {Clementini}(2003)}]{2003LNP...635..105C}
{Cacciari}, C., \& {Clementini}, G. 2003, {Globular Cluster Distances from RR
  Lyrae Stars}, ed. D.~{Alloin} \& W.~{Gieren}, Vol. 635, 105--122,
  \dodoi{10.1007/978-3-540-39882-0\_6}

\bibitem[{{Cao} \& {Gu}(2017)}]{2017RAA....17...55C}
{Cao}, D.-T., \& {Gu}, S.-H. 2017, Research in Astronomy and Astrophysics, 17,
  055, \dodoi{10.1088/1674-4527/17/6/55}

\bibitem[{{Cassisi} {et~al.}(2003){Cassisi}, {Schlattl}, {Salaris}, \&
  {Weiss}}]{2003ApJ...582L..43C}
{Cassisi}, S., {Schlattl}, H., {Salaris}, M., \& {Weiss}, A. 2003, \apjl, 582,
  L43, \dodoi{10.1086/346200}

\bibitem[{Castro(2013)}]{Castro2013}
Castro, A. I. G.~d. 2013, Young Stellar Objects and Protostellar Disks, ed.
  T.~D. Oswalt \& M.~A. Barstow (Dordrecht: Springer Netherlands), 279--335,
  \dodoi{10.1007/978-94-007-5615-1_6}

\bibitem[{{Chambers} \& {et al.}(2017)}]{2017yCat.2349....0C}
{Chambers}, K.~C., \& {et al.} 2017, VizieR Online Data Catalog, II/349

\bibitem[{{Chambers} {et~al.}(2016){Chambers}, {Magnier}, {Metcalfe},
  {Flewelling}, {Huber}, {Waters}, {Denneau}, {Draper}, {Farrow}, {Finkbeiner},
  {Holmberg}, {Koppenhoefer}, {Price}, {Rest}, {Saglia}, {Schlafly}, {Smartt},
  {Sweeney}, {Wainscoat}, {Burgett}, {Chastel}, {Grav}, {Heasley}, {Hodapp},
  {Jedicke}, {Kaiser}, {Kudritzki}, {Luppino}, {Lupton}, {Monet}, {Morgan},
  {Onaka}, {Shiao}, {Stubbs}, {Tonry}, {White}, {Ba{\~n}ados}, {Bell},
  {Bender}, {Bernard}, {Boegner}, {Boffi}, {Botticella}, {Calamida},
  {Casertano}, {Chen}, {Chen}, {Cole}, {Deacon}, {Frenk}, {Fitzsimmons},
  {Gezari}, {Gibbs}, {Goessl}, {Goggia}, {Gourgue}, {Goldman}, {Grant},
  {Grebel}, {Hambly}, {Hasinger}, {Heavens}, {Heckman}, {Henderson}, {Henning},
  {Holman}, {Hopp}, {Ip}, {Isani}, {Jackson}, {Keyes}, {Koekemoer}, {Kotak},
  {Le}, {Liska}, {Long}, {Lucey}, {Liu}, {Martin}, {Masci}, {McLean}, {Mindel},
  {Misra}, {Morganson}, {Murphy}, {Obaika}, {Narayan}, {Nieto-Santisteban},
  {Norberg}, {Peacock}, {Pier}, {Postman}, {Primak}, {Rae}, {Rai}, {Riess},
  {Riffeser}, {Rix}, {R{\"o}ser}, {Russel}, {Rutz}, {Schilbach}, {Schultz},
  {Scolnic}, {Strolger}, {Szalay}, {Seitz}, {Small}, {Smith}, {Soderblom},
  {Taylor}, {Thomson}, {Taylor}, {Thakar}, {Thiel}, {Thilker}, {Unger},
  {Urata}, {Valenti}, {Wagner}, {Walder}, {Walter}, {Watters}, {Werner},
  {Wood-Vasey}, \& {Wyse}}]{2016arXiv161205560C}
{Chambers}, K.~C., {Magnier}, E.~A., {Metcalfe}, N., {et~al.} 2016, arXiv
  e-prints, arXiv:1612.05560.
\newblock \doarXiv{1612.05560}

\bibitem[{Chandola {et~al.}(2009)Chandola, Banerjee, \&
  Kumar}]{10.1145/1541880.1541882}
Chandola, V., Banerjee, A., \& Kumar, V. 2009, ACM Comput. Surv., 41,
  \dodoi{10.1145/1541880.1541882}

\bibitem[{Chen {et~al.}(2016)Chen, Karl, \& van~der Smagt}]{7803340}
Chen, N., Karl, M., \& van~der Smagt, P. 2016, in 2016 IEEE-RAS 16th
  International Conference on Humanoid Robots (Humanoids), 629--636,
  \dodoi{10.1109/HUMANOIDS.2016.7803340}

\bibitem[{{Chen} {et~al.}(2006){Chen}, {Li}, \& {Qian}}]{2006ApJ...649..973C}
{Chen}, W.-C., {Li}, X.-D., \& {Qian}, S.-B. 2006, \apj, 649, 973,
  \dodoi{10.1086/506433}

\bibitem[{{Chen} {et~al.}(2018){Chen}, {Wang}, {Deng}, {de Grijs}, \&
  {Yang}}]{2018ApJS..237...28C}
{Chen}, X., {Wang}, S., {Deng}, L., {de Grijs}, R., \& {Yang}, M. 2018, \apjs,
  237, 28, \dodoi{10.3847/1538-4365/aad32b}

\bibitem[{Chen {et~al.}(2020)Chen, Wang, Deng, de~Grijs, Yang, \&
  Tian}]{Chen_2020}
Chen, X., Wang, S., Deng, L., {et~al.} 2020, The Astrophysical Journal
  Supplement Series, 249, 18, \dodoi{10.3847/1538-4365/ab9cae}

\bibitem[{{Choi} {et~al.}(2016){Choi}, {Dotter}, {Conroy}, {Cantiello},
  {Paxton}, \& {Johnson}}]{2016ApJ...823..102C}
{Choi}, J., {Dotter}, A., {Conroy}, C., {et~al.} 2016, \apj, 823, 102,
  \dodoi{10.3847/0004-637X/823/2/102}

\bibitem[{{Cole} \& {Deupree}(1981)}]{1981ApJ...247..607C}
{Cole}, P.~W., \& {Deupree}, R.~G. 1981, \apj, 247, 607, \dodoi{10.1086/159071}

\bibitem[{{Cosner} {et~al.}(1984){Cosner}, {Despain}, \&
  {Truran}}]{1984ApJ...283..313C}
{Cosner}, K.~R., {Despain}, K.~H., \& {Truran}, J.~W. 1984, \apj, 283, 313,
  \dodoi{10.1086/162308}

\bibitem[{{Cutri} {et~al.}(2021){Cutri}, {Wright}, {Conrow}, {Fowler},
  {Eisenhardt}, {Grillmair}, {Kirkpatrick}, {Masci}, {McCallon}, {Wheelock},
  {Fajardo-Acosta}, {Yan}, {Benford}, {Harbut}, {Jarrett}, {Lake}, {Leisawitz},
  {Ressler}, {Stanford}, {Tsai}, {Liu}, {Helou}, {Mainzer}, {Gettngs},
  {Gonzalez}, {Hoffman}, {Marsh}, {Padgett}, {Skrutskie}, {Beck}, {Papin}, \&
  {Wittman}}]{2014yCat.2328....0C}
{Cutri}, R.~M., {Wright}, E.~L., {Conrow}, T., {et~al.} 2021, VizieR Online
  Data Catalog, II/328

\bibitem[{{Damian} {et~al.}(2021){Damian}, {Jose}, {Samal}, {Moraux}, {Das}, \&
  {Patra}}]{2021MNRAS.504.2557D}
{Damian}, B., {Jose}, J., {Samal}, M.~R., {et~al.} 2021, \mnras, 504, 2557,
  \dodoi{10.1093/mnras/stab194}

\bibitem[{{de Jager} {et~al.}(1988){de Jager}, {Nieuwenhuijzen}, \& {van der
  Hucht}}]{1988A&AS...72..259D}
{de Jager}, C., {Nieuwenhuijzen}, H., \& {van der Hucht}, K.~A. 1988, \aaps,
  72, 259

\bibitem[{{de Silva} {et~al.}(2009){de Silva}, {Gibson}, {Lattanzio}, \&
  {Asplund}}]{2009A&A...500L..25D}
{de Silva}, G.~M., {Gibson}, B.~K., {Lattanzio}, J., \& {Asplund}, M. 2009,
  \aap, 500, L25, \dodoi{10.1051/0004-6361/200912279}

\bibitem[{{D{\'e}k{\'a}ny} {et~al.}(2019){D{\'e}k{\'a}ny}, {Hajdu}, {Grebel},
  \& {Catelan}}]{2019ApJ...883...58D}
{D{\'e}k{\'a}ny}, I., {Hajdu}, G., {Grebel}, E.~K., \& {Catelan}, M. 2019,
  \apj, 883, 58, \dodoi{10.3847/1538-4357/ab3b60}

\bibitem[{{Di Benedetto}(2013)}]{2013MNRAS.430..546D}
{Di Benedetto}, G.~P. 2013, \mnras, 430, 546, \dodoi{10.1093/mnras/sts655}

\bibitem[{{Dotter}(2016)}]{2016ApJS..222....8D}
{Dotter}, A. 2016, \apjs, 222, 8, \dodoi{10.3847/0067-0049/222/1/8}

\bibitem[{Drake {et~al.}(2014)Drake, Graham, Djorgovski, Catelan, Mahabal,
  Torrealba, Garc{\'{\i}}a-{\'{A}}lvarez, Donalek, Prieto, Williams, Larson,
  sen, Belokurov, Koposov, Beshore, Boattini, Gibbs, Hill, Kowalski, Johnson,
  \& Shelly}]{Drake_2014}
Drake, A.~J., Graham, M.~J., Djorgovski, S.~G., {et~al.} 2014, The
  Astrophysical Journal Supplement Series, 213, 9,
  \dodoi{10.1088/0067-0049/213/1/9}

\bibitem[{{Drake} {et~al.}(2017){Drake}, {Djorgovski}, {Catelan}, {Graham},
  {Mahabal}, {Larson}, {Christensen}, {Torrealba}, {Beshore}, {McNaught},
  {Garradd}, {Belokurov}, \& {Koposov}}]{2017MNRAS.469.3688D}
{Drake}, A.~J., {Djorgovski}, S.~G., {Catelan}, M., {et~al.} 2017, \mnras, 469,
  3688, \dodoi{10.1093/mnras/stx1085}

\bibitem[{Eyer \& Mowlavi(2008)}]{Eyer_2008}
Eyer, L., \& Mowlavi, N. 2008, Journal of Physics: Conference Series, 118,
  012010, \dodoi{10.1088/1742-6596/118/1/012010}

\bibitem[{{Farmer} {et~al.}(2015){Farmer}, {Fields}, \&
  {Timmes}}]{2015ApJ...807..184F}
{Farmer}, R., {Fields}, C.~E., \& {Timmes}, F.~X. 2015, \apj, 807, 184,
  \dodoi{10.1088/0004-637X/807/2/184}

\bibitem[{{Feast} {et~al.}(2002){Feast}, {Whitelock}, \&
  {Menzies}}]{2002MNRAS.329L...7F}
{Feast}, M., {Whitelock}, P., \& {Menzies}, J. 2002, \mnras, 329, L7,
  \dodoi{10.1046/j.1365-8711.2002.05126.x}

\bibitem[{Foreman-Mackey {et~al.}(2014)Foreman-Mackey, Hoyer, Bernhard, \&
  Angus}]{dan_foreman_mackey_2014_11989}
Foreman-Mackey, D., Hoyer, S., Bernhard, J., \& Angus, R. 2014, george: George
  (v0.2.0), v0.2.0,  Zenodo, \dodoi{10.5281/zenodo.11989}

\bibitem[{{Fuentes-Morales} \& {Vogt}(2014)}]{2014AN....335.1072F}
{Fuentes-Morales}, I., \& {Vogt}, N. 2014, Astronomische Nachrichten, 335,
  1072, \dodoi{10.1002/asna.201412117}

\bibitem[{Fustes {et~al.}(2013)Fustes, Dafonte, Arcay, Manteiga, Smith,
  Vallenari, \& Luri}]{FUSTES20131530}
Fustes, D., Dafonte, C., Arcay, B., {et~al.} 2013, Expert Systems with
  Applications, 40, 1530, \dodoi{https://doi.org/10.1016/j.eswa.2012.08.069}

\bibitem[{{Gaia Collaboration} {et~al.}(2019){Gaia Collaboration}, {Eyer, L.},
  {Rimoldini, L.}, {Audard, M.}, {Anderson, R. I.}, {Nienartowicz, K.}, {Glass,
  F.}, {Marchal, O.}, {Grenon, M.}, {Mowlavi, N.}, {Holl, B.}, {Clementini,
  G.}, {Aerts, C.}, {Mazeh, T.}, {Evans, D. W.}, {Szabados, L.}, {Brown, A. G.
  A.}, {Vallenari, A.}, {Prusti, T.}, {de Bruijne, J. H. J.}, {Babusiaux, C.},
  {Bailer-Jones, C. A. L.}, {Biermann, M.}, {Jansen, F.}, {Jordi, C.},
  {Klioner, S. A.}, {Lammers, U.}, {Lindegren, L.}, {Luri, X.}, {Mignard, F.},
  {Panem, C.}, {Pourbaix, D.}, {Randich, S.}, {Sartoretti, P.}, {Siddiqui, H.
  I.}, {Soubiran, C.}, {van Leeuwen, F.}, {Walton, N. A.}, {Arenou, F.},
  {Bastian, U.}, {Cropper, M.}, {Drimmel, R.}, {Katz, D.}, {Lattanzi, M. G.},
  {Bakker, J.}, {Cacciari, C.}, {Casta\~neda, J.}, {Chaoul, L.}, {Cheek, N.},
  {De Angeli, F.}, {Fabricius, C.}, {Guerra, R.}, {Masana, E.}, {Messineo, R.},
  {Panuzzo, P.}, {Portell, J.}, {Riello, M.}, {Seabroke, G. M.}, {Tanga, P.},
  {Th\'evenin, F.}, {Gracia-Abril, G.}, {Comoretto, G.}, {Garcia-Reinaldos,
  M.}, {Teyssier, D.}, {Altmann, M.}, {Andrae, R.}, {Bellas-Velidis, I.},
  {Benson, K.}, {Berthier, J.}, {Blomme, R.}, {Burgess, P.}, {Busso, G.},
  {Carry, B.}, {Cellino, A.}, {Clotet, M.}, {Creevey, O.}, {Davidson, M.}, {De
  Ridder, J.}, {Delchambre, L.}, {Dell\'{}Oro, A.}, {Ducourant, C.},
  {Fern\'andez-Hern\'andez, J.}, {Fouesneau, M.}, {Fr\'emat, Y.}, {Galluccio,
  L.}, {Garc\'{\i}a-Torres, M.}, {Gonz\'alez-N\'u\~nez, J.}, {Gonz\'alez-Vidal,
  J. J.}, {Gosset, E.}, {Guy, L. P.}, {Halbwachs, J.-L.}, {Hambly, N. C.},
  {Harrison, D. L.}, {Hern\'andez, J.}, {Hestroffer, D.}, {Hodgkin, S. T.},
  {Hutton, A.}, {Jasniewicz, G.}, {Jean-Antoine-Piccolo, A.}, {Jordan, S.},
  {Korn, A. J.}, {Krone-Martins, A.}, {Lanzafame, A. C.}, {Lebzelter, T.},
  {L\"offler, W.}, {Manteiga, M.}, {Marrese, P. M.}, {Mart\'{\i}n-Fleitas, J.
  M.}, {Moitinho, A.}, {Mora, A.}, {Muinonen, K.}, {Osinde, J.}, {Pancino, E.},
  {Pauwels, T.}, {Petit, J.-M.}, {Recio-Blanco, A.}, {Richards, P. J.}, {Robin,
  A. C.}, {Sarro, L. M.}, {Siopis, C.}, {Smith, M.}, {Sozzetti, A.},
  {S\"uveges, M.}, {Torra, J.}, {van Reeven, W.}, {Abbas, U.}, {Abreu Aramburu,
  A.}, {Accart, S.}, {Altavilla, G.}, {\'Alvarez, M. A.}, {Alvarez, R.},
  {Alves, J.}, {Andrei, A. H.}, {Anglada Varela, E.}, {Antiche, E.}, {Antoja,
  T.}, {Arcay, B.}, {Astraatmadja, T. L.}, {Bach, N.}, {Baker, S. G.},
  {Balaguer-N\'u\~nez, L.}, {Balm, P.}, {Barache, C.}, {Barata, C.}, {Barbato,
  D.}, {Barblan, F.}, {Barklem, P. S.}, {Barrado, D.}, {Barros, M.}, {Barstow,
  M. A.}, {Bartholom\'e Mu\~noz, S.}, {Bassilana, J.-L.}, {Becciani, U.},
  {Bellazzini, M.}, {Berihuete, A.}, {Bertone, S.}, {Bianchi, L.}, {Bienaym\'e,
  O.}, {Blanco-Cuaresma, S.}, {Boch, T.}, {Boeche, C.}, {Bombrun, A.},
  {Borrachero, R.}, {Bossini, D.}, {Bouquillon, S.}, {Bourda, G.}, {Bragaglia,
  A.}, {Bramante, L.}, {Breddels, M. A.}, {Bressan, A.}, {Brouillet, N.},
  {Br\"usemeister, T.}, {Brugaletta, E.}, {Bucciarelli, B.}, {Burlacu, A.},
  {Busonero, D.}, {Butkevich, A. G.}, {Buzzi, R.}, {Caffau, E.}, {Cancelliere,
  R.}, {Cannizzaro, G.}, {Cantat-Gaudin, T.}, {Carballo, R.}, {Carlucci, T.},
  {Carrasco, J. M.}, {Casamiquela, L.}, {Castellani, M.}, {Castro-Ginard, A.},
  {Charlot, P.}, {Chemin, L.}, {Chiavassa, A.}, {Cocozza, G.}, {Costigan, G.},
  {Cowell, S.}, {Crifo, F.}, {Crosta, M.}, {Crowley, C.}, {Cuypers, J.},
  {Dafonte, C.}, {Damerdji, Y.}, {Dapergolas, A.}, {David, P.}, {David, M.},
  {de Laverny, P.}, {De Luise, F.}, {De March, R.}, {de Martino, D.}, {de
  Souza, R.}, {de Torres, A.}, {Debosscher, J.}, {del Pozo, E.}, {Delbo, M.},
  {Delgado, A.}, {Delgado, H. E.}, {Diakite, S.}, {Diener, C.}, {Distefano,
  E.}, {Dolding, C.}, {Drazinos, P.}, {Dur\'an, J.}, {Edvardsson, B.}, {Enke,
  H.}, {Eriksson, K.}, {Esquej, P.}, {Eynard Bontemps, G.}, {Fabre, C.},
  {Fabrizio, M.}, {Faigler, S.}, {Falc\~ao, A. J.}, {Farr\`as Casas, M.},
  {Federici, L.}, {Fedorets, G.}, {Fernique, P.}, {Figueras, F.}, {Filippi,
  F.}, {Findeisen, K.}, {Fonti, A.}, {Fraile, E.}, {Fraser, M.}, {Fr\'ezouls,
  B.}, {Gai, M.}, {Galleti, S.}, {Garabato, D.}, {Garc\'{\i}a-Sedano, F.},
  {Garofalo, A.}, {Garralda, N.}, {Gavel, A.}, {Gavras, P.}, {Gerssen, J.},
  {Geyer, R.}, {Giacobbe, P.}, {Gilmore, G.}, {Girona, S.}, {Giuffrida, G.},
  {Gomes, M.}, {Granvik, M.}, {Gueguen, A.}, {Guerrier, A.}, {Guiraud, J.},
  {Guti\'errez-S\'anchez, R.}, {Haigron, R.}, {Hatzidimitriou, D.}, {Hauser,
  M.}, {Haywood, M.}, {Heiter, U.}, {Helmi, A.}, {Heu, J.}, {Hilger, T.},
  {Hobbs, D.}, {Hofmann, W.}, {Holland, G.}, {Huckle, H. E.}, {Hypki, A.},
  {Icardi, V.}, {Jan\ss{}en, K.}, {Jevardat de Fombelle, G.}, {Jonker, P. G.},
  {Juh\'asz, \'A. L.}, {Julbe, F.}, {Karampelas, A.}, {Kewley, A.}, {Klar, J.},
  {Kochoska, A.}, {Kohley, R.}, {Kolenberg, K.}, {Kontizas, M.}, {Kontizas,
  E.}, {Koposov, S. E.}, {Kordopatis, G.}, {Kostrzewa-Rutkowska, Z.}, {Koubsky,
  P.}, {Lambert, S.}, {Lanza, A. F.}, {Lasne, Y.}, {Lavigne, J.-B.}, {Le
  Fustec, Y.}, {Le Poncin-Lafitte, C.}, {Lebreton, Y.}, {Leccia, S.}, {Leclerc,
  N.}, {Lecoeur-Taibi, I.}, {Lenhardt, H.}, {Leroux, F.}, {Liao, S.}, {Licata,
  E.}, {Lindstr\o{}m, H. E. P.}, {Lister, T. A.}, {Livanou, E.}, {Lobel, A.},
  {L\'opez, M.}, {Lorenz, D.}, {Managau, S.}, {Mann, R. G.}, {Mantelet, G.},
  {Marchant, J. M.}, {Marconi, M.}, {Marinoni, S.}, {Marschalk\'o, G.},
  {Marshall, D. J.}, {Martino, M.}, {Marton, G.}, {Mary, N.}, {Massari, D.},
  {Matijevic, G.}, {McMillan, P. J.}, {Messina, S.}, {Michalik, D.}, {Millar,
  N. R.}, {Molina, D.}, {Molinaro, R.}, {Moln\'ar, L.}, {Montegriffo, P.},
  {Mor, R.}, {Morbidelli, R.}, {Morel, T.}, {Morgenthaler, S.}, {Morris, D.},
  {Mulone, A. F.}, {Muraveva, T.}, {Musella, I.}, {Nelemans, G.}, {Nicastro,
  L.}, {Noval, L.}, {O\'{}Mullane, W.}, {Ord\'enovic, C.}, {Ord\'o\~nez-Blanco,
  D.}, {Osborne, P.}, {Pagani, C.}, {Pagano, I.}, {Pailler, F.}, {Palacin, H.},
  {Palaversa, L.}, {Panahi, A.}, {Pawlak, M.}, {Piersimoni, A. M.}, {Pineau,
  F.-X.}, {Plachy, E.}, {Plum, G.}, {Poggio, E.}, {Poujoulet, E.}, {Prsa, A.},
  {Pulone, L.}, {Racero, E.}, {Ragaini, S.}, {Rambaux, N.}, {Ramos-Lerate, M.},
  {Regibo, S.}, {Reyl\'e, C.}, {Riclet, F.}, {Ripepi, V.}, {Riva, A.}, {Rivard,
  A.}, {Rixon, G.}, {Roegiers, T.}, {Roelens, M.}, {Romero-G\'omez, M.},
  {Rowell, N.}, {Royer, F.}, {Ruiz-Dern, L.}, {Sadowski, G.}, {Sagrist\`a
  Sell\'es, T.}, {Sahlmann, J.}, {Salgado, J.}, {Salguero, E.}, {Sanna, N.},
  {Santana-Ros, T.}, {Sarasso, M.}, {Savietto, H.}, {Schultheis, M.}, {Sciacca,
  E.}, {Segol, M.}, {Segovia, J. C.}, {S\'egransan, D.}, {Shih, I.-C.},
  {Siltala, L.}, {Silva, A. F.}, {Smart, R. L.}, {Smith, K. W.}, {Solano, E.},
  {Solitro, F.}, {Sordo, R.}, {Soria Nieto, S.}, {Souchay, J.}, {Spagna, A.},
  {Spoto, F.}, {Stampa, U.}, {Steele, I. A.}, {Steidelm\"uller, H.},
  {Stephenson, C. A.}, {Stoev, H.}, {Suess, F. F.}, {Surdej, J.},
  {Szegedi-Elek, E.}, {Tapiador, D.}, {Taris, F.}, {Tauran, G.}, {Taylor, M.
  B.}, {Teixeira, R.}, {Terrett, D.}, {Teyssandier, P.}, {Thuillot, W.},
  {Titarenko, A.}, {Torra Clotet, F.}, {Turon, C.}, {Ulla, A.}, {Utrilla, E.},
  {Uzzi, S.}, {Vaillant, M.}, {Valentini, G.}, {Valette, V.}, {van Elteren,
  A.}, {Van Hemelryck, E.}, {van Leeuwen, M.}, {Vaschetto, M.}, {Vecchiato,
  A.}, {Veljanoski, J.}, {Viala, Y.}, {Vicente, D.}, {Vogt, S.}, {von Essen,
  C.}, {Voss, H.}, {Votruba, V.}, {Voutsinas, S.}, {Walmsley, G.}, {Weiler,
  M.}, {Wertz, O.}, {Wevers, T.}, {Wyrzykowski, L.}, {Yoldas, A.}, {Zerjal,
  M.}, {Ziaeepour, H.}, {Zorec, J.}, {Zschocke, S.}, {Zucker, S.}, {Zurbach,
  C.}, \& {Zwitter, T.}}]{gaiacollaboration}
{Gaia Collaboration}, {Eyer, L.}, {Rimoldini, L.}, {et~al.} 2019, A\&A, 623,
  A110, \dodoi{10.1051/0004-6361/201833304}

\bibitem[{{Gaia Collaboration} {et~al.}(2021){Gaia Collaboration}, {Brown},
  {Vallenari}, {Prusti}, {de Bruijne}, {Babusiaux}, {Biermann}, {Creevey},
  {Evans}, {Eyer}, {Hutton}, {Jansen}, {Jordi}, {Klioner}, {Lammers},
  {Lindegren}, {Luri}, {Mignard}, {Panem}, {Pourbaix}, {Randich}, {Sartoretti},
  {Soubiran}, {Walton}, {Arenou}, {Bailer-Jones}, {Bastian}, {Cropper},
  {Drimmel}, {Katz}, {Lattanzi}, {van Leeuwen}, {Bakker}, {Cacciari},
  {Casta{\~n}eda}, {De Angeli}, {Ducourant}, {Fabricius}, {Fouesneau},
  {Fr{\'e}mat}, {Guerra}, {Guerrier}, {Guiraud}, {Jean-Antoine Piccolo},
  {Masana}, {Messineo}, {Mowlavi}, {Nicolas}, {Nienartowicz}, {Pailler},
  {Panuzzo}, {Riclet}, {Roux}, {Seabroke}, {Sordo}, {Tanga}, {Th{\'e}venin},
  {Gracia-Abril}, {Portell}, {Teyssier}, {Altmann}, {Andrae}, {Bellas-Velidis},
  {Benson}, {Berthier}, {Blomme}, {Brugaletta}, {Burgess}, {Busso}, {Carry},
  {Cellino}, {Cheek}, {Clementini}, {Damerdji}, {Davidson}, {Delchambre},
  {Dell'Oro}, {Fern{\'a}ndez-Hern{\'a}ndez}, {Galluccio}, {Garc{\'\i}a-Lario},
  {Garcia-Reinaldos}, {Gonz{\'a}lez-N{\'u}{\~n}ez}, {Gosset}, {Haigron},
  {Halbwachs}, {Hambly}, {Harrison}, {Hatzidimitriou}, {Heiter},
  {Hern{\'a}ndez}, {Hestroffer}, {Hodgkin}, {Holl}, {Jan{\ss}en}, {Jevardat de
  Fombelle}, {Jordan}, {Krone-Martins}, {Lanzafame}, {L{\"o}ffler}, {Lorca},
  {Manteiga}, {Marchal}, {Marrese}, {Moitinho}, {Mora}, {Muinonen}, {Osborne},
  {Pancino}, {Pauwels}, {Petit}, {Recio-Blanco}, {Richards}, {Riello},
  {Rimoldini}, {Robin}, {Roegiers}, {Rybizki}, {Sarro}, {Siopis}, {Smith},
  {Sozzetti}, {Ulla}, {Utrilla}, {van Leeuwen}, {van Reeven}, {Abbas}, {Abreu
  Aramburu}, {Accart}, {Aerts}, {Aguado}, {Ajaj}, {Altavilla}, {{\'A}lvarez},
  {{\'A}lvarez Cid-Fuentes}, {Alves}, {Anderson}, {Anglada Varela}, {Antoja},
  {Audard}, {Baines}, {Baker}, {Balaguer-N{\'u}{\~n}ez}, {Balbinot}, {Balog},
  {Barache}, {Barbato}, {Barros}, {Barstow}, {Bartolom{\'e}}, {Bassilana},
  {Bauchet}, {Baudesson-Stella}, {Becciani}, {Bellazzini}, {Bernet}, {Bertone},
  {Bianchi}, {Blanco-Cuaresma}, {Boch}, {Bombrun}, {Bossini}, {Bouquillon},
  {Bragaglia}, {Bramante}, {Breedt}, {Bressan}, {Brouillet}, {Bucciarelli},
  {Burlacu}, {Busonero}, {Butkevich}, {Buzzi}, {Caffau}, {Cancelliere},
  {C{\'a}novas}, {Cantat-Gaudin}, {Carballo}, {Carlucci}, {Carnerero},
  {Carrasco}, {Casamiquela}, {Castellani}, {Castro-Ginard}, {Castro Sampol},
  {Chaoul}, {Charlot}, {Chemin}, {Chiavassa}, {Cioni}, {Comoretto}, {Cooper},
  {Cornez}, {Cowell}, {Crifo}, {Crosta}, {Crowley}, {Dafonte}, {Dapergolas},
  {David}, {David}, {de Laverny}, {De Luise}, {De March}, {De Ridder}, {de
  Souza}, {de Teodoro}, {de Torres}, {del Peloso}, {del Pozo}, {Delbo},
  {Delgado}, {Delgado}, {Delisle}, {Di Matteo}, {Diakite}, {Diener},
  {Distefano}, {Dolding}, {Eappachen}, {Edvardsson}, {Enke}, {Esquej}, {Fabre},
  {Fabrizio}, {Faigler}, {Fedorets}, {Fernique}, {Fienga}, {Figueras},
  {Fouron}, {Fragkoudi}, {Fraile}, {Franke}, {Gai}, {Garabato},
  {Garcia-Gutierrez}, {Garc{\'\i}a-Torres}, {Garofalo}, {Gavras}, {Gerlach},
  {Geyer}, {Giacobbe}, {Gilmore}, {Girona}, {Giuffrida}, {Gomel}, {Gomez},
  {Gonzalez-Santamaria}, {Gonz{\'a}lez-Vidal}, {Granvik},
  {Guti{\'e}rrez-S{\'a}nchez}, {Guy}, {Hauser}, {Haywood}, {Helmi}, {Hidalgo},
  {Hilger}, {H{\l}adczuk}, {Hobbs}, {Holland}, {Huckle}, {Jasniewicz},
  {Jonker}, {Juaristi Campillo}, {Julbe}, {Karbevska}, {Kervella}, {Khanna},
  {Kochoska}, {Kontizas}, {Kordopatis}, {Korn}, {Kostrzewa-Rutkowska},
  {Kruszy{\'n}ska}, {Lambert}, {Lanza}, {Lasne}, {Le Campion}, {Le Fustec},
  {Lebreton}, {Lebzelter}, {Leccia}, {Leclerc}, {Lecoeur-Taibi}, {Liao},
  {Licata}, {Lindstr{\o}m}, {Lister}, {Livanou}, {Lobel}, {Madrero Pardo},
  {Managau}, {Mann}, {Marchant}, {Marconi}, {Marcos Santos}, {Marinoni},
  {Marocco}, {Marshall}, {Martin Polo}, {Mart{\'\i}n-Fleitas}, {Masip},
  {Massari}, {Mastrobuono-Battisti}, {Mazeh}, {McMillan}, {Messina},
  {Michalik}, {Millar}, {Mints}, {Molina}, {Molinaro}, {Moln{\'a}r},
  {Montegriffo}, {Mor}, {Morbidelli}, {Morel}, {Morris}, {Mulone}, {Munoz},
  {Muraveva}, {Murphy}, {Musella}, {Noval}, {Ord{\'e}novic}, {Orr{\`u}},
  {Osinde}, {Pagani}, {Pagano}, {Palaversa}, {Palicio}, {Panahi}, {Pawlak},
  {Pe{\~n}alosa Esteller}, {Penttil{\"a}}, {Piersimoni}, {Pineau}, {Plachy},
  {Plum}, {Poggio}, {Poretti}, {Poujoulet}, {Pr{\v{s}}a}, {Pulone}, {Racero},
  {Ragaini}, {Rainer}, {Raiteri}, {Rambaux}, {Ramos}, {Ramos-Lerate}, {Re
  Fiorentin}, {Regibo}, {Reyl{\'e}}, {Ripepi}, {Riva}, {Rixon}, {Robichon},
  {Robin}, {Roelens}, {Rohrbasser}, {Romero-G{\'o}mez}, {Rowell}, {Royer},
  {Rybicki}, {Sadowski}, {Sagrist{\`a} Sell{\'e}s}, {Sahlmann}, {Salgado},
  {Salguero}, {Samaras}, {Sanchez Gimenez}, {Sanna}, {Santove{\~n}a},
  {Sarasso}, {Schultheis}, {Sciacca}, {Segol}, {Segovia}, {S{\'e}gransan},
  {Semeux}, {Shahaf}, {Siddiqui}, {Siebert}, {Siltala}, {Slezak}, {Smart},
  {Solano}, {Solitro}, {Souami}, {Souchay}, {Spagna}, {Spoto}, {Steele},
  {Steidelm{\"u}ller}, {Stephenson}, {S{\"u}veges}, {Szabados}, {Szegedi-Elek},
  {Taris}, {Tauran}, {Taylor}, {Teixeira}, {Thuillot}, {Tonello}, {Torra},
  {Torra}, {Turon}, {Unger}, {Vaillant}, {van Dillen}, {Vanel}, {Vecchiato},
  {Viala}, {Vicente}, {Voutsinas}, {Weiler}, {Wevers}, {Wyrzykowski}, {Yoldas},
  {Yvard}, {Zhao}, {Zorec}, {Zucker}, {Zurbach}, \&
  {Zwitter}}]{2021A&A...649A...1G}
{Gaia Collaboration}, {Brown}, A.~G.~A., {Vallenari}, A., {et~al.} 2021, \aap,
  649, A1, \dodoi{10.1051/0004-6361/202039657}

\bibitem[{{Gail} {et~al.}(2009){Gail}, {Zhukovska}, {Hoppe}, \&
  {Trieloff}}]{2009ApJ...698.1136G}
{Gail}, H.~P., {Zhukovska}, S.~V., {Hoppe}, P., \& {Trieloff}, M. 2009, \apj,
  698, 1136, \dodoi{10.1088/0004-637X/698/2/1136}

\bibitem[{Gautschy \& Saio(2017)}]{10.1093/mnras/stx811}
Gautschy, A., \& Saio, H. 2017, Monthly Notices of the Royal Astronomical
  Society, 468, 4419, \dodoi{10.1093/mnras/stx811}

\bibitem[{{Giardino} {et~al.}(2007){Giardino}, {Favata}, {Micela}, {Sciortino},
  \& {Winston}}]{2007A&A...463..275G}
{Giardino}, G., {Favata}, F., {Micela}, G., {Sciortino}, S., \& {Winston}, E.
  2007, \aap, 463, 275, \dodoi{10.1051/0004-6361:20066424}

\bibitem[{{Ginsburg} {et~al.}(2019){Ginsburg}, {Sip{\H{o}}cz}, {Brasseur},
  {Cowperthwaite}, {Craig}, {Deil}, {Guillochon}, {Guzman}, {Liedtke}, {Lian
  Lim}, {Lockhart}, {Mommert}, {Morris}, {Norman}, {Parikh}, {Persson},
  {Robitaille}, {Segovia}, {Singer}, {Tollerud}, {de Val-Borro}, {Valtchanov},
  {Woillez}, {Astroquery Collaboration}, \& {a subset of astropy
  Collaboration}}]{2019AJ....157...98G}
{Ginsburg}, A., {Sip{\H{o}}cz}, B.~M., {Brasseur}, C.~E., {et~al.} 2019, \aj,
  157, 98, \dodoi{10.3847/1538-3881/aafc33}

\bibitem[{{Girin} {et~al.}(2020){Girin}, {Leglaive}, {Bie}, {Diard}, {Hueber},
  \& {Alameda-Pineda}}]{2020arXiv200812595G}
{Girin}, L., {Leglaive}, S., {Bie}, X., {et~al.} 2020, arXiv e-prints,
  arXiv:2008.12595.
\newblock \doarXiv{2008.12595}

\bibitem[{{Goldman}(2020)}]{2020JOSS....5.2554G}
{Goldman}, S. 2020, The Journal of Open Source Software, 5, 2554,
  \dodoi{10.21105/joss.02554}

\bibitem[{{Graham}(2019)}]{2019eeu..confE..23G}
{Graham}, M. 2019, in The Extragalactic Explosive Universe: the New Era of
  Transient Surveys and Data-Driven Discovery, 23,
  \dodoi{10.5281/zenodo.3478038}

\bibitem[{{Green}(2018)}]{2018JOSS....3..695M}
{Green}, G. 2018, The Journal of Open Source Software, 3, 695,
  \dodoi{10.21105/joss.00695}

\bibitem[{{Green} {et~al.}(2019){Green}, {Schlafly}, {Zucker}, {Speagle}, \&
  {Finkbeiner}}]{2019ApJ...887...93G}
{Green}, G.~M., {Schlafly}, E., {Zucker}, C., {Speagle}, J.~S., \&
  {Finkbeiner}, D. 2019, \apj, 887, 93, \dodoi{10.3847/1538-4357/ab5362}

\bibitem[{{Groenewegen}(2012)}]{2012A&A...543A..36G}
{Groenewegen}, M.~A.~T. 2012, \aap, 543, A36,
  \dodoi{10.1051/0004-6361/201218965}

\bibitem[{{Groenewegen} \& {Jurkovic}(2017)}]{2017A&A...604A..29G}
{Groenewegen}, M.~A.~T., \& {Jurkovic}, M.~I. 2017, \aap, 604, A29,
  \dodoi{10.1051/0004-6361/201730946}

\bibitem[{{Groenewegen} {et~al.}(2009){Groenewegen}, {Sloan}, {Soszy{\'n}ski},
  \& {Petersen}}]{2009A&A...506.1277G}
{Groenewegen}, M.~A.~T., {Sloan}, G.~C., {Soszy{\'n}ski}, I., \& {Petersen},
  E.~A. 2009, \aap, 506, 1277, \dodoi{10.1051/0004-6361/200912678}

\bibitem[{{Guillochon} {et~al.}(2017){Guillochon}, {Parrent}, {Kelley}, \&
  {Margutti}}]{2017ApJ...835...64G}
{Guillochon}, J., {Parrent}, J., {Kelley}, L.~Z., \& {Margutti}, R. 2017, \apj,
  835, 64, \dodoi{10.3847/1538-4357/835/1/64}

\bibitem[{{Hajdu} {et~al.}(2021){Hajdu}, {Pietrzy{\'n}ski}, {Jurcsik},
  {Catelan}, {Karczmarek}, {Pilecki}, {Soszy{\'n}ski}, {Udalski}, \&
  {Thompson}}]{2021ApJ...915...50H}
{Hajdu}, G., {Pietrzy{\'n}ski}, G., {Jurcsik}, J., {et~al.} 2021, \apj, 915,
  50, \dodoi{10.3847/1538-4357/abff4b}

\bibitem[{{Hallakoun} \& {Maoz}(2021)}]{2021MNRAS.507..398H}
{Hallakoun}, N., \& {Maoz}, D. 2021, \mnras, 507, 398,
  \dodoi{10.1093/mnras/stab2145}

\bibitem[{{Heinze} {et~al.}(2018){Heinze}, {Tonry}, {Denneau}, {Flewelling},
  {Stalder}, {Rest}, {Smith}, {Smartt}, \& {Weiland}}]{2018AJ....156..241H}
{Heinze}, A.~N., {Tonry}, J.~L., {Denneau}, L., {et~al.} 2018, \aj, 156, 241,
  \dodoi{10.3847/1538-3881/aae47f}

\bibitem[{Henrion {et~al.}(2013)Henrion, Mortlock, Hand, \&
  Gandy}]{Henrion2013}
Henrion, M., Mortlock, D.~J., Hand, D.~J., \& Gandy, A. 2013, Classification
  and Anomaly Detection for Astronomical Survey Data, ed. J.~M. Hilbe (New
  York, NY: Springer New York), 149--184, \dodoi{10.1007/978-1-4614-3508-2_8}

\bibitem[{{Herbst}(2018)}]{2018JAVSO..46...83H}
{Herbst}, W. 2018, \jaavso, 46, 83

\bibitem[{H{\"o}fner \& Olofsson(2018)}]{Hofner2018}
H{\"o}fner, S., \& Olofsson, H. 2018, The Astronomy and Astrophysics Review,
  26, 1, \dodoi{10.1007/s00159-017-0106-5}

\bibitem[{{Iben}(1999)}]{1999IAUS..191..591I}
{Iben}, I., J. 1999, in Asymptotic Giant Branch Stars, ed. T.~{Le Bertre},
  A.~{Lebre}, \& C.~{Waelkens}, Vol. 191, 591

\bibitem[{Ichinohe \& Yamada(2019)}]{10.1093/mnras/stz1528}
Ichinohe, Y., \& Yamada, S. 2019, Monthly Notices of the Royal Astronomical
  Society, 487, 2874, \dodoi{10.1093/mnras/stz1528}

\bibitem[{Ivezi{\'{c}} {et~al.}(2019)Ivezi{\'{c}}, Kahn, Tyson, Abel, Acosta,
  Allsman, Alonso, AlSayyad, Anderson, Andrew, Angel, Angeli, Ansari,
  Antilogus, Araujo, Armstrong, Arndt, Astier, Aubourg, Auza, Axelrod, Bard,
  Barr, Barrau, Bartlett, Bauer, Bauman, Baumont, Bechtol, Bechtol, Becker,
  Becla, Beldica, Bellavia, Bianco, Biswas, Blanc, Blazek, Blandford, Bloom,
  Bogart, Bond, Booth, Borgland, Borne, Bosch, Boutigny, Brackett, Bradshaw,
  Brandt, Brown, Bullock, Burchat, Burke, Cagnoli, Calabrese, Callahan, Callen,
  Carlin, Carlson, Chandrasekharan, Charles-Emerson, Chesley, Cheu, Chiang,
  Chiang, Chirino, Chow, Ciardi, Claver, Cohen-Tanugi, Cockrum, Coles,
  Connolly, Cook, Cooray, Covey, Cribbs, Cui, Cutri, Daly, Daniel, Daruich,
  Daubard, Daues, Dawson, Delgado, Dellapenna, de~Peyster, de~Val-Borro, Digel,
  Doherty, Dubois, Dubois-Felsmann, Durech, Economou, Eifler, Eracleous,
  Emmons, Neto, Ferguson, Figueroa, Fisher-Levine, Focke, Foss, Frank, Freemon,
  Gangler, Gawiser, Geary, Gee, Geha, Gessner, Gibson, Gilmore, Glanzman,
  Glick, Goldina, Goldstein, Goodenow, Graham, Gressler, Gris, Guy, Guyonnet,
  Haller, Harris, Hascall, Haupt, Hernandez, Herrmann, Hileman, Hoblitt,
  Hodgson, Hogan, Howard, Huang, Huffer, Ingraham, Innes, Jacoby, Jain, Jammes,
  Jee, Jenness, Jernigan, Jevremovi{\'{c}}, Johns, Johnson, Johnson, Jones,
  Juramy-Gilles, Juri{\'{c}}, Kalirai, Kallivayalil, Kalmbach, Kantor, Karst,
  Kasliwal, Kelly, Kessler, Kinnison, Kirkby, Knox, Kotov, Krabbendam,
  Krughoff, Kub{\'{a}}nek, Kuczewski, Kulkarni, Ku, Kurita, Lage, Lambert,
  Lange, Langton, Guillou, Levine, Liang, Lim, Lintott, Long, Lopez, Lotz,
  Lupton, Lust, MacArthur, Mahabal, Mandelbaum, Markiewicz, Marsh, Marshall,
  Marshall, May, McKercher, McQueen, Meyers, Migliore, Miller, Mills, Miraval,
  Moeyens, Moolekamp, Monet, Moniez, Monkewitz, Montgomery, Morrison, Mueller,
  Muller, Arancibia, Neill, Newbry, Nief, Nomerotski, Nordby, O'Connor, Oliver,
  Olivier, Olsen, O'Mullane, Ortiz, Osier, Owen, Pain, Palecek, Parejko,
  Parsons, Pease, Peterson, Peterson, Petravick, Petrick, Petry, Pierfederici,
  Pietrowicz, Pike, Pinto, Plante, Plate, Plutchak, Price, Prouza, Radeka,
  Rajagopal, Rasmussen, Regnault, Reil, Reiss, Reuter, Ridgway, Riot, Ritz,
  Robinson, Roby, Roodman, Rosing, Roucelle, Rumore, Russo, Saha, Sassolas,
  Schalk, Schellart, Schindler, Schmidt, Schneider, Schneider, Schoening,
  Schumacher, Schwamb, Sebag, Selvy, Sembroski, Seppala, Serio, Serrano, Shaw,
  Shipsey, Sick, Silvestri, Slater, Smith, Smith, Sobhani, Soldahl,
  Storrie-Lombardi, Stover, Strauss, Street, Stubbs, Sullivan, Sweeney,
  Swinbank, Szalay, Takacs, Tether, Thaler, Thayer, Thomas, Thornton, Thukral,
  Tice, Trilling, Turri, Berg, Berk, Vetter, Virieux, Vucina, Wahl, Walkowicz,
  Walsh, Walter, Wang, Wang, Warner, Wiecha, Willman, Winters, Wittman, Wolff,
  Wood-Vasey, Wu, Xin, Yoachim, \& Zhan}]{Ivezi__2019}
Ivezi{\'{c}}, {\v{Z}}., Kahn, S.~M., Tyson, J.~A., {et~al.} 2019, The
  Astrophysical Journal, 873, 111, \dodoi{10.3847/1538-4357/ab042c}

\bibitem[{Jacklin {et~al.}(2017)Jacklin, Lund, Pepper, \&
  Stassun}]{Jacklin_2017}
Jacklin, S.~R., Lund, M.~B., Pepper, J., \& Stassun, K.~G. 2017, The
  Astronomical Journal, 153, 186, \dodoi{10.3847/1538-3881/aa64d1}

\bibitem[{Jamal \& Bloom(2020)}]{Jamal_2020}
Jamal, S., \& Bloom, J.~S. 2020, The Astrophysical Journal Supplement Series,
  250, 30, \dodoi{10.3847/1538-4365/aba8ff}

\bibitem[{{Jayasinghe} {et~al.}(2018){Jayasinghe}, {Kochanek}, {Stanek},
  {Shappee}, {Holoien}, {Thompson}, {Prieto}, {Dong}, {Pawlak}, {Shields},
  {Pojmanski}, {Otero}, {Britt}, \& {Will}}]{2018MNRAS.477.3145J}
{Jayasinghe}, T., {Kochanek}, C.~S., {Stanek}, K.~Z., {et~al.} 2018, \mnras,
  477, 3145, \dodoi{10.1093/mnras/sty838}

\bibitem[{{Jayasinghe} {et~al.}(2019){Jayasinghe}, {Stanek}, {Kochanek},
  {Shappee}, {Holoien}, {Thompson}, {Prieto}, {Dong}, {Pawlak}, {Pejcha},
  {Shields}, {Pojmanski}, {Otero}, {Britt}, \& {Will}}]{2019MNRAS.486.1907J}
{Jayasinghe}, T., {Stanek}, K.~Z., {Kochanek}, C.~S., {et~al.} 2019, \mnras,
  486, 1907, \dodoi{10.1093/mnras/stz844}

\bibitem[{{Jurkovic}(2018)}]{2018SerAJ.197...13J}
{Jurkovic}, M.~I. 2018, Serbian Astronomical Journal, 197, 13,
  \dodoi{10.2298/SAJ180316002J}

\bibitem[{{Kains} {et~al.}(2012){Kains}, {Bramich}, {Figuera Jaimes}, {Arellano
  Ferro}, {Giridhar}, \& {Kuppuswamy}}]{2012A&A...548A..92K}
{Kains}, N., {Bramich}, D.~M., {Figuera Jaimes}, R., {et~al.} 2012, \aap, 548,
  A92, \dodoi{10.1051/0004-6361/201220217}

\bibitem[{{Kimura} {et~al.}(2021){Kimura}, {Yamada}, {Nakaniwa}, {Makita},
  {Negoro}, {Shidatsu}, {Kato}, {Enoto}, {Isogai}, {Mihara}, {Akazawa},
  {Gendreau}, {Hambsch}, {Dubovsky}, {Kudzej}, {Kasai}, {Tordai}, {Pavlenko},
  {Sosnovskij}, {Babina}, {Antonyuk}, {Itoh}, \&
  {Maehara}}]{2021arXiv210615756K}
{Kimura}, M., {Yamada}, S., {Nakaniwa}, N., {et~al.} 2021, arXiv e-prints,
  arXiv:2106.15756.
\newblock \doarXiv{2106.15756}

\bibitem[{{Kingma} \& {Ba}(2014)}]{2014arXiv1412.6980K}
{Kingma}, D.~P., \& {Ba}, J. 2014, arXiv e-prints, arXiv:1412.6980.
\newblock \doarXiv{1412.6980}

\bibitem[{{Kingma} \& {Welling}(2013)}]{2013arXiv1312.6114K}
{Kingma}, D.~P., \& {Welling}, M. 2013, arXiv e-prints, arXiv:1312.6114.
\newblock \doarXiv{1312.6114}

\bibitem[{Kingma \& Welling(2019)}]{MAL-056}
Kingma, D.~P., \& Welling, M. 2019, Foundations and Trends® in Machine
  Learning, 12, 307, \dodoi{10.1561/2200000056}

\bibitem[{{Kiss} {et~al.}(2007){Kiss}, {Derekas}, {Szab{\'o}}, {Bedding}, \&
  {Szabados}}]{2007MNRAS.375.1338K}
{Kiss}, L.~L., {Derekas}, A., {Szab{\'o}}, G.~M., {Bedding}, T.~R., \&
  {Szabados}, L. 2007, \mnras, 375, 1338,
  \dodoi{10.1111/j.1365-2966.2006.11387.x}

\bibitem[{{Kiss} {et~al.}(1999){Kiss}, {Szatm{\'a}ry}, {Cadmus}, \&
  {Mattei}}]{1999A&A...346..542K}
{Kiss}, L.~L., {Szatm{\'a}ry}, K., {Cadmus}, R.~R., J., \& {Mattei}, J.~A.
  1999, \aap, 346, 542.
\newblock \doarXiv{astro-ph/9904128}

\bibitem[{{Kiss, L. L.} {et~al.}(2000){Kiss, L. L.}, {Szatm\'ary, K.},
  {Szab\'o, G.}, \& {Mattei, J. A.}}]{sr1999}
{Kiss, L. L.}, {Szatm\'ary, K.}, {Szab\'o, G.}, \& {Mattei, J. A.} 2000,
  Astron. Astrophys. Suppl. Ser., 145, 283, \dodoi{10.1051/aas:2000353}

\bibitem[{{Kochanek} {et~al.}(2017){Kochanek}, {Shappee}, {Stanek}, {Holoien},
  {Thompson}, {Prieto}, {Dong}, {Shields}, {Will}, {Britt}, {Perzanowski}, \&
  {Pojma{\'n}ski}}]{2017PASP..129j4502K}
{Kochanek}, C.~S., {Shappee}, B.~J., {Stanek}, K.~Z., {et~al.} 2017, \pasp,
  129, 104502, \dodoi{10.1088/1538-3873/aa80d9}

\bibitem[{{Kovtyukh} {et~al.}(2018){Kovtyukh}, {Yegorova}, {Andrievsky},
  {Korotin}, {Saviane}, {Lemasle}, {Chekhonadskikh}, \&
  {Belik}}]{2018MNRAS.477.2276K}
{Kovtyukh}, V., {Yegorova}, I., {Andrievsky}, S., {et~al.} 2018, \mnras, 477,
  2276, \dodoi{10.1093/mnras/sty671}

\bibitem[{{Lagadec} \& {Zijlstra}(2008)}]{2008MNRAS.390L..59L}
{Lagadec}, E., \& {Zijlstra}, A.~A. 2008, \mnras, 390, L59,
  \dodoi{10.1111/j.1745-3933.2008.00535.x}

\bibitem[{{Lagadec} {et~al.}(2008){Lagadec}, {Zijlstra}, {Matsuura}, {Menzies},
  {van Loon}, \& {Whitelock}}]{2008MNRAS.383..399L}
{Lagadec}, E., {Zijlstra}, A.~A., {Matsuura}, M., {et~al.} 2008, \mnras, 383,
  399, \dodoi{10.1111/j.1365-2966.2007.12561.x}

\bibitem[{Law {et~al.}(2009)Law, Kulkarni, Dekany, Ofek, Quimby, Nugent,
  Surace, Grillmair, Bloom, Kasliwal, Bildsten, Brown, Cenko, Ciardi, Croner,
  Djorgovski, van Eyken, Filippenko, Fox, Gal-Yam, Hale, Hamam, Helou, Henning,
  Howell, Jacobsen, Laher, Mattingly, McKenna, Pickles, Poznanski, Rahmer, Rau,
  Rosing, Shara, Smith, Starr, Sullivan, Velur, Walters, \&
  Zolkower}]{Law_2009}
Law, N.~M., Kulkarni, S.~R., Dekany, R.~G., {et~al.} 2009, Publications of the
  Astronomical Society of the Pacific, 121, 1395, \dodoi{10.1086/648598}

\bibitem[{LeCun {et~al.}(1989)LeCun, Boser, Denker, Henderson, Howard, Hubbard,
  \& Jackel}]{10.1162/neco.1989.1.4.541}
LeCun, Y., Boser, B., Denker, J.~S., {et~al.} 1989, Neural Computation, 1, 541,
  \dodoi{10.1162/neco.1989.1.4.541}

\bibitem[{{Leung} \& {Nomoto}(2019)}]{2019PASA...36....6L}
{Leung}, S.-C., \& {Nomoto}, K. 2019, \pasa, 36, e006,
  \dodoi{10.1017/pasa.2018.49}

\bibitem[{{Leung} {et~al.}(2020){Leung}, {Nomoto}, \&
  {Suzuki}}]{2020ApJ...889...34L}
{Leung}, S.-C., {Nomoto}, K., \& {Suzuki}, T. 2020, \apj, 889, 34,
  \dodoi{10.3847/1538-4357/ab5d2f}

\bibitem[{Li {et~al.}(2001)Li, Filippenko, Gates, Chornock, Gal-Yam, Ofek,
  Leonard, Modjaz, Rich, Riess, \& Treffers}]{Li_2001}
Li, W., Filippenko, A.~V., Gates, E., {et~al.} 2001, Publications of the
  Astronomical Society of the Pacific, 113, 1178, \dodoi{10.1086/323355}

\bibitem[{{Lindegren} {et~al.}(2018){Lindegren}, {Hern{\'a}ndez}, {Bombrun},
  {Klioner}, {Bastian}, {Ramos-Lerate}, {de Torres}, {Steidelm{\"u}ller},
  {Stephenson}, {Hobbs}, {Lammers}, {Biermann}, {Geyer}, {Hilger}, {Michalik},
  {Stampa}, {McMillan}, {Casta{\~n}eda}, {Clotet}, {Comoretto}, {Davidson},
  {Fabricius}, {Gracia}, {Hambly}, {Hutton}, {Mora}, {Portell}, {van Leeuwen},
  {Abbas}, {Abreu}, {Altmann}, {Andrei}, {Anglada}, {Balaguer-N{\'u}{\~n}ez},
  {Barache}, {Becciani}, {Bertone}, {Bianchi}, {Bouquillon}, {Bourda},
  {Br{\"u}semeister}, {Bucciarelli}, {Busonero}, {Buzzi}, {Cancelliere},
  {Carlucci}, {Charlot}, {Cheek}, {Crosta}, {Crowley}, {de Bruijne}, {de
  Felice}, {Drimmel}, {Esquej}, {Fienga}, {Fraile}, {Gai}, {Garralda},
  {Gonz{\'a}lez-Vidal}, {Guerra}, {Hauser}, {Hofmann}, {Holl}, {Jordan},
  {Lattanzi}, {Lenhardt}, {Liao}, {Licata}, {Lister}, {L{\"o}ffler},
  {Marchant}, {Martin-Fleitas}, {Messineo}, {Mignard}, {Morbidelli}, {Poggio},
  {Riva}, {Rowell}, {Salguero}, {Sarasso}, {Sciacca}, {Siddiqui}, {Smart},
  {Spagna}, {Steele}, {Taris}, {Torra}, {van Elteren}, {van Reeven}, \&
  {Vecchiato}}]{2018A&A...616A...2L}
{Lindegren}, L., {Hern{\'a}ndez}, J., {Bombrun}, A., {et~al.} 2018, \aap, 616,
  A2, \dodoi{10.1051/0004-6361/201832727}

\bibitem[{Liu {et~al.}(2008)Liu, Ting, \& Zhou}]{4781136}
Liu, F.~T., Ting, K.~M., \& Zhou, Z.-H. 2008, in 2008 Eighth IEEE International
  Conference on Data Mining, 413--422, \dodoi{10.1109/ICDM.2008.17}

\bibitem[{Liu {et~al.}(2012)Liu, Ting, \& Zhou}]{10.1145/2133360.2133363}
Liu, F.~T., Ting, K.~M., \& Zhou, Z.-H. 2012, ACM Trans. Knowl. Discov. Data,
  6, \dodoi{10.1145/2133360.2133363}

\bibitem[{{L{\'o}pez-Morales} {et~al.}(2006){L{\'o}pez-Morales}, {Morrell},
  {Butler}, \& {Seager}}]{2006PASP..118.1506L}
{L{\'o}pez-Morales}, M., {Morrell}, N.~I., {Butler}, R.~P., \& {Seager}, S.
  2006, \pasp, 118, 1506, \dodoi{10.1086/508904}

\bibitem[{Madore {et~al.}(1998)Madore, Freedman, Silbermann, Harding, Huchra,
  Mould, Graham, Ferrarese, Gibson, Han, Hoessel, Hughes, Illingworth, Phelps,
  Sakai, \& Stetson}]{Madore1998}
Madore, B.~F., Freedman, W.~L., Silbermann, N., {et~al.} 1998, Nature, 395, 47,
  \dodoi{10.1038/25678}

\bibitem[{{Maercker} {et~al.}(2012){Maercker}, {Mohamed}, {Vlemmings},
  {Ramstedt}, {Groenewegen}, {Humphreys}, {Kerschbaum}, {Lindqvist},
  {Olofsson}, {Paladini}, {Wittkowski}, {de Gregorio-Monsalvo}, \&
  {Nyman}}]{2012Natur.490..232M}
{Maercker}, M., {Mohamed}, S., {Vlemmings}, W.~H.~T., {et~al.} 2012, \nat, 490,
  232, \dodoi{10.1038/nature11511}

\bibitem[{{Malanchev} {et~al.}(2021){Malanchev}, {Pruzhinskaya}, {Korolev},
  {Aleo}, {Kornilov}, {Ishida}, {Krushinsky}, {Mondon}, {Sreejith}, {Volnova},
  {Belinski}, {Dodin}, {Tatarnikov}, {Zheltoukhov}, \& {(The SNAD
  Team)}}]{2021MNRAS.502.5147M}
{Malanchev}, K.~L., {Pruzhinskaya}, M.~V., {Korolev}, V.~S., {et~al.} 2021,
  \mnras, 502, 5147, \dodoi{10.1093/mnras/stab316}

\bibitem[{Margalef-Bentabol {et~al.}(2020)Margalef-Bentabol, Huertas-Company,
  Charnock, Margalef-Bentabol, Bernardi, Dubois, Storey-Fisher, \&
  Zanisi}]{10.1093/mnras/staa1647}
Margalef-Bentabol, B., Huertas-Company, M., Charnock, T., {et~al.} 2020,
  Monthly Notices of the Royal Astronomical Society, 496, 2346,
  \dodoi{10.1093/mnras/staa1647}

\bibitem[{{Mart{\'\i}nez-Palomera} {et~al.}(2020){Mart{\'\i}nez-Palomera},
  {Bloom}, \& {Abrahams}}]{2020arXiv200507773M}
{Mart{\'\i}nez-Palomera}, J., {Bloom}, J.~S., \& {Abrahams}, E.~S. 2020, arXiv
  e-prints, arXiv:2005.07773.
\newblock \doarXiv{2005.07773}

\bibitem[{Massey {et~al.}(2019)Massey, Neugent, \& Levesque}]{Massey_2019}
Massey, P., Neugent, K.~F., \& Levesque, E.~M. 2019, The Astronomical Journal,
  157, 227, \dodoi{10.3847/1538-3881/ab1aa1}

\bibitem[{Masuda {et~al.}(2019)Masuda, Kawahara, Latham, Bieryla, Kunitomo,
  MacLeod, \& Aoki}]{Masuda_2019}
Masuda, K., Kawahara, H., Latham, D.~W., {et~al.} 2019, The Astrophysical
  Journal, 881, L3, \dodoi{10.3847/2041-8213/ab321b}

\bibitem[{{Mauerhan} {et~al.}(2015){Mauerhan}, {Smith}, {Van Dyk}, {Morzinski},
  {Close}, {Hinz}, {Males}, \& {Rodigas}}]{2015MNRAS.450.2551M}
{Mauerhan}, J., {Smith}, N., {Van Dyk}, S.~D., {et~al.} 2015, \mnras, 450,
  2551, \dodoi{10.1093/mnras/stv257}

\bibitem[{{McDonald} \& {Zijlstra}(2015)}]{2015MNRAS.448..502M}
{McDonald}, I., \& {Zijlstra}, A.~A. 2015, \mnras, 448, 502,
  \dodoi{10.1093/mnras/stv007}

\bibitem[{{Miglio} {et~al.}(2009){Miglio}, {Montalb{\'a}n}, {Baudin},
  {Eggenberger}, {Noels}, {Hekker}, {De Ridder}, {Weiss}, \&
  {Baglin}}]{2009A&A...503L..21M}
{Miglio}, A., {Montalb{\'a}n}, J., {Baudin}, F., {et~al.} 2009, \aap, 503, L21,
  \dodoi{10.1051/0004-6361/200912822}

\bibitem[{{Moe} \& {Di Stefano}(2017)}]{2017ApJS..230...15M}
{Moe}, M., \& {Di Stefano}, R. 2017, \apjs, 230, 15,
  \dodoi{10.3847/1538-4365/aa6fb6}

\bibitem[{{Moln{\'a}r} {et~al.}(2019){Moln{\'a}r}, {Joyce}, \&
  {Kiss}}]{2019ApJ...879...62M}
{Moln{\'a}r}, L., {Joyce}, M., \& {Kiss}, L.~L. 2019, \apj, 879, 62,
  \dodoi{10.3847/1538-4357/ab22a5}

\bibitem[{{Mowlavi} {et~al.}(2018){Mowlavi}, {Lecoeur-Ta{\"\i}bi}, {Lebzelter},
  {Rimoldini}, {Lorenz}, {Audard}, {De Ridder}, {Eyer}, {Guy}, {Holl},
  {Jevardat de Fombelle}, {Marchal}, {Nienartowicz}, {Regibo}, {Roelens}, \&
  {Sarro}}]{2018A&A...618A..58M}
{Mowlavi}, N., {Lecoeur-Ta{\"\i}bi}, I., {Lebzelter}, T., {et~al.} 2018, \aap,
  618, A58, \dodoi{10.1051/0004-6361/201833366}

\bibitem[{{Mullan} \& {MacDonald}(2019)}]{2019ApJ...885..113M}
{Mullan}, D.~J., \& {MacDonald}, J. 2019, \apj, 885, 113,
  \dodoi{10.3847/1538-4357/ab4658}

\bibitem[{{Munari} {et~al.}(2008){Munari}, {Siviero}, {Ochner}, {Dallaporta},
  \& {Simoncelli}}]{2008BaltA..17..223M}
{Munari}, U., {Siviero}, A., {Ochner}, P., {Dallaporta}, S., \& {Simoncelli},
  C. 2008, Baltic Astronomy, 17, 223.
\newblock \doarXiv{0810.1375}

\bibitem[{{Naul} {et~al.}(2018){Naul}, {Bloom}, {P{\'e}rez}, \& {van der
  Walt}}]{2018NatAs...2..151N}
{Naul}, B., {Bloom}, J.~S., {P{\'e}rez}, F., \& {van der Walt}, S. 2018, Nature
  Astronomy, 2, 151, \dodoi{10.1038/s41550-017-0321-z}

\bibitem[{{Neilson} {et~al.}(2016){Neilson}, {Percy}, \&
  {Smith}}]{2016JAVSO..44..179N}
{Neilson}, H.~R., {Percy}, J.~R., \& {Smith}, H.~A. 2016, \jaavso, 44, 179.
\newblock \doarXiv{1611.03030}

\bibitem[{{Nenkova} {et~al.}(2000){Nenkova}, {Ivezi{\'c}}, \&
  {Elitzur}}]{2000ASPC..196...77N}
{Nenkova}, M., {Ivezi{\'c}}, {\v{Z}}., \& {Elitzur}, M. 2000, in Astronomical
  Society of the Pacific Conference Series, Vol. 196, Thermal Emission
  Spectroscopy and Analysis of Dust, Disks, and Regoliths, ed. M.~L. {Sitko},
  A.~L. {Sprague}, \& D.~K. {Lynch}, 77--82

\bibitem[{{Ness} {et~al.}(2019){Ness}, {Johnston}, {Blancato}, {Rix}, {Beane},
  {Bird}, \& {Hawkins}}]{2019ApJ...883..177N}
{Ness}, M.~K., {Johnston}, K.~V., {Blancato}, K., {et~al.} 2019, \apj, 883,
  177, \dodoi{10.3847/1538-4357/ab3e3c}

\bibitem[{{Nicholls} {et~al.}(2009){Nicholls}, {Wood}, {Cioni}, \&
  {Soszy{\'n}ski}}]{2009MNRAS.399.2063N}
{Nicholls}, C.~P., {Wood}, P.~R., {Cioni}, M. R.~L., \& {Soszy{\'n}ski}, I.
  2009, \mnras, 399, 2063, \dodoi{10.1111/j.1365-2966.2009.15401.x}

\bibitem[{Niemczura {et~al.}(2017)Niemczura, H{\"u}mmerich, Castelli, Paunzen,
  Bernhard, Hambsch, \& He{\l}miniak}]{Niemczura2017}
Niemczura, E., H{\"u}mmerich, S., Castelli, F., {et~al.} 2017, Scientific
  Reports, 7, 5906, \dodoi{10.1038/s41598-017-05987-6}

\bibitem[{{Nieuwenhuijzen} \& {de Jager}(1990)}]{1990A&A...231..134N}
{Nieuwenhuijzen}, H., \& {de Jager}, C. 1990, \aap, 231, 134

\bibitem[{Nun {et~al.}(2016)Nun, Protopapas, Sim, \& Chen}]{Nun_2016}
Nun, I., Protopapas, P., Sim, B., \& Chen, W. 2016, The Astronomical Journal,
  152, 71, \dodoi{10.3847/0004-6256/152/3/71}

\bibitem[{{Odaibo}(2019)}]{2019arXiv190708956O}
{Odaibo}, S. 2019, arXiv e-prints, arXiv:1907.08956.
\newblock \doarXiv{1907.08956}

\bibitem[{{O'Grady} {et~al.}(2021){O'Grady}, {Drout}, {Shappee}, {Bauer},
  {Fuller}, {Kochanek}, {Jayasinghe}, {Gaensler}, {Stanek}, {Holoien},
  {Prieto}, \& {Thompson}}]{2021AAS...23812103O}
{O'Grady}, A.~J.~G., {Drout}, M.~R., {Shappee}, B.~J., {et~al.} 2021, in
  American Astronomical Society Meeting Abstracts, Vol.~53, American
  Astronomical Society Meeting Abstracts, 121.03

\bibitem[{{Oring} {et~al.}(2020){Oring}, {Yakhini}, \&
  {Hel-Or}}]{2020arXiv200801487O}
{Oring}, A., {Yakhini}, Z., \& {Hel-Or}, Y. 2020, arXiv e-prints,
  arXiv:2008.01487.
\newblock \doarXiv{2008.01487}

\bibitem[{{Ossenkopf} {et~al.}(1992){Ossenkopf}, {Henning}, \&
  {Mathis}}]{1992A&A...261..567O}
{Ossenkopf}, V., {Henning}, T., \& {Mathis}, J.~S. 1992, \aap, 261, 567

\bibitem[{{Paxton} {et~al.}(2011){Paxton}, {Bildsten}, {Dotter}, {Herwig},
  {Lesaffre}, \& {Timmes}}]{2011ApJS..192....3P}
{Paxton}, B., {Bildsten}, L., {Dotter}, A., {et~al.} 2011, \apjs, 192, 3,
  \dodoi{10.1088/0067-0049/192/1/3}

\bibitem[{{Paxton} {et~al.}(2013){Paxton}, {Cantiello}, {Arras}, {Bildsten},
  {Brown}, {Dotter}, {Mankovich}, {Montgomery}, {Stello}, {Timmes}, \&
  {Townsend}}]{2013ApJS..208....4P}
{Paxton}, B., {Cantiello}, M., {Arras}, P., {et~al.} 2013, \apjs, 208, 4,
  \dodoi{10.1088/0067-0049/208/1/4}

\bibitem[{{Paxton} {et~al.}(2015){Paxton}, {Marchant}, {Schwab}, {Bauer},
  {Bildsten}, {Cantiello}, {Dessart}, {Farmer}, {Hu}, {Langer}, {Townsend},
  {Townsley}, \& {Timmes}}]{2015ApJS..220...15P}
{Paxton}, B., {Marchant}, P., {Schwab}, J., {et~al.} 2015, \apjs, 220, 15,
  \dodoi{10.1088/0067-0049/220/1/15}

\bibitem[{{Percy}(2020)}]{2020JAVSO..48...50P}
{Percy}, J.~R. 2020, \jaavso, 48, 50

\bibitem[{{Percy} {et~al.}(2001){Percy}, {Soondarsingh}, \&
  {Velocci}}]{2001JAVSO..29...82P}
{Percy}, J.~R., {Soondarsingh}, D., \& {Velocci}, V. 2001, \jaavso, 29, 82

\bibitem[{{Percy} \& {Tan}(2013)}]{2013JAVSO..41...75P}
{Percy}, J.~R., \& {Tan}, P.~J. 2013, \jaavso, 41, 75

\bibitem[{{Petersen} \& {Christensen-Dalsgaard}(1999)}]{1999A&A...352..547P}
{Petersen}, J.~O., \& {Christensen-Dalsgaard}, J. 1999, \aap, 352, 547

\bibitem[{Pierce {et~al.}(1994)Pierce, Welch, McClure, van~den Bergh, Racine,
  \& Stetson}]{Pierce1994}
Pierce, M.~J., Welch, D.~L., McClure, R.~D., {et~al.} 1994, Nature, 371, 385,
  \dodoi{10.1038/371385a0}

\bibitem[{Pietrzy{\'{n}}ski {et~al.}(2010)Pietrzy{\'{n}}ski, Thompson, Gieren,
  Graczyk, Bono, Udalski, Soszy{\'{n}}ski, Minniti, \&
  Pilecki}]{Pietrzynski2010}
Pietrzy{\'{n}}ski, G., Thompson, I.~B., Gieren, W., {et~al.} 2010, Nature, 468,
  542, \dodoi{10.1038/nature09598}

\bibitem[{Pruzhinskaya {et~al.}(2019)Pruzhinskaya, Malanchev, Kornilov, Ishida,
  Mondon, Volnova, \& Korolev}]{10.1093/mnras/stz2362}
Pruzhinskaya, M.~V., Malanchev, K.~L., Kornilov, M.~V., {et~al.} 2019, Monthly
  Notices of the Royal Astronomical Society, 489, 3591,
  \dodoi{10.1093/mnras/stz2362}

\bibitem[{Qian {et~al.}(2017)Qian, He, Zhang, Zhu, Shi, Zhao, \&
  Zhou}]{Qian_2017}
Qian, S.-B., He, J.-J., Zhang, J., {et~al.} 2017, Research in Astronomy and
  Astrophysics, 17, 087, \dodoi{10.1088/1674-4527/17/8/87}

\bibitem[{{Qu} {et~al.}(2021){Qu}, {Sako}, {M{\"o}ller}, \&
  {Doux}}]{2021arXiv210604370Q}
{Qu}, H., {Sako}, M., {M{\"o}ller}, A., \& {Doux}, C. 2021, arXiv e-prints,
  arXiv:2106.04370.
\newblock \doarXiv{2106.04370}

\bibitem[{{Ramsey}(1990)}]{1990ASPC....9..195R}
{Ramsey}, L.~W. 1990, in Astronomical Society of the Pacific Conference Series,
  Vol.~9, Cool Stars, Stellar Systems, and the Sun, ed. G.~{Wallerstein}, 195

\bibitem[{Rasmussen \& Williams(2006)}]{books/lib/RasmussenW06}
Rasmussen, C.~E., \& Williams, C. K.~I. 2006, Gaussian processes for machine
  learning., Adaptive computation and machine learning (MIT Press), I--XVIII,
  1--248

\bibitem[{Rodrigues(2008)}]{PhysRevD.77.023534}
Rodrigues, D.~C. 2008, Phys. Rev. D, 77, 023534,
  \dodoi{10.1103/PhysRevD.77.023534}

\bibitem[{{Rosenfield} {et~al.}(2016){Rosenfield}, {Marigo}, {Girardi},
  {Dalcanton}, {Bressan}, {Williams}, \& {Dolphin}}]{2016ApJ...822...73R}
{Rosenfield}, P., {Marigo}, P., {Girardi}, L., {et~al.} 2016, \apj, 822, 73,
  \dodoi{10.3847/0004-637X/822/2/73}

\bibitem[{{Rosenfield} {et~al.}(2014){Rosenfield}, {Marigo}, {Girardi},
  {Dalcanton}, {Bressan}, {Gullieuszik}, {Weisz}, {Williams}, {Dolphin}, \&
  {Aringer}}]{2014ApJ...790...22R}
---. 2014, \apj, 790, 22, \dodoi{10.1088/0004-637X/790/1/22}

\bibitem[{{Rubin} \& {Ford}(1970)}]{1970ApJ...159..379R}
{Rubin}, V.~C., \& {Ford}, W.~Kent, J. 1970, \apj, 159, 379,
  \dodoi{10.1086/150317}

\bibitem[{{Saio} {et~al.}(2015){Saio}, {Wood}, {Takayama}, \&
  {Ita}}]{2015MNRAS.452.3863S}
{Saio}, H., {Wood}, P.~R., {Takayama}, M., \& {Ita}, Y. 2015, \mnras, 452,
  3863, \dodoi{10.1093/mnras/stv1587}

\bibitem[{Samus' {et~al.}(2017)Samus', Kazarovets, Durlevich, Kireeva, \&
  Pastukhova}]{Samus2017}
Samus', N.~N., Kazarovets, E.~V., Durlevich, O.~V., Kireeva, N.~N., \&
  Pastukhova, E.~N. 2017, Astronomy Reports, 61, 80,
  \dodoi{10.1134/S1063772917010085}

\bibitem[{{S{\'a}nchez-S{\'a}ez} {et~al.}(2021){S{\'a}nchez-S{\'a}ez}, {Lira},
  {Mart{\'\i}}, {S{\'a}nchez-Pi}, {Arredondo}, {Bauer}, {Bayo},
  {Cabrera-Vives}, {Donoso-Oliva}, {Est{\'e}vez}, {Eyheramendy}, {F{\"o}rster},
  {Hern{\'a}ndez-Garc{\'\i}a}, {Mu{\~n}oz Arancibia}, {P{\'e}rez-Carrasco},
  {Sep{\'u}lveda}, \& {Vergara}}]{2021arXiv210607660S}
{S{\'a}nchez-S{\'a}ez}, P., {Lira}, H., {Mart{\'\i}}, L., {et~al.} 2021, arXiv
  e-prints, arXiv:2106.07660.
\newblock \doarXiv{2106.07660}

\bibitem[{{Sargent} {et~al.}(2011){Sargent}, {Srinivasan}, \&
  {Meixner}}]{2011ApJ...728...93S}
{Sargent}, B.~A., {Srinivasan}, S., \& {Meixner}, M. 2011, \apj, 728, 93,
  \dodoi{10.1088/0004-637X/728/2/93}

\bibitem[{{Scaringi} {et~al.}(2015){Scaringi}, {Maccarone}, {Kording},
  {Knigge}, {Vaughan}, {Marsh}, {Aranzana}, {Dhillon}, \&
  {Barros}}]{2015SciA....1E0686S}
{Scaringi}, S., {Maccarone}, T.~J., {Kording}, E., {et~al.} 2015, Science
  Advances, 1, e1500686, \dodoi{10.1126/sciadv.1500686}

\bibitem[{{Selam, S. O.}(2004)}]{selam}
{Selam, S. O.} 2004, A\&A, 416, 1097, \dodoi{10.1051/0004-6361:20034578}

\bibitem[{{Shappee} {et~al.}(2014){Shappee}, {Prieto}, {Grupe}, {Kochanek},
  {Stanek}, {De Rosa}, {Mathur}, {Zu}, {Peterson}, {Pogge}, {Komossa}, {Im},
  {Jencson}, {Holoien}, {Basu}, {Beacom}, {Szczygie{\l}}, {Brimacombe},
  {Adams}, {Campillay}, {Choi}, {Contreras}, {Dietrich}, {Dubberley},
  {Elphick}, {Foale}, {Giustini}, {Gonzalez}, {Hawkins}, {Howell}, {Hsiao},
  {Koss}, {Leighly}, {Morrell}, {Mudd}, {Mullins}, {Nugent}, {Parrent},
  {Phillips}, {Pojmanski}, {Rosing}, {Ross}, {Sand}, {Terndrup}, {Valenti},
  {Walker}, \& {Yoon}}]{2014ApJ...788...48S}
{Shappee}, B.~J., {Prieto}, J.~L., {Grupe}, D., {et~al.} 2014, \apj, 788, 48,
  \dodoi{10.1088/0004-637X/788/1/48}

\bibitem[{{Skiff}(2014)}]{2014yCat....1.2023S}
{Skiff}, B.~A. 2014, VizieR Online Data Catalog, B/mk

\bibitem[{{Skowron} {et~al.}(2019){Skowron}, {Skowron}, {Mr{\'o}z}, {Udalski},
  {Pietrukowicz}, {Soszy{\'n}ski}, {Szyma{\'n}ski}, {Poleski}, {Koz{\l}owski},
  {Ulaczyk}, {Rybicki}, \& {Iwanek}}]{2019Sci...365..478S}
{Skowron}, D.~M., {Skowron}, J., {Mr{\'o}z}, P., {et~al.} 2019, Science, 365,
  478, \dodoi{10.1126/science.aau3181}

\bibitem[{{Skrutskie} {et~al.}(2006){Skrutskie}, {Cutri}, {Stiening},
  {Weinberg}, {Schneider}, {Carpenter}, {Beichman}, {Capps}, {Chester},
  {Elias}, {Huchra}, {Liebert}, {Lonsdale}, {Monet}, {Price}, {Seitzer},
  {Jarrett}, {Kirkpatrick}, {Gizis}, {Howard}, {Evans}, {Fowler}, {Fullmer},
  {Hurt}, {Light}, {Kopan}, {Marsh}, {McCallon}, {Tam}, {Van Dyk}, \&
  {Wheelock}}]{2006AJ....131.1163S}
{Skrutskie}, M.~F., {Cutri}, R.~M., {Stiening}, R., {et~al.} 2006, \aj, 131,
  1163, \dodoi{10.1086/498708}

\bibitem[{{Smith}(2004)}]{2004rrls.book.....S}
{Smith}, H.~A. 2004, {RR Lyrae Stars}

\bibitem[{{Smith}(2012)}]{2012JAVSO..40..327S}
---. 2012, \jaavso, 40, 327

\bibitem[{{Smith} {et~al.}(2011){Smith}, {Gehrz}, {Campbell}, {Kassis}, {Le
  Mignant}, {Kuluhiwa}, \& {Filippenko}}]{2011MNRAS.418.1959S}
{Smith}, N., {Gehrz}, R.~D., {Campbell}, R., {et~al.} 2011, \mnras, 418, 1959,
  \dodoi{10.1111/j.1365-2966.2011.19614.x}

\bibitem[{{Spina} {et~al.}(2016){Spina}, {Mel{\'e}ndez}, {Karakas},
  {Ram{\'\i}rez}, {Monroe}, {Asplund}, \& {Yong}}]{2016A&A...593A.125S}
{Spina}, L., {Mel{\'e}ndez}, J., {Karakas}, A.~I., {et~al.} 2016, \aap, 593,
  A125, \dodoi{10.1051/0004-6361/201628557}

\bibitem[{{Srinivasan} {et~al.}(2011){Srinivasan}, {Sargent}, \&
  {Meixner}}]{2011A&A...532A..54S}
{Srinivasan}, S., {Sargent}, B.~A., \& {Meixner}, M. 2011, \aap, 532, A54,
  \dodoi{10.1051/0004-6361/201117033}

\bibitem[{{Srinivasan} {et~al.}(2010){Srinivasan}, {Sargent}, {Matsuura},
  {Meixner}, {Kemper}, {Tielens}, {Volk}, {Speck}, {Woods}, {Gordon},
  {Marengo}, \& {Sloan}}]{2010A&A...524A..49S}
{Srinivasan}, S., {Sargent}, B.~A., {Matsuura}, M., {et~al.} 2010, \aap, 524,
  A49, \dodoi{10.1051/0004-6361/201014991}

\bibitem[{{Stancliffe} {et~al.}(2004){Stancliffe}, {Tout}, \&
  {Pols}}]{2004MNRAS.352..984S}
{Stancliffe}, R.~J., {Tout}, C.~A., \& {Pols}, O.~R. 2004, \mnras, 352, 984,
  \dodoi{10.1111/j.1365-2966.2004.07987.x}

\bibitem[{{Storey-Fisher} {et~al.}(2021){Storey-Fisher}, {Huertas-Company},
  {Ramachandra}, {Lanusse}, {Leauthaud}, {Luo}, {Huang}, \&
  {Prochaska}}]{2021arXiv210502434S}
{Storey-Fisher}, K., {Huertas-Company}, M., {Ramachandra}, N., {et~al.} 2021,
  arXiv e-prints, arXiv:2105.02434.
\newblock \doarXiv{2105.02434}

\bibitem[{{Templeton} {et~al.}(2005){Templeton}, {Mattei}, \&
  {Willson}}]{2005AJ....130..776T}
{Templeton}, M.~R., {Mattei}, J.~A., \& {Willson}, L.~A. 2005, \aj, 130, 776,
  \dodoi{10.1086/431740}

\bibitem[{{Udalski} {et~al.}(2008){Udalski}, {Szymanski}, {Soszynski}, \&
  {Poleski}}]{2008AcA....58...69U}
{Udalski}, A., {Szymanski}, M.~K., {Soszynski}, I., \& {Poleski}, R. 2008,
  \actaa, 58, 69.
\newblock \doarXiv{0807.3884}

\bibitem[{{Udalski} {et~al.}(2015){Udalski}, {Szyma{\'n}ski}, \&
  {Szyma{\'n}ski}}]{2015AcA....65....1U}
{Udalski}, A., {Szyma{\'n}ski}, M.~K., \& {Szyma{\'n}ski}, G. 2015, \actaa, 65,
  1.
\newblock \doarXiv{1504.05966}

\bibitem[{{Ueta} \& {Meixner}(2003)}]{2003ApJ...586.1338U}
{Ueta}, T., \& {Meixner}, M. 2003, \apj, 586, 1338, \dodoi{10.1086/367818}

\bibitem[{{Uttenthaler} {et~al.}(2016){Uttenthaler}, {Greimel}, \&
  {Templeton}}]{2016AN....337..293U}
{Uttenthaler}, S., {Greimel}, R., \& {Templeton}, M. 2016, Astronomische
  Nachrichten, 337, 293, \dodoi{10.1002/asna.201512296}

\bibitem[{van Dokkum {et~al.}(2018)van Dokkum, Danieli, Cohen, Merritt,
  Romanowsky, Abraham, Brodie, Conroy, Lokhorst, Mowla, O'Sullivan, \&
  Zhang}]{vanDokkum2018}
van Dokkum, P., Danieli, S., Cohen, Y., {et~al.} 2018, Nature, 555, 629,
  \dodoi{10.1038/nature25767}

\bibitem[{{van Loon} {et~al.}(2005){van Loon}, {Cioni}, {Zijlstra}, \&
  {Loup}}]{2005A&A...438..273V}
{van Loon}, J.~T., {Cioni}, M. R.~L., {Zijlstra}, A.~A., \& {Loup}, C. 2005,
  \aap, 438, 273, \dodoi{10.1051/0004-6361:20042555}

\bibitem[{{Venuti} {et~al.}(2021){Venuti}, {Cody}, {Rebull}, {Beccari},
  {Irwin}, {Thanvantri}, {Howell}, \& {Barentsen}}]{2021AJ....162..101V}
{Venuti}, L., {Cody}, A.~M., {Rebull}, L.~M., {et~al.} 2021, \aj, 162, 101,
  \dodoi{10.3847/1538-3881/ac0536}

\bibitem[{{Villar} {et~al.}(2021){Villar}, {Cranmer}, {Berger}, {Contardo},
  {Ho}, {Hosseinzadeh}, \& {Yao-Yu Lin}}]{2021arXiv210312102V}
{Villar}, V.~A., {Cranmer}, M., {Berger}, E., {et~al.} 2021, arXiv e-prints,
  arXiv:2103.12102.
\newblock \doarXiv{2103.12102}

\bibitem[{Villar {et~al.}(2020)Villar, Hosseinzadeh, Berger, Ntampaka, Jones,
  Challis, Chornock, Drout, Foley, Kirshner, Lunnan, Margutti, Milisavljevic,
  Sanders, Pan, Rest, Scolnic, Magnier, Metcalfe, Wainscoat, \&
  Waters}]{Villar_2020}
Villar, V.~A., Hosseinzadeh, G., Berger, E., {et~al.} 2020, The Astrophysical
  Journal, 905, 94, \dodoi{10.3847/1538-4357/abc6fd}

\bibitem[{{Wachter} {et~al.}(2008){Wachter}, {Winters}, {Schr{\"o}der}, \&
  {Sedlmayr}}]{2008A&A...486..497W}
{Wachter}, A., {Winters}, J.~M., {Schr{\"o}der}, K.~P., \& {Sedlmayr}, E. 2008,
  \aap, 486, 497, \dodoi{10.1051/0004-6361:200809893}

\bibitem[{{Wang} \& {Chen}(2019)}]{2019ApJ...877..116W}
{Wang}, S., \& {Chen}, X. 2019, \apj, 877, 116,
  \dodoi{10.3847/1538-4357/ab1c61}

\bibitem[{{Wenger, M.} {et~al.}(2000){Wenger, M.}, {Ochsenbein, F.}, {Egret,
  D.}, {Dubois, P.}, {Bonnarel, F.}, {Borde, S.}, {Genova, F.}, {Jasniewicz,
  G.}, {Lalo\"e, S.}, {Lesteven, S.}, \& {Monier, R.}}]{simbad}
{Wenger, M.}, {Ochsenbein, F.}, {Egret, D.}, {et~al.} 2000, Astron. Astrophys.
  Suppl. Ser., 143, 9, \dodoi{10.1051/aas:2000332}

\bibitem[{{Whitelock} {et~al.}(1994){Whitelock}, {Menzies}, {Feast}, {Marang},
  {Carter}, {Roberts}, {Catchpole}, \& {Chapman}}]{1994MNRAS.267..711W}
{Whitelock}, P., {Menzies}, J., {Feast}, M., {et~al.} 1994, \mnras, 267, 711,
  \dodoi{10.1093/mnras/267.3.711}

\bibitem[{{Winters} {et~al.}(2003){Winters}, {Le Bertre}, {Jeong}, {Nyman}, \&
  {Epchtein}}]{2003A&A...409..715W}
{Winters}, J.~M., {Le Bertre}, T., {Jeong}, K.~S., {Nyman}, L.~{\r{A}}., \&
  {Epchtein}, N. 2003, \aap, 409, 715, \dodoi{10.1051/0004-6361:20031110}

\bibitem[{{Wood} {et~al.}(2004){Wood}, {Olivier}, \&
  {Kawaler}}]{2004ApJ...604..800W}
{Wood}, P.~R., {Olivier}, E.~A., \& {Kawaler}, S.~D. 2004, \apj, 604, 800,
  \dodoi{10.1086/382123}

\bibitem[{{Xue} {et~al.}(2016){Xue}, {Jiang}, {Gao}, {Liu}, {Wang}, \&
  {Li}}]{2016ApJS..224...23X}
{Xue}, M., {Jiang}, B.~W., {Gao}, J., {et~al.} 2016, \apjs, 224, 23,
  \dodoi{10.3847/0067-0049/224/2/23}

\bibitem[{{Yao} {et~al.}(2017){Yao}, {Liu}, {Deng}, {de Grijs}, \&
  {Matsunaga}}]{2017ApJS..232...16Y}
{Yao}, Y., {Liu}, C., {Deng}, L., {de Grijs}, R., \& {Matsunaga}, N. 2017,
  \apjs, 232, 16, \dodoi{10.3847/1538-4365/aa88a9}

\bibitem[{{Zacharias} {et~al.}(2015){Zacharias}, {Finch}, {Subasavage},
  {Bredthauer}, {Crockett}, {Divittorio}, {Ferguson}, {Harris}, {Harris},
  {Henden}, {Kilian}, {Munn}, {Rafferty}, {Rhodes}, {Schultheiss}, {Tilleman},
  \& {Wieder}}]{2015AJ....150..101Z}
{Zacharias}, N., {Finch}, C., {Subasavage}, J., {et~al.} 2015, \aj, 150, 101,
  \dodoi{10.1088/0004-6256/150/4/101}

\bibitem[{Zgirski {et~al.}(2017)Zgirski, Gieren, Pietrzy{\'{n}}ski, Karczmarek,
  Gorski, Wielgorski, Narloch, Graczyk, Kudritzki, \& Bresolin}]{Zgirski_2017}
Zgirski, B., Gieren, W., Pietrzy{\'{n}}ski, G., {et~al.} 2017, The
  Astrophysical Journal, 847, 88, \dodoi{10.3847/1538-4357/aa88c4}

\bibitem[{{Zha} {et~al.}(2019){Zha}, {Leung}, {Suzuki}, \&
  {Nomoto}}]{2019ApJ...886...22Z}
{Zha}, S., {Leung}, S.-C., {Suzuki}, T., \& {Nomoto}, K. 2019, \apj, 886, 22,
  \dodoi{10.3847/1538-4357/ab4b4b}

\bibitem[{Zhang {et~al.}(2017)Zhang, Childress, Davis, Karpenka, Lidman,
  Schmidt, \& Smith}]{10.1093/mnras/stx1600}
Zhang, B.~R., Childress, M.~J., Davis, T.~M., {et~al.} 2017, Monthly Notices of
  the Royal Astronomical Society, 471, 2254, \dodoi{10.1093/mnras/stx1600}

\bibitem[{Zhang \& Bloom(2021)}]{10.1093/mnras/stab1248}
Zhang, K., \& Bloom, J.~S. 2021, Monthly Notices of the Royal Astronomical
  Society, 505, 515, \dodoi{10.1093/mnras/stab1248}

\bibitem[{Zhang \& Zou(2018)}]{Zhang_2018}
Zhang, R., \& Zou, Q. 2018, Journal of Physics: Conference Series, 1061,
  012012, \dodoi{10.1088/1742-6596/1061/1/012012}

\bibitem[{{Zubko} {et~al.}(1996){Zubko}, {Mennella}, {Colangeli}, \&
  {Bussoletti}}]{1996MNRAS.282.1321Z}
{Zubko}, V.~G., {Mennella}, V., {Colangeli}, L., \& {Bussoletti}, E. 1996,
  \mnras, 282, 1321, \dodoi{10.1093/mnras/282.4.1321}

\end{thebibliography}
\bibliographystyle{aasjournal}

\end{document}